\documentclass[a4paper,11pt]{article}

\usepackage{jheppub}
\usepackage{amsfonts}
\usepackage{amssymb}
\usepackage{slashed}
\usepackage{bm}

\def\sss{\scriptscriptstyle}
\def\Re{{\rm Re}}

\def \cf {C_F}
\def \nc {N_c}
\def \ca {C_A}
\def \nf {N_f}

\def \bp {\mathbf{p}}

\newcommand{\Tint}[1]{{\hbox{$\sum$}\!\!\!\!\!\!\!\int\,}_{\!\!\!\!\raise-0.9ex\hbox{$\scriptstyle{#1}$}}}

\def\siml{{\ \lower-1.2pt\vbox{\hbox{\rlap{$<$}\lower6pt\vbox{\hbox{$\sim$}}}}\ }}
\def\simg{{\ \lower-1.2pt\vbox{\hbox{\rlap{$>$}\lower6pt\vbox{\hbox{$\sim$}}}}\ }}

\def\naBla{{\bm \nabla}}

\def \bp {\mathbf{p}}

\def \als {\alpha_{\mathrm{s}}}

\def \m2   {\mu^{2 \epsilon}}

\newcommand{\MS}{{\overline{\rm MS}}}
\def\siml{{\ \lower-1.2pt\vbox{\hbox{\rlap{$<$}\lower6pt\vbox{\hbox{$\sim$}}}}\ }}
\def\simg{{\ \lower-1.2pt\vbox{\hbox{\rlap{$>$}\lower6pt\vbox{\hbox{$\sim$}}}}\ }}

\def\nn {\nonumber}

\def \bff {\mathbf{F}}

\def\qp {q_\perp}

\def\bqp {\mathbf{q}_\perp}
\def\bh {\mathbf{h}}

\def\p{{\bm p}}
\def\ph{\hat{\bm p}}
\def\q{{\bm q}}
\def\k{{\bm k}}
\def\kh{\hat{\bm k}}
\def\v{{\bm v}}
\def\x{{\bm x}}
\def\F{{\bm F}}
\def\n{{\bm n}}
\def \OO {\mathcal{O}}
\def\Fext{\F_{\rm ext}}
\def\ij{{i \cdots j}}
\def\lra{\leftrightarrow}
\def\onetwo{{1\lra2}}
\def\twotwo{{2\lra2}}
\def\GammaC{\Gamma^{\mathrm{conv}}}
\def\C{{\cal C}}
\def\M{{\mathcal M}}
\def\Ctt{\mathcal{C}^{\twotwo}}
\def\Cot{\mathcal{C}^{\onetwo}}
\def\S{{\cal S}}

\def\pressure{{\cal P}}
\def\u{{\bm u}}
\def\j{{\bm j}}

\def\degen{\nu}
\def\dpslash{\frac{d^3\p}{(2\pi)^3} \>}
\def\half{{\textstyle{\frac 12}}}
\def\LO {\mathrm{LO}}
\def\NLO {\mathrm{NLO}}
\def\AMY{\mathrm{AMY}}
\def\ql {\hat{q}_{\sss L}}
\def\qhat {\hat{q}}
\def\loss{diffusion }

\def\gain{gain }
\def\lossshort{diff }
\def\lossf{conversion }
\def\gainf{gain }
\def\lossfshort{conv }
\def\mmf {m_\infty^2}
\def\mmg {M_\infty^2}
\def\md {m_{\sss D}}
\def\semi {\mathrm{semi}}
\def\st{\begin{equation}}
\def\stp{\end{equation}}
\def\Eq#1{Eq.~\eqref{#1}}
\def\A{{\mathcal C}}
\def\App#1{Appendix~\ref{#1}}
\def\Sec#1{Section~\ref{#1}}

\def\Fig#1{Fig.~\ref{#1}}

\title{QCD Shear Viscosity at (almost) NLO}
\author[1]{Jacopo Ghiglieri,}
\author[2]{Guy D. Moore,}
\author[3]{and Derek Teaney}

\affiliation[1]{Theoretical Physics Department, CERN, Geneva, Switzerland}
\affiliation[2]{Institut f\"ur Kernphysik, Technische Universit\"at Darmstadt\\
Schlossgartenstra{\ss}e 2, D-64289 Darmstadt, Germany}
\affiliation[3]{Department of Physics and Astronomy, Stony Brook University,\\
Stony Brook, New York 11794-3800, United States}
\emailAdd{jacopo.ghiglieri@cern.ch}
\emailAdd{guymoore@theorie.ikp.physik.tu-darmstadt.de}
\emailAdd{derek.teaney@stonybrook.edu}

\abstract{We compute the shear viscosity 
  of QCD with matter, including almost all
  next-to-leading order corrections -- that is, corrections suppressed
  by one power of $g$ relative to leading order.  We argue that the
  still missing terms are small.  The next-to-leading order
  corrections are large and
  bring $\eta/s$ down by more than a factor of 3 at physically
  relevant couplings.  The perturbative expansion is problematic
  even at $T \simeq 100$ GeV.  The largest next-to-leading order 
  correction to $\eta/s$ arises from modifications to the $\hat q$ parameter, 
  which determines the rate of transverse momentum diffusion. 
  We also explore quark number diffusion,
  and shear viscosity in pure-glue QCD and in QED.}


\preprint{CERN-TH-2018-009}

\begin{document}
\maketitle

\section{Introduction}
\label{sec:intro}

The original idea of the Quark-Gluon Plasma phase
\cite{Shuryak:1977ut,Polyakov:1978vu,Susskind:1979up} was that it would
consist of weakly-interacting, nearly-free quarks and gluons
(this assumption is implicit, for instance, in treatments of the
cosmological QCD phase transition \cite{Witten:1984rs}).
This picture was naive, since the QCD coupling varies only
logarithmically with scale \cite{Politzer:1973fx,Gross:1973id}, so the
coupling is in fact quite large at any achievable temperature.
Although thermodynamical quantities approach the expected
weak-coupling values rather quickly
\cite{Borsanyi:2013bia,Bazavov:2014pvz,Borsanyi:2011sw,Bazavov:2012jq},
this does not necessarily indicate weak coupling; even in the limit of
infinite coupling, analogue theories display 3/4 of the free theory
value for the pressure, for instance
\cite{Gubser:1998nz}.

Weak coupling would imply large transport coefficients, characterized
for instance by a large ratio of the shear viscosity to the entropy
density, $\eta/s \gg 1$.  In fact, leading-order (LO) perturbative
calculations of $\eta$
\cite{Arnold:2000dr,Arnold:2003zc}
find $\eta/s \sim 0.5$ for coupling values of physical relevance for
achievable temperatures.  Unfortunately, the presentation in
Ref.~\cite{Arnold:2003zc} has led to frequent misinterpretation of the
results, such as using the next-to-leading-log
(NLL) pocket formulae in regimes where the paper cautions that they
are not applicable.

But in any case, experimental results at RHIC
\cite{Adler:2003kt,Adams:2004bi}
and the LHC
\cite{ALICE:2011ab,Chatrchyan:2013nka,Aad:2014fla,ALICE:2016kpq}
indicate that the shear viscosity is even
smaller:  numerous authors have found that the experimental data on
angular correlations and other experimental measurables are fit very
well by relativistic, viscous hydrodynamics, but only if the shear
viscosity to entropy ratio is quite small,
$\eta/s \sim 0.1$--$0.3$
(see \cite{Heinz:2013th,Gale:2013da} for reviews).
This would indicate that $\eta/s$ is quite close to the value in
extremely strongly coupled theories with holographic duals
\cite{Policastro:2001yc,Kovtun:2003wp,Kovtun:2004de}. First attempts
at non-perturbative QCD determinations from the lattice, which require
a highly non-trivial analytical continuation, also point towards
small values
\cite{Nakamura:2004sy,Meyer:2007ic,Astrakhantsev:2015jta,%
  Astrakhantsev:2017nrs,Mages:2015rea,Pasztor:2016wxq,Pasztor:2018yae}.
So do FRG analyses, which also require analytical continuation and
other truncations \cite{Haas:2013hpa,Christiansen:2014ypa}.

So how should we understand this discrepancy with the weak-coupling
calculations?  The best way to address this question is to compute the
next-to-leading order (NLO) corrections to the shear viscosity.  We finally
have the technology to do so.  One key breakthrough, due to
Caron-Huot, was the development of a technique to understand how
particles are ``kicked'' transversely as they move through the plasma,
at next-to-leading order \cite{CaronHuot:2008ni}.
Then there was the development of sum-rule tools for next-to-leading
order longitudinal momentum diffusion, identity change, and collinear emission,
developed to study photon production
\cite{Ghiglieri:2013gia,Ghiglieri:2014kma}
and recently extended to treat jet energy modification at
subleading order
\cite{Ghiglieri:2015zma,Ghiglieri:2015ala}.
With rather modest modifications, we can apply this technology to
perform an ``almost'' next-to-leading order 
calculation of $\eta$, and of quark number
diffusion $D_q$, in a hot QCD plasma.

In the following sections, we will give a rather detailed explanation
of how one computes $\eta/s$ perturbatively in QCD, and of what is and
is not included in our ``almost'' NLO calculation.  But for the
impatient reader, we will give a short summary of the procedure, of
what is included and what is missing, why we think the remaining
``missing'' parts should give only a small correction, and of our
final results.

Shear viscosity describes the persistence of any anisotropy in the
stress tensor $T_{\mu\nu}$.  When a fluid flows in a nonuniform way,
such anisotropy constantly develops from the fluid flow, and
constantly disappears due to dissipative physics.  Shear viscosity
measures the inefficiency of that dissipation.  It can also be studied
by using random thermal fluctuations, through which $T_{\mu\nu}$
accidentally becomes anisotropic.  The fluctuation-dissipation theorem
says that the persistence of these fluctuations also determines the
shear viscosity.  These concepts are well defined in any theory with
well defined thermodynamics, whether or not the stress tensor can be
understood in terms of some ``particle'' degrees of freedom.

The perturbative picture is that the plasma is made up primarily of
quasiparticle excitations with momenta of the order of the temperature,
 and these are responsible for carrying the
stress tensor $T_{\mu\nu}$ of the plasma.  An anisotropic $T_{\mu\nu}$
arises when the quasiparticles are distributed anisotropically in
momentum space.  Their scattering relaxes
$T_{\mu\nu}$ towards its equilibrium value.  This description is
sufficient at both leading order and $\OO(g)$ NLO order.  The
challenge is to determine the exact form of the collision operator
which relaxes the particles towards equilibrium.  The LO
calculation \cite{Arnold:2003zc} requires two sorts of scattering
process, the $\twotwo$ scatterings with all hard ($\OO(T)$) external
particles and $\onetwo$ effective splitting processes between hard
participants.  There are two features in the calculation.  First,
there is the momentum a particle carries into a scattering and the
momentum it carries out.  Second, there is the effect of the momentum
which it ``dumps'' into the other particle in the scattering process.
While the first effect always makes the momentum distribution more
isotropic, the second effect can make it more or less isotropic,
depending on the relative angles of the participants.

To treat the problem at NLO, we need to find all new scattering
processes, and corrections to the already-considered processes, which
are suppressed by a single power of $g$.  No other corrections are
needed because the quasiparticle picture first needs amending at
$\OO(g^2)$ or higher.  As we shall show in detail, there are only a
few such $\OO(g)$ subleading effects.  First, the rate of soft
$\twotwo$ scattering is modified; this can be described as an
additional momentum-diffusion coefficient $\delta\qhat$.  This
modification, and an $\OO(g)$ correction to the medium corrections to
dispersion, also provide an $\OO(g)$ shift in the $\onetwo$ splitting
rate.  Next, the $\onetwo$ splitting rate must be corrected wherever
one participant becomes ``soft'' ($p \sim gT$) or when the opening
angle becomes less collinear.  And finally, the numerical
implementation of the LO scattering kernel \cite{Arnold:2003zc}
already resums a small amount of these NLO effects, requiring a
subtraction (or ``counterterm'') to $\delta \qhat$ and
$\delta \ql$ (longitudinal momentum diffusion).

We are able to give a relatively simple determination of these effects
by the use of light-cone techniques.  Unfortunately, these methods
typically keep track of the incoming and outgoing momentum of a
particle, but lose track of the momentum which it transfers to the
other participants.  This momentum transfer also affects the departure
from equilibrium of the other particle or particles which receive the
momentum, an effect which we will fail to account for at NLO.
Therefore our treatment is only ``almost'' NLO.  However we compute
the importance of this effect in the leading-order  case and
use it to make an estimate for this incomplete treatment.  The
associated errors turn out to be small, much smaller than the
difference between LO and NLO, and therefore presumably smaller than
still-uncomputed NNLO effects.

\begin{figure}
	\begin{center}
		\includegraphics[width=0.49\textwidth]{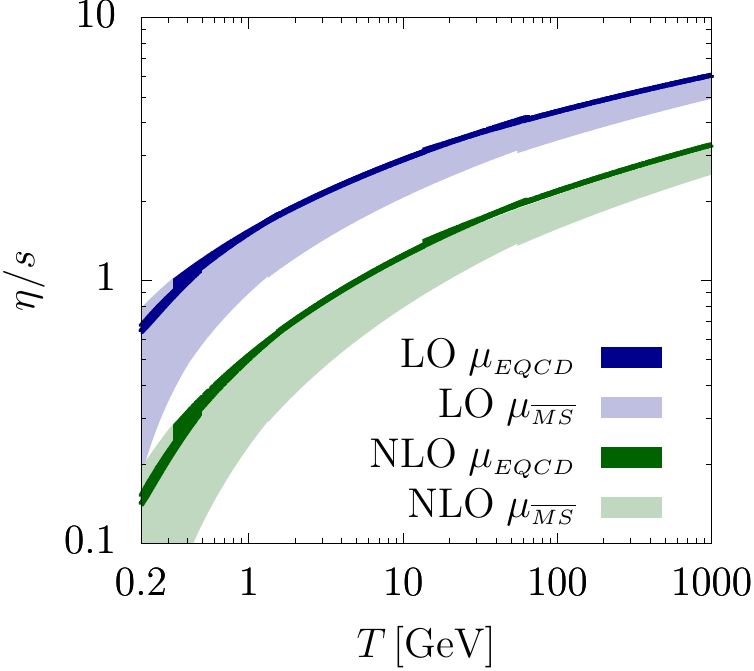}
      \hfill
		\includegraphics[width=0.49\textwidth]{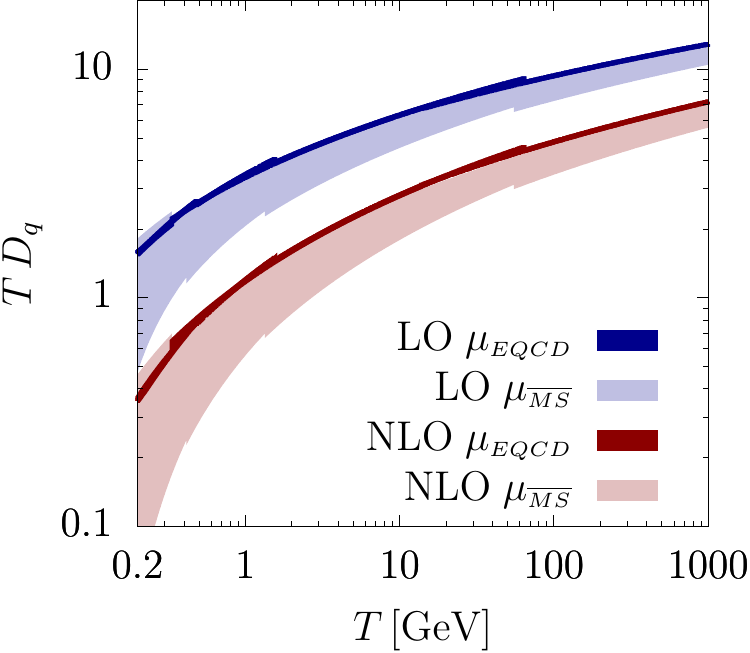}
	\end{center}
   \caption{(a) The shear viscosity to entropy density ratio and (b) baryon number diffusion coefficient as a
          function of temperature, at leading-order (LO) 
           and at next-to-leading order (NLO)  for
          different choices of the running coupling. 
          The solid band
          fixes the coupling using the two-loop EQCD value with
          $\mu_{EQCD}=(2.7\leftrightarrow 4\pi) T$, while the
          shaded band 
          uses the standard $\overline{\mathrm{MS}}$
        two-loop coupling with $\mu_{\overline{\mathrm{MS}}}=(\pi \leftrightarrow 4\pi)T$ ;  the corresponding values of $\alpha_s(T)$ are presented in 
        \Fig{fig_grun}. The dominant NLO correction arises from NLO modifications of $\hat{q}$ as is 
        illustrated  \Fig{fig_visc_md}. The uncertainty arising from gain terms which are only estimated (and not computed) is shown in \Fig{fig_run}.
        \label{fig:intro} }
\end{figure}

Our main results are presented in Section \ref{sec_results}, but we
will present one ``summary'' result right away in Figure
\ref{fig:intro}.  The figure shows the ratio of the shear viscosity to
the entropy density, computed at LO and NLO.  The temperature enters
in the choice of renormalized coupling and the number of quark species
(there are slight discontinuities where we cross quark-number
``thresholds'').  The solid thinner band represents our ``best
estimate'' based on 2-loop renormalization group flow from the
$Z$-pole and the coupling fixed via the EQCD choice of Laine and
Schr\"oder \cite{Laine:2005ai}. The renormalization
uncertainty is estimated by varying the scale $\mu_{EQCD}$ over
the range $\mu_{EQCD}=(2.7\leftrightarrow4\pi) T$.
The wider bands represent fixing $g^2$ from the
scale $\pi T$ to $4\pi T$ with the standard $\MS$
approach, to indicate the importance of the
renormalization uncertainty.
The plot shows that next-to-leading order corrections lower the shear
viscosity by a factor of two at high temperatures $T\sim 1000\,{\rm GeV}$, and
by a factor of four for physically relevant temperatures,
$T \sim 250\,{\rm MeV}$.  This large change
is suggestive that the true value of $\eta/s$ is smaller than the
leading-order perturbative estimate, but it also signals severe
convergence problems in the perturbative expansion, even for
surprisingly large temperatures or, equivalently, small values
of $g$.  The figure also shows the analogous result for the (light) quark
diffusion coefficient, which displays very similar coupling dependence.
We present more results and discussion in later sections, but we point
out now that the largest NLO correction arises from NLO modifications
of $\hat{q}$  -- see \Fig{fig_visc_md}. 
Accurate fits of or NLO results for $\eta/s$ as a 
function of coupling are provided in an appendix.

Having finished a quick summary of the problem, our approach, and our
main conclusions, we now summarize the content
of the remainder of the paper.  In Section \ref{sec:ingredients} we
review the definition of transport coefficients and their calculation
within the kinetic theory of Arnold, Moore, and Yaffe
\cite{Arnold:2002zm}.  In Section \ref{sec_nlo} we show how to
interpret parts of the leading-order calculation in terms of
transverse and longitudinal momentum diffusion and of
identity-changing processes.  The NLO effects take the form of these
three effects, plus a shift in the rate of $\onetwo$ splittings, and
can therefore be efficiently included once we express the problem in
terms of these pieces.  We do this in Section \ref{sec_nlo2}, with
special attention to the ``overlap'' regions between these processes.
With all pieces available, we present the main results in Section
\ref{sec_results}.  We also decompose the NLO correction into the
respective pieces to see which are most influential.  Some technical
details, together with fits for our NLO
results as a function of the coupling, are postponed to the  appendices.

\section{Ingredients}
\label{sec:ingredients}

Let us start by briefly summarizing how the transport coefficients
we investigate are defined and how they have been computed to leading
order in the Effective Kinetic Theory (EKT). Transport coefficients
characterize a system's response to weak, slowly varying
inhomogeneities or external forces.  In the case of the viscosity, if
the flow velocity of the plasma is nonuniform,  then the
stress-energy tensor (which defines the flux of momentum density)
departs from its perfect fluid form.
In the local (Landau-Lifshitz) fluid rest frame at a point $x$,
the stress tensor, to first order in the velocity gradient,
has the form
\begin {equation}
    \langle T_{ij}(x) \rangle
    =
    \delta_{ij} \, \langle {\pressure} \rangle
    - \eta \sigma_{ij}
    - \zeta \, \delta_{ij} \, \nabla^l \, u_l \,,
	\quad \sigma_{ij}\equiv  \nabla_i \, u_j + \nabla_j \, u_i
		- {\textstyle \frac 23} \, \delta_{ij} \, \nabla^l \, u_l\,,
\label {eq:Tij}
\end {equation}
where the metric is the ``mostly-plus'' one,
 ${\pressure}$ is the equilibrium pressure associated with
the energy density $\langle T_{00}(x) \rangle=\epsilon$,
and the coefficients $\eta$ and $\zeta$ are known as
the shear and bulk viscosities, respectively.
The flow velocity $\u$ equals the momentum density divided by the
enthalpy density $\epsilon+\pressure=sT$.
We will only be concerned with the shear viscosity in this paper;
the bulk viscosity requires a more complicated analysis, which has been
carried out at leading order in \cite{Jeon:1994if,Jeon:1995zm} for
a scalar theory and in \cite{Arnold:2006fz} for a
gauge theory with massless quarks. The charm
contribution has been computed in \cite{Laine:2014hba}.
Additional coefficients such as $\tau_\pi$
would appear at higher order in the gradient expansion
\cite{Baier:2007ix,Bhattacharyya:2008jc,%
  Romatschke:2009kr,Haehl:2015pja},
but we leave their evaluation for a future investigation.

In the presence of further conserved global charges beyond four-momentum,
such as baryon or lepton number, the associated charge density
$n \equiv j^0$ and current density $\j$
satisfy a diffusion equation,
\begin {equation}
    \langle {\j} \rangle
    = -D \> \naBla \langle n \rangle \,,
\label {eq:jia}
\end {equation}
in the local (Landau-Lifshitz) rest frame of the medium.
The coefficient $D$ is called the diffusion constant.

When (some of) the diffusing species of excitations carry electric charge,
as is the case for baryon and lepton number, the diffusion constants for these
charged species determine
the electric conductivity $\sigma$ through an Einstein relation
(see Refs.~\cite{Arnold:2000dr,Arnold:2003zc}).
If the net number of each species of charge carriers is conserved, then
\begin {equation}
    \sigma = \sum_a \, e_a^2 \, D_a \, \frac {\partial n_a}{\partial \mu_a} \,,
\label {eq:einstein}
\end {equation}
where the sum runs over the different species or flavors of excitations
with $e_a$, $D_a$ and $\mu_a$ the corresponding electric charge, diffusion constant
and chemical potential, respectively.

These transport coefficients
 all find a field-theoretical definition through Kubo-type formulae
relating them to the zero-frequency (transport) limit of the spectral functions
of two-point correlators of the appropriate operators (the stress-energy tensor
or other conserved currents). However, for a leading- and next-to-leading-order
 perturbative evaluation, the diagrammatic approach that would result from a direct
 application of these Kubo formulae would require cumbersome resummations
 to all orders
 of many classes of sub-diagrams. It is thus more convenient to use the
 linearized version \cite{Arnold:2003zc}
of the EKT developed in \cite{Arnold:2002zm}.
Solving the linearized theory automatically accounts for
the needed resummations. The leading-order equivalence between the diagrammatic and
kinetic approaches has been proven in \cite{Jeon:1994if} for a scalar theory and
in \cite{Aarts:2002tn,Aarts:2004sd,Aarts:2005vc,Gagnon:2006hi,Gagnon:2007qt}
for gauge theories.

The leading-order
EKT introduced in \cite{Arnold:2002zm} is given by this Boltzmann equation
\begin {equation}
    \left[
	\frac{\partial}{\partial t}
	+
	\v_\p \cdot \frac{\partial}{\partial \x}
	+
	\Fext^a \cdot \frac{\partial}{\partial \p}
    \right]
    f^a(\p,\x,t)
    =
    -C_a[f]\stackrel{\LO}{=}-\bigg(C_a^{\twotwo}[f] + C_a^{\onetwo}[f] \bigg) \, ,
\label {eq:Boltz}
\end {equation}
where $f^a(\p,\x,t)=dN^a/d^3\x d^3\p$ is the phase space distribution function for
the excitation (gluon, quark, antiquark ) of index $a$. The
leading-order collision operator encodes the contribution of tree-level $\twotwo$
scattering processes, with Hard-Loop resummed
propagators in the soft-sensitive channels,
as well as collinear, effective $\onetwo$ processes resumming the effect
of an infinite number of soft scatterings. Both processes
contribute to order $g^4T$ to the collision operator; a subset of $C_a^{\twotwo}[f]$
is logarithmically enhanced, $g^4T\ln(1/g)$, due to the aforementioned sensitivity
to the soft scale $gT$.
$C_a^{\twotwo}[f]$ and $C_a^{\onetwo}[f]$ are described in detail
in \cite{Arnold:2002zm,Arnold:2003zc}.

We now proceed to the gradient expansion of \Eq{eq:Boltz},
following the notation of \cite{Arnold:2000dr}.
Schematically, $f^a=f_0^a+f_0(1{\pm} f_0)[f_1^a+f_2^a+\ldots]$, where $f_0^a$
 is the equilibrium
distribution\footnote{
   We use capital letters for four-vectors, bold lowercase ones
for three-vectors and italic lowercase for the modulus of the
latter. We work in the ``mostly plus'' metric, so that
$P^2=-p_0^2+p^2$. The upper sign is for bosons, and the lower sign is
for fermions.  The full collision operator is $C_a$; the collision
operator linearized in the departure from equilibrium is $\C_a$.},
$f_0^a=(\exp(-\beta u^\mu P_\mu - q^a_{\alpha} \beta\mu)\mp 1)^{-1}$,
which  is determined by the Boltzmann equation at zeroth order in
the gradients, $C_a[f_0]=0$.
The inverse
temperature $\beta$, the chemical potential $\mu$,  and the flow velocity $u^\mu$ are
functions of $(t,\x)$ and obey the equations of ideal hydrodynamics.
We will consider $u^{i}(t,\x)$ and $\mu(t,\x)$ to be small perturbations
on top of an approximately homogeneous background. The charge
of species $a$ under conserved charge $Q_\alpha$ is $q_{\alpha}^a$, where $\alpha$ is a label for the
flavor symmetry of interest (i.e.\ quark number in our case).

Substituting  $f_0^a$ into the lefthand side of the Boltzmann
equation, \Eq{eq:Boltz},
yields a source 
for the first dissipative correction,
which (after using the hydrodynamic equations of motion)
is proportional to  the strains~\cite{Teaney:2009qa}
\begin {equation}
    X_\ij
    \equiv
    \begin {cases}
	\;\nabla_i \, \mu_\alpha \,, & \mbox{(diffusion),} \cr
\frac{1}{\sqrt{6}} \left( \nabla_i u_j + \nabla_j u_i
	- \frac{2}{3} \, \delta_{ij} \nabla \cdot \u \right) \,,
& \mbox{(shear viscosity),} \cr
    \end {cases}
\label {eq:Tdrive}
\end {equation}
depending on whether we are considering chemical
potential fluctuations (diffusion) or velocity fluctuations (shear viscosity).
The source takes the form
\begin{align}
\nonumber
  \S^a(\p)\equiv &  \beta^2 \S^{a}_{\ij}(\p) X_{\ij}\,, \\
   =& - \beta q^a f_0^a(p)[1{\pm}f_0^a(p)] \, I_\ij(\hat\p) X_{\ij}\, .
\label{eq:Boltzsource}
\end{align}
Here $q^a$ denotes the relevant conserved charge
carried by species $a$ associated with
the transport coefficient of interest,
\begin{equation}
q^a \equiv \begin {cases}

	q^a_\alpha \, , & \mbox{(diffusion),} \cr
	|\p| \, , & \mbox{(shear viscosity).} \cr
	\end {cases}
\label{eq:qcases}
\end{equation}
$I_\ij$ is the unique $\ell=1$ or $\ell=2$ rotationally covariant
tensor depending only on $\hat\p$,
\begin {equation}
    I_\ij(\hat \p)
    \equiv
    \begin {cases}
	\;\hat p_i \,, & \ell = 1 \mbox{ (conductivity/diffusion),} \cr
	\sqrt {\frac 32} \, (\hat p_i \hat p_j - {\frac 13} \delta_{ij}) \,,
	& \ell = 2 \mbox{ (shear viscosity).} \cr
    \end {cases}
\label {eq:Iij}
\end {equation}
The normalization on $I_\ij$ was chosen so that
\begin {equation}
   I_\ij(\hat\p) \, I_\ij(\hat\p) = 1,
\label {eq:Inorm}
\end {equation}
and more generally,
\begin {equation}
   I_\ij(\hat \p) \, I_\ij(\hat \k) = P_\ell^{}(\hat\p\cdot\hat\k) ,
\label {eq:II}
\end {equation}
where $P_\ell(x)$ is the $\ell$'th Legendre polynomial.

The linearized kinetic equation may be written compactly as
\begin{equation}
  \label{eq:Boltz1}
  \S^a(\p) = (\C f_1)^a(\p) \,,
\end{equation}
where $\C$ is a linearized collision operator defined  below.
To linear order the first dissipative correction
must be proportional to the driving term $X$, allowing us to define
the proportionality coefficient $\chi_\ij(\p)$,
\begin{equation}
  f_1^a(\p) \equiv \beta^2 X_{\ij} \, \chi_{\ij}^a(\p)
  \equiv \beta^2 X_\ij \, I_\ij(\hat\p) \: \chi^a(p) \,,
\label {eq:f1}
\end{equation}
where we have also used rotational invariance of the collision
operator (in the rest frame) to define a scalar proportionality
coefficient $\chi(p)$, which describes how the departure from
equilibrium varies as a function of the magnitude of the momentum.

The first-order transport coefficients are then obtained
from the kinetic-theory expressions for $T^{\mu\nu}$ and
for the conserved current $J^\mu_\alpha$ associated to the conserved
charge $Q_\alpha$, \textsl{i.e.},
\begin{equation}
	T^{\mu\nu}(\x,t)=\int\frac{d^3\p}{(2\pi)^3}\frac{p^\mu p^\nu}{p}\sum_a\nu_a
	f^a(\p,\x,t),\quad J_\alpha^\mu(\x,t)=\int\frac{d^3\p}{(2\pi)^3}\frac{p^\mu}
	{p}\sum_a\nu_a q^a_\alpha f^a(\p,\x,t),
	\label{tj}
\end{equation}
where $\nu_a$ is the spin and color degeneracy of the excitation $a$
($2\nc$ for quarks and antiquarks, $2(\nc^2-1)$ for gluons).
Upon inserting the first-order
deviation $f^1$ in these equations, the first-order coefficients
are easily recovered.
Hence, the solution of the first-order, linear \Eq{eq:Boltz1}
yields $\eta$ and $D$.

As in \cite{Arnold:2000dr,Arnold:2003zc},
we will solve \Eq{eq:Boltz1} and its NLO extension by means
of a variational method. To this end, we introduce the inner
product
\begin {equation}
\label{eq:inner_product}
    \Big( f,g \Big) \equiv
    \beta^3 \sum_a \, \degen_a \! \int\dpslash \, f^a(\p) \, g^a(\p) \,.
\end {equation}
The linearized collision operator $\C$ is symmetric with respect to
this inner product, and is given by variation  of the  quadratic form
\st
   \label{variation}
   (\C f_1)^a(\p) = \frac{(2\pi)^3}{\beta^3 \nu_a}
   \frac{\delta \,}{\delta f_1^a(\p) } \half (f_1, \C f_1) \, .
\stp
$\C$ is a positive semi-definite operator, and is strictly positive
definite in the $\ell {=} 1$ and $\ell{=}2$ channels relevant
for diffusion or shear viscosity. As we will show, some
NLO contributions are negative, so some care will be needed
in defining a positive definite $\mathcal{C}_\NLO$. Once that
is taken care of,
the linearized Boltzmann equation (\ref{eq:Boltz1}) at LO and
NLO is precisely the condition for maximizing the functional
\begin {equation}
    \beta^4 \mathcal{Q}[\chi]
    \equiv
    \Big( f_1, \S\Big)
    - \half \, \Big( f_1, \C \, f_1 \Big).
\label{eq:Q1}
\end{equation}
Note that the maximized $\mathcal{Q}$ determines the rate per volume at which work is
dissipated into heat; $\beta \mathcal{Q}$ then gives the rate per
volume of entropy production.
This structure is valid at LO and NLO, so we
have not explicitly labeled $\C$ and $f_1$ in that respect.
The strains $X_{\ij}$ may be pulled out of the integrals,  and then
rotational invariance of the measure and collision operator
guarantees that
\begin{align}
  T^4 \Big( f_1, \C f_1 \Big)  =& \: \frac{X^2}{2\ell + 1 }
  \Big( \chi_{\ij}, \C \, \chi_{\ij} \Big) , \label{rotinvar} \\
  T^4 \Big( f_1, \S \Big) =& \: \frac{X^2}{2\ell + 1 }
  \Big( \chi_{\ij}, S_{\ij} \Big) ,
  \label{rotinvar2}
\end{align}
where $X^2 = X_\ij \, X_\ij$.

The explicit forms of the source and LO collision parts of this quadratic
functional are
\begin {align}
   T^4 \Big( f_1, \S \Big)
    &{}=
     \frac{X^2}{2\ell +1}
    \sum_a \beta^2 \nu_a
    \int \dpslash
	f_0(p) \, [1\pm f_0(p)] \>
   q^a \, \chi^a(p) \,,
\label {eq:Sij}
\\
\noalign {\hbox {and}}
    \Big( f_1, \C_\LO \, f_1 \Big)
    &{}=
    \Big( f_1, \Ctt \, f_1 \Big) +
    \Big( f_1, \Cot \, f_1 \Big) \,.
\label {eq:Q2}
\end{align}
The $\twotwo$ contribution reads, after symmetrization
of the departures from equilibrium \cite{Arnold:2003zc}
\begin{align}
\nn \Big( f_1, \C^\twotwo \, f_1 \Big)
    \equiv&\,
    \frac {\beta^3}8
    \sum_{abcd}
    \int_{\p\k\p'\k'}
    \frac{\bigl| \M^{ab}_{cd}(\p,\k;\p',\k') \bigr|^2 }{(2p) (2k) (2p') (2k')}
	\, (2\pi)^4\,\delta^{(4)}(P{+}K{-}P'{-}K')\\
	& { }\times	
f^a_0(p) \, f^b_0(k) \, [1{\pm}f^c_0(p')] \, [1{\pm}f^d_0(k')]
	\,  \Bigl[
	    f_1^a(\p) + f_1^b(\k) -
	    f_1^c(\p') - f_1^d(\k')
	\Bigr]^2_{\strut} \,,
\label {eq:C22}
\end{align}
where $\int_\p$ is shorthand for $\int d^3\p/(2\pi)^3$ and
$\bigl| \M^{ab}_{cd}(\p,\k;\p',\k') \bigr|^2 $ is the matrix element
squared for the $ab\lra cd$ process, summed over all
spins polarizations and colors and Hard Thermal Loop (HTL) resummed
\cite{Braaten:1989mz,Frenkel:1989br}
in the IR-sensitive cases. A complete
list of these matrix elements appears in \cite{Arnold:2002zm,Arnold:2003zc}.
The $\onetwo$ contribution reads instead \cite{Arnold:2003zc}
\begin{align}
    \Big( f_1, \C^\onetwo \, f_1 \Big)
    \equiv&\,
       \frac{\beta^3}{2}
       \sum_{abc}
       \int d\Omega_{\hat \n}
       \int_0^\infty dp\int_0^p dk \;
       \gamma^a_{bc}(p;p-k,k) \;
\nonumber\\ &  \times
          f_0^a(p) \, [1 {\pm} f_0^b(p-k)] \, [1 {\pm} f_0^c(k)] \>
	\Bigl[
      f_1^a(p \hat \n) - f_1^b((p-k)\hat\n) - f_1^c(k\hat \n)
	\Bigr]^2 \,.
\label {eq:C12}
\end{align}
The splitting rate is given by
\begin{equation}
	\gamma^{a}_{bc}(p;p-k,k)=\frac{g^2d_{R_b}C_{R_b}}{64\pi^4}\left\{
\begin{array}{cc}
	\frac{p^4+k^4+(p-k)^4}{p^3k^3(p-k)^3}& g\leftrightarrow gg\\
	 \frac{p^2+(p-k)^2}{p^2(p-k)^2k^3} & q\leftrightarrow q g\\
	\frac{(p-k)^2+k^2}{(p-k)^2k^2p^3} & g\leftrightarrow q\bar q
\end{array}
\right.
\int\frac{d^2h}{(2\pi)^2}\; 2\bh \cdot \Re\, \bff_b(\bh)\,,
	\label{onetworate}
\end{equation}
where $\bh=\p\times\k$ is a transverse, two-dimensional vector
related to the transverse momentum picked up during the splitting
process. $d_{R_b}$ and $C_{R_b}$ are the dimension and quadratic Casimir
operator of the representation $R$ of the particle $b$.
 A complete leading-order treatment of collinear radiation must consistently
resum the effect of the many soft, transverse collisions
to account for the \emph{Landau-Pomeranchuk-Migdal}
(LPM) effect
\cite{Baier:1994bd,Baier:1996kr,Zakharov:1996fv,Zakharov:1997uu,Arnold:2002ja}.
This is achieved through the following integral equation for  $\bff_b(\bh)$:
\begin{eqnarray}
\nn2\bh&=&i\delta E(\bh)\bff_b(\bh)+\int\frac{d^2\qp}{(2\pi)^2}
\bar C(\qp)
\bigg\{\left(C_{R_b}-\frac{\ca}{2}\right)[\bff_b(\bh)-\bff_b(\bh-k\bqp)]\\
&&+\frac{\ca}{2}[
\bff_b(\bh)-\bff_b(\bh+p\bqp)]+\frac{\ca}{2}[
\bff_b(\bh)-\bff_b(\bh-(p-k)\bqp)]\bigg\}.
\label{defimplfull}
\end{eqnarray}
For the case of
$g\to q\bar q$, $C_{R_b}-\ca/2=\cf-\ca/2$ multiplies the term with $\bff_b(\bh-p\bqp)$ rather than
$\bff_b(\bh-k\bqp)$. The equation depends on two inputs,
$\bar C(\qp)$  and $\delta E(\bh,p,k)$. The former is the leading-order
 transverse scattering kernel $C(\qp)$ in units of the Casimir factor,
\cite{Arnold:2001ba,Aurenche:2002pd}
\begin{equation}
	\bar C(\qp)\equiv \frac{C_R(\qp)}{C_R}=\frac{g^2 T
	\md^2}{\qp^2(\qp^2+\md^2)}.
\label{Cbar}
\end{equation}
$\delta E$ is the energy difference between the initial and final collinear
particles. It reads
\begin{equation}
\label{defdeltaE}
\delta E(\bh)=\frac{h^2}{2pk(p-k)}+\frac{m^2_{\infty\,c}}{2k}
+\frac{m^2_{\infty\,b}}{2(p-k)}
-\frac{m^2_{\infty\,a}}{2p},
\end{equation}
where $m_{\infty\,a}^2$ is the asymptotic mass of the particle $a$. For gluons
$m_{\infty\,g}^2\equiv\mmg=\md^2/2$ (with $\md^2=g^2T^2/3(T_A+T_FN_f)$),
 for quarks $m_{\infty\,q}^2\equiv\mmf=\cf g^2T^2/4$.

The maximization of the functional~\eqref{eq:Q1} is then
carried out using a variational Ansatz of form
\begin{equation}
\label{Ansatz}
  \chi(p) = \sum_{m=1}^{K} c_m \phi^{(m)}(p) \,, \qquad
  \phi^{(m)}(p) = \frac{p^{\ell-1} (p/T)^{m-1}}{(1+p/T)^{K-3/2}} \,,
\end{equation}
which, when substituted into \Eq{eq:Q1}, transforms its extremization
into a matrix algebra problem \cite{Arnold:2000dr}.  Our choice of functional basis
is motivated by the need \cite{Arnold:2003zc} to allow infrared
behavior of form $\propto p^{\ell-1}$ and ultraviolet behavior of form
$\propto p^{\ell-1/2}$.  The
variational procedure is only guaranteed to converge to the right
answer as the functional basis becomes complete, but in practice we
see good convergence above 4 functions.  Later, in our numerical
results, we will use 6 basis functions, but our results shift by less
than $10^{-4}$ when we make the basis still larger.

\section{Reorganization of the LO quadratic functional}
\label{sec_nlo}

The effective kinetic theory introduced in \cite{Arnold:2002zm} has
been extended to next-to-leading order in \cite{Ghiglieri:2015ala}
 for the case where one follows
the evolution of a dilute set of high-energy particles of typical energy $E$
interacting with an
equilibrated medium at a temperature $T$ such that $\exp(-E/T)\ll 1$,
which is a sensible approximation for the evolution of the
leading partons in a jet. There, we found that a reorganization
of the form of the LO collision operator was necessary
to systematically compute
NLO corrections. Specifically, $\OO(g)$ NLO corrections can only occur
where one or more lines carry a soft $\OO(gT)$ momentum, because only
there do statistical functions give rise to a $1/g$ enhancement of
loop level effects.  But transport coefficients are only sensitive to
hard $\OO(T)$ momenta.  So NLO corrections only occur when there is a
momentum hierarchy within a diagram.  In such cases one can always
re-express the diagram as an effective process.  When the soft
particle is a gluon and does not change particle identity, the process
can be understood as giving rise to momentum \textsl{diffusion}; when
the exchanged particle is a quark and therefore changes quantum
numbers, it is a \textsl{conversion} process.  One can already isolate
such processes at the leading order.  Doing so will make it easier to
see how to incorporate NLO processes.

In the diffusion case, the action of the soft gluon exchange is to randomize (diffuse)
the momentum of the hard particles by small, $\OO(g)$ amounts that can
be described by a Fokker-Planck equation. The drag, longitudinal and
transverse momentum diffusion coefficients appearing in the Fokker-Planck equation can be defined
field-theoretically in terms of Wilson-line operators supported on light fronts,
which can in turn be evaluated in analytical form using the light-cone techniques
mentioned in the introduction. In the conversion
case, the soft quark exchange converts a hard quark (gluon) into a gluon (quark)
of the same momentum, up to $\OO(gT)$. Again, a light-front Wilson line definition
for the conversion rate was introduced in \cite{Ghiglieri:2015ala}, leading
to a simple closed form expression. At NLO, the diffusion and conversion rates
receive $\OO(g)$ corrections, which were computed in
\cite{CaronHuot:2008ni,Ghiglieri:2013gia,Ghiglieri:2015ala}. These,
together with corrections to the collinear $\onetwo$ rate and a new,
\emph{semi-collinear} process (which only contributes starting from NLO),
constitute the entirety of the NLO corrections to the EKT in the ``dilute-hard''
approximation appropriate for energy loss.

For computing the transport coefficients we will first show (again) how the effect
of soft-gluon exchange can be  reorganized
into a Fokker-Planck equation in \Sec{sec:softglue}.  However, in order to conserve
energy and momentum, the Fokker-Planck equation must be supplemented by
gain terms, which describe precisely how the momentum lost by a parton
in the bath is redistributed. This redistribution
of energy and momentum is unimportant for determining the energy loss, but plays
an essential role in determining the transport coefficients.
The computation of these gain terms is not amenable to an evaluation using light-cone techniques
 since more than one light-like particle is involved, and therefore computing the gain terms constitutes a major obstacle to
computing transport coefficients at NLO.  We will use the LO analysis
in this section to motivate a NLO Ansatz for the gain terms in \Sec{sec_nlo2}.
The treatment of soft fermion exchange and conversions is analogous and
will be discussed in \Sec{sec:softquark}.

\subsection{Soft gluon exchange}
\label{sec:softglue}

\begin{figure}[ht]
	\begin{center}
		\includegraphics[width=7cm]{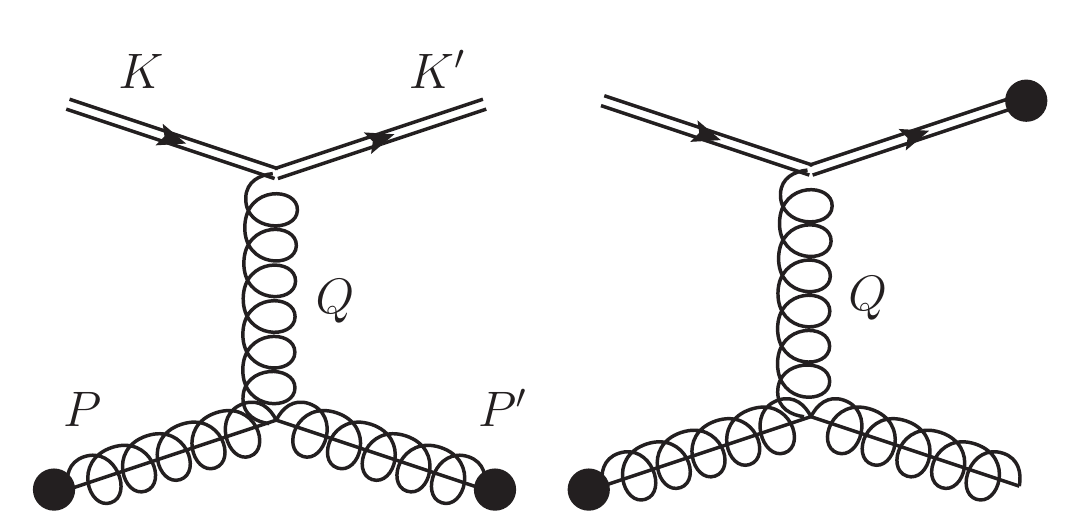}
	\end{center}
	\caption{Diagrammatic representation of the
\loss and \gain terms in soft gluon exchange processes. Double lines are
hard quarks or gluons and the intermediate gluon is soft. The blobs represent
   the insertion of the deviation from equilibrium $f_1(\p)$. The diagram
on the left is a \loss term, entering in $(f_1^g(\p)-f_1^g(\p'))^2$,
whereas the one on the right is a \gain term recording the correlations
   between momentum $\p$ and $\k$.}
   \label{fig_softgluon}
\end{figure}

We will now analyze soft gluon exchange shown in \Fig{fig_softgluon}.
Intuitively,
the effect of soft gluon exchange on the evolution
of the system can be summarized by
a Fokker-Planck equation. Anticipating the results of this section,
the Fokker-Planck collision kernel can be written
 \begin{equation}
    \label{FPall}
	(\C_{\mathrm{\lossshort}}^{\twotwo} f_1)^a(\p) =
    -\frac{1}{2} \frac{\partial}{\partial p^i} \,
    \hat q^{ij}_a \,  f_0^a(p) (1 {\pm} f_0^a(p))
    \frac{\partial f_1^a(\p)}{\partial p^j}  + \mbox{gain-terms} \, ,
 \end{equation}
where
\begin{equation}
  \label{Eq:qhatLT}
  \hat q^{ij}_a = \hat q_{a,L}  \hat p^i \hat p^j
  +  \tfrac{1}{2} \hat q_a \left(\delta^{ij} - \hat p^i \hat p^j \right)
\end{equation}
records the momentum diffusion parallel and perpendicular
to the particle's momentum through $\hat q_{L}$ and $\hat q$ respectively.
The gain terms are necessary to conserve energy an momentum, and
record how the energy lost by a parton with momentum $\k$ is redistributed
to particles with momentum $\p$.
The gain terms will take the following form:
\begin{equation}
   \label{FPgain}
   \mbox{gain-terms}  = 
   \frac{1}{2} \frac{\partial}{\partial p^i } \left(
   f_0^a(p) (1 {\pm} f_0^a(p))  \, \sum_{b} \nu_b \!\int_\k
   \A^{ij}_{ab} (\hat \p \cdot \hat \k) \, f_0^b(k) (1 \pm f_0^{b}(k))
   \frac{\partial f_1^b(\k)}{ \partial k^j } \right) \, .
\end{equation}
Here the angular function $\A^{ij}_{ab}(\hat \p \cdot \hat \k)$
determines $\hat q^{ij}_a$
\st
 \label{qhatintegral}
 \hat q^{ij}_a =  \sum_b \nu_b \int_\k f_0^b(k) (1 \pm f_0^{b}(k)) \, \A_{ab}^{ij}(\hat \p \cdot\hat \k)  \,,
\stp
and its  explicit form given in \Eq{eq:qhatkernel}.
It is easily verified that energy and momentum are
conserved under the time evolution
$(\partial_t + v_\p\partial_\x) f(\p) = -(\C f_{1,{\rm diff}})(\p)$. A
simulation and discussion of a similar Fokker-Planck equation (with
the gain terms) is given in \cite{Hong:2010at}.

Now we will derive these equations by analyzing the $\twotwo$
collision integral with soft gluon exchange recorded in \Eq{eq:C22}
and illustrated \Fig{fig_softgluon}.
The relevant processes are $gg\lra gg$, $q_1q_2\lra q_1 q_2$
 (and similar ones where $q_1$ and/or
$q_2$ are replaced by their antiquarks),
$q_1q_1\lra q_1q_1$ and $\bar q_1\bar q_1\lra \bar q_1\bar q_1$,
 $q_1\bar q_1\lra q_1\bar q_1$ and finally
 $q_1g\lra q_1 g$ and $\bar q_1g\lra \bar q_1 g$.
The LO contribution from soft gluon exchange is obtained by expanding
\Eq{eq:C22} for $\omega,q\sim gT$, $p,k\sim T$, where $Q=(\omega,\q)=P'-P$ is
the  momentum exchange shown in \Fig{fig_softgluon}. In more detail,
the phase space integration in \Eq{eq:C22} is approximated by (see \App{sub_phase}) 
 \begin{align}
    \int_{PS} \equiv \int_{\p\k\p'\k'} (2\pi)^4 \delta(P + K - P' - K')
    \simeq  \int_{\p\k} \int \frac{d^4Q}{(2\pi)^4} \, 2\pi \delta(v_\p \cdot Q) \, 2\pi  \delta(v_\k \cdot Q) \, ,
    \label{eq:softps}
 \end{align}
 where $v_\p^{\mu}=(1,\hat\p)$ and $v_\k^{\mu}=(1,\hat\k)$ are light-like vectors in the direction
 of $\hat\p$ and $\hat\k$.
The $t$-channel matrix element in the soft approximation reads
\cite{Arnold:2002zm,Arnold:2003zc}
\begin{eqnarray}
	\label{softmatgluegeneric}
   \frac{\vert\mathcal{M}^{ab}_{ab}\vert^2_{\mathrm{soft\,g}\,t }}{(2p)^2 (2k)^2} 
   &=& \frac{\nu_a C_{R_a} \nu_b C_{R_b}g^4}{d_A} \left\vert G_{\mu\nu}^R(Q)v_\p^\mu v_\k^\nu\right\vert^2 \,.
	 \label{softmatglue}
 \end{eqnarray}
$G_{\mu\nu}^R(Q)$
is the retarded, HTL-resummed propagator
\cite{Braaten:1989mz,Frenkel:1989br} in Coulomb gauge (see App.~\ref{app_props}). 
For process with identical
particles in the initial or final state, the $u$-channel
exchange is equivalent in the soft limit. 
Finally, in a soft
(or \emph{diffusive}) expansion we may approximate the departures from
equilibrium appearing in \Eq{eq:C22}
\st
\left[ f_1^a(\p) + f_1^b(\k) - f_1^a(\p') - f_1^b(\k') \right]^2
=  \left[ q^i \frac{\partial f_1^a(\p) }{\partial p^i} -
q^j \frac{\partial f_1^b(\k) }{\partial k^j} \right]^2   .
\label{eq:fdiff}
\stp
With these approximations the $\twotwo$ collision operator in
a diffusive approximation takes the form
\st
\label{lossgain}
 (f_1, \C^{2 \rightarrow 2}_{\mathrm{\lossshort}} f_1)  \equiv
 \left. (f_1, \C^{2 \rightarrow 2}_{\mathrm{\lossshort}} f_1)\right|_{\rm loss}
 +
 \left. (f_1, \C^{2 \rightarrow 2}_{\mathrm{\lossshort}} f_1)\right|_{\rm gain} \, ,
 \stp
 where
 \st
 \label{FPloss}
 \left. (f_1, \C^{2 \rightarrow 2}_{\scriptscriptstyle\rm diff} f_1) \right|_{\rm loss}
= \frac{1}{2} \beta^3 \sum_{a} \nu_a \,
\int_\p  f_0^a(p)(1 \pm f_0^a(p)) \,  \hat q^{ij}_{a} \, \frac{\partial f_1^a(\p) }{\partial p^i} \frac{\partial f_1^a(\p) }{\partial p^j } \, ,
\stp
and the gain terms take the form
\st
\label{gain}
\left. (f_1, \Ctt_{\mathrm{\lossshort}} f_1)\right|_{\rm gain} \!\!
=  -\frac{\beta^3}{2}  \sum_{ab} \nu_a \nu_b \!\int_{\p\k} \!\!
f_0^a(p)(1 \pm f_{0}^a(p)) f^b_0(k) (1 \pm f_{0}^b(k))
 \, \A_{ab}^{ij} (\ph\cdot\kh) \frac{\partial f_1^a(\p) }{\partial p^i}
  \frac{\partial f_1^b(\k)}{\partial k^j},
\stp
where the angular function is
\st
\label{eq:qhatkernel}
\A_{ab}^{ij}(\ph\cdot\kh)
=  \frac{g^4 C_{R_a} C_{R_b}}{d_A} \int \frac{d^4Q}{(2\pi)^4}   |G_{\mu\nu}^R(Q) v_{\p}^\mu v_\k^{\nu}|^2  \,  2\pi \delta(v_\p\cdot Q) 2\pi\delta(v_\k \cdot Q) q^i q^j \, ,
\stp
and $\hat q^{ij}_a$ is given by \Eq{qhatintegral}.
Varying the quadratic functional $(f_1,
\C_{\mathrm{\lossshort}}f_1)$ according to \Eq{variation} we see the
Fokker-Planck evolution equations, Eqs.~\eqref{FPall} and
\eqref{FPgain}, emerge.

At a technical level, the loss terms arise when the deviations from equilibrium
are on the same side of the gluon exchange diagram, and their contribution to
the quadratic functional therefore involves
$(f_1(\p) - f_1(\p'))^2 \sim q^2 (f_1'(\p))^2$. This is illustrated by
the black dots in \Fig{fig_softgluon} (left). The gain terms describe the correlation
between the momenta \emph{across} the exchange diagram (illustrated by
the dots in \Fig{fig_softgluon} (right)), and the quadratic functional
involves
$(f_1(\p)-f_1(\p'))(f_1(\k')-f_1(\k)) \sim q^2 f_1'(\p) f_1'(\k)$.
From the point of view of the Fokker-Planck equations this term gives
rise to a gain term.  But from the point of view of the original
Boltzmann equation, this is a cross-correlation between the departures
from equilibrium of the two particles.  Therefore we will refer to
these contributions both a \textsl{gain} terms and as \textsl{cross}
terms, depending on the context.

Examining the expression for
$\left. (f_1, \C_{\mathrm{\lossshort}}^{\twotwo}f_1) \right|_{\rm loss}$,
we see that it involves one light-like vector, $v_\p$. Indeed, the
expression for $\hat q^{ij}$ can be rewritten as
the Wightman correlator of soft thermal gauge fields along this light-like
direction.
Using the causality and KMS properties of such light-like correlators~\cite{CaronHuot:2008ni}, these
soft contributions to $\qhat$ and $\ql$  can be evaluated
in closed form \cite{Aurenche:2002pd,Ghiglieri:2015zma,Ghiglieri:2015ala},
\begin{equation}
	\hat {q}^a\bigg\vert_\mathrm{soft}=\frac{g^2C_{R_a} T
	\md^2}{2\pi}\ln\frac{\mu_\perp}{\md},	\qquad
	\hat {q}_{\sss L}^a\bigg\vert_\mathrm{soft}=\frac{g^2C_{R_a} T
	\md^2}{4\pi}\ln\frac{\sqrt{2}\mu_\perp}{\md}.
	\label{qhatql}
\end{equation}
Here $gT\ll\mu_\perp\ll T$ is a cutoff on the $\qp \equiv \sqrt{q^2 - \omega^2}$
integration separating the soft from the hard scale\footnote{\label{foot_coords}
We have performed the change of integration variables
$\int_0^{\infty} dq \int_{-q}^{+q}d\omega\to\int_0^{\mu_\perp}d\qp \int_{-\infty}^{+\infty}
d\omega \qp/q$. }.
The dependence on this cutoff cancels against the region where
$\omega,q\simg T$, where the bare matrix elements can be used
to evaluate the hard contribution to $\C^{\twotwo}$.
The simple form of $\qhat$ and $\ql$ is a consequence of the fact that
light-like separated points are effectively causally disconnected
as far as the soft gauge fields are concerned.
Using the explicit form of the angular dependence of
$f_1^a(\p) = \beta^2 \chi_{\ij}(\p) X_{\ij}$ and \Eq{rotinvar},
straightforward analysis shows that the loss term  reduces to
\begin{multline}
    \left. \Big( \chi_\ij, \C^\twotwo_{\mathrm{\lossshort}} \, \chi_\ij \Big) \right|_{\rm loss} =
 	 \frac{\beta^3}{2}
 	  \sum_a \nu_a\int_\p f^a_0(p)
 	 (1\pm f_0^a(p)) \times
         \\
 	\left[(\chi^{a}(p)')^2  \hat {q}^a_{\sss L}\bigg\vert_\mathrm{soft}
 	+
 	\frac{\ell(\ell+1)\chi^a(p)^2}{2p^2}
        \hat {q}^a\bigg\vert_\mathrm{soft}\right] \, ,
\label{diffterms}
\end{multline}
which is the most useful form for evaluating the transport coefficients
numerically.

The gain terms (\Eq{gain}) intrinsically involve two light-like
momenta $v_\p$ and $v_\k$ associated with $f_1(\p)$
and $f_1(\k)$.  The points on these  two light-like rays
are causally connected by soft gauge fields,
thus the analyticity techniques used for $\hat q$ cannot be expected to work.
All attempts to extend these techniques to  two light-like rays have
met with frustration, and $\A_{ab}^{ij}(\hat\p\cdot\hat\k)$ and its
moments must be computed numerically.
 To this end, using \Eq{rotinvar} and  the angular dependence  of
 $f_1(\p) = \beta^2 \chi(p) \,I_{\ij}(\hat\p) \, X_{\ij}$, we may rewrite the
 gain terms as
 \begin{multline}
 	\label{gainterms}
\hspace{-5mm}\left. \Big(\chi_\ij, \C^\twotwo_{\mathrm{\lossshort}} \chi_\ij\Big)\right|_{\rm gain}
    =  -   \, \sum_{ab} \frac{g^4 \nu_a C_{R_a} \nu_b C_{R_b}}{8\pi^2d_AT^3}
	  \int_{\p}
	  \int_0^\infty dk\,k^2\,  
	 f_0^a(p) (1 \pm f_{0}^a(p)) f_0^b(k) (1 \pm f_{0}^b(k))   \\
    \times \left[
    c_1\chi^{a}(p)'\chi^{b}(k)'  +
    c_2 \left(\chi^{a}(p)'\frac{\chi^b(k)}{k} +	\chi^{b}(k)' \frac{\chi^a(p)}{p}\right) +
   c_3 \frac{\chi^a(p)}{p}\frac{\chi^b(k)}{k}  \right],
 \end{multline}
where
\begin{subequations}
  \label{eq:cigluongain}
  \begin{align}
   c_1 \equiv&  \int \frac{d\Omega_{\k}}{4\pi}\int \frac{d^4Q}{(2\pi)^4}  \, |G_{\mu\nu}^R(Q) v_\p^{\mu} v_{\k}^{\nu} |^2  \,
   2\pi \delta(v_\p \cdot Q) \, 2\pi  \delta(v_{\k} \cdot Q)\, \omega^2 P_\ell(\hat\p\cdot\hat\k)\,,  \\
   c_2 \equiv&
    \int \frac{d\Omega_\k}{4\pi}\int \frac{d^4Q}{(2\pi)^4}  \, |G_{\mu\nu}^R(Q) v_\p^{\mu} v_\k^{\nu} |^2
   \, 2\pi \delta(v_\p \cdot Q) \, 2\pi  \delta(v_\k \cdot Q)\, \omega^2 (1-\hat\p\cdot\hat\k)P'_\ell(\hat\p\cdot\hat\k)\,,  \\
   c_3 \equiv&
    \int \frac{d\Omega_\k}{4\pi}\int \frac{d^4Q}{(2\pi)^4}  \, |G_{\mu\nu}^R(Q) v_\p^{\mu} v_\k^{\nu} |^2
  \, 2\pi \delta(v_\p \cdot Q) \, 2\pi  \delta(v_\k \cdot Q)\nn\\
   &\hspace{2.8cm}\times\left[\qp^2P'_\ell(\hat\p\cdot\hat\k)
  + \omega^2 (1-\hat\p\cdot\hat\k)\big((1-\hat\p\cdot\hat\k)
   P'_\ell(\hat\p\cdot\hat\k)\big)' \right]
\end{align}
\end{subequations}
are coefficients which must be evaluated numerically. The complicated weights involving $P_{\ell}(\hat \p \cdot\kh)$
multiplying the  matrix elements reflect the angular structure of the collision kernel. 

When computing the diffusion coefficient ($\ell=1$),
the gluon-mediated gain terms described by \Eq{gainterms} actually vanish. This is because the
 gluons carry no charge and quarks and antiquarks
have opposite charges, so that $\chi^g(p)=0$, $\chi^q(p)=-\chi^{\bar q}(p)$. Thus,
the only processes that can give rise to gluon-mediated \gain terms are
$qq\lra qq$, $\bar q\bar q\lra \bar q \bar q$ and $q\bar q\lra q\bar q$.
However, due to their opposite signs, the quark
and antiquark contributions end up canceling in the sum
of these processes in \Eq{gainterms}
(see \cite{Hong:2010at} for further discussion).

When computing the shear viscosity ($\ell=2$), these integrals are convergent, and
$(c_1,c_2,c_3)$ may be evaluated directly
(see \App{sub_gain} for further details).
The UV finiteness of the gain terms
was discussed  previously in \cite{Arnold:2000dr,Hong:2010at} where
it was noted that
in a leading-log approximation (where the HTL propagator in \Eq{eq:cigluongain} is
replaced with the bare propagator)
the coefficients $c_1, c_2, c_3$ vanish for $\ell\geq 2$.

As discussed in \Sec{sec_nlo2}, we expect that the functional
form of the gain terms in \Eq{gainterms} will remain valid at NLO  but
the coefficients $c_1,c_2$ and $c_3$ will be modified by order $g$
corrections. We will only be able to estimate these modifications and
their associated (numerically small) contributions to the NLO shear viscosity.

\subsection{Soft quark exchange}
\label{sec:softquark}

\begin{figure}[ht]
	\begin{center}
		\includegraphics[width=8cm]{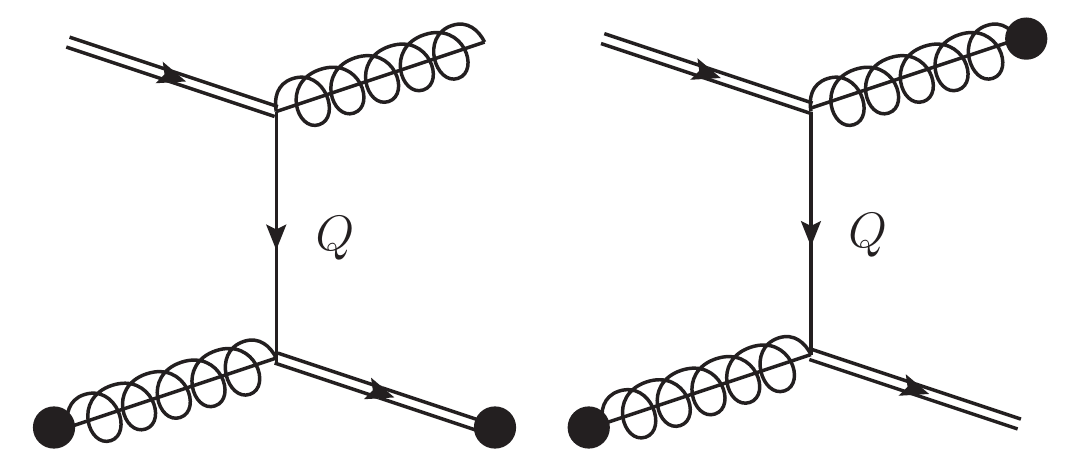}
	\end{center}
	\caption{Diagrammatic representation of the
	\lossf and \gainf terms in soft quark exchange processes. The
	graphical notation is the same as in Fig.~\ref{fig_softgluon} and the
	intermediate quark is soft (single line). The diagram
	on the left is a \lossf term, entering in
$(f_1^q(\p)-f_1^g(\p))^2$,
	whereas the one on the right is a \gainf term.}
		\label{fig_softquark}
\end{figure}

We will now analyze soft fermion exchange shown in \Fig{fig_softquark}, which
parallels the soft gluon exchange described in the previous section. In
this case, a hard quark with momentum $\p$ is \emph{converted} into
a hard gluon with approximately the same momentum through
the soft fermion exchange. (The reverse process is also
possible, and the set of matrix
elements involved in this process are  $q\bar q \lra gg$, $qg \lra qg$, and $\bar q g \lra \bar{q} g$.)

The dynamics of the conversion process are summarized 
by a set of  rate equations~\cite{Hong:2010at,Ghiglieri:2015ala}
\begin{subequations}
   \label{eq:kinetic}
\begin{align}
   \left(\partial_t + v_\p \cdot \partial_\x\right) f^q(\p) =& -\Gamma^{\rm conv}_{q\rightarrow g}(p) \;  f_0^q(p) (1 + f_0^g(p)) \; \left[ f_1^q(\p) - f_1^g(\p) \right]  + \mbox{gain-term} \, , \\
   \left(\partial_t + v_\p \cdot \partial_\x\right) f^{\bar q}(\p) =& -\Gamma^{\rm conv}_{q\rightarrow g}(p) \;  f_0^q(p) (1 + f_0^g(p)) \; \left[ f_1^{\bar q}(\p) - f_1^g(\p) \right]  - \mbox{gain-term} \, ,\\
   \left(\partial_t + v_\p \cdot \partial_\x \right) f^g(\p) =&
   - \sum_{q} \Gamma^{\rm conv}_{g\rightarrow q}(p)  \;f_0^g (p) (1 -  f_0^q(p))  \; \left[ f_1^g(\p) - f_1^q(\p) \right] \\ \nonumber
   & \qquad
   - \sum_{\bar q} \Gamma^{\rm conv}_{g\rightarrow \bar q}(p) \;  f_0^g(p) (1 - f_0^q(p))  \; \left[ f_1^g(\p) - f_1^{\bar q}(\p) \right] \, .
\end{align}
\end{subequations}
The conversion rates $\Gamma_{q \rightarrow g}^{\rm conv}(p)$ at leading
and next-to-leading order are given by \Eq{defconvrate} and
\Eq{nloconvrate} respectively~\cite{Ghiglieri:2015ala}, and the gluon
conversion rate  is
\st
\label{gqqg}
\Gamma^{\rm conv}_{g \rightarrow q} = \frac{\nu_q}{\nu_g} \Gamma_{q \rightarrow g}^{\rm conv} \, .
\stp
The gain term is  necessary to conserve baryon number under time evolution.
Indeed, the gain term records how the baryon charge associated with conversion
of a quark of momentum $\k$  to a gluon is balanced by an
increase of quarks (or decrease of anti-quarks) of momentum $\p$. We will
show that the gain term takes the form
\st
\label{eq:conv-gain}
\mbox{gain-term} =  \frac{f_0^q(p)(1 + f_0^g(p))}{2p} \int_\k \frac{f_0^q(k) (1 + f_0^g(k))}{k} \, {\mathcal C}_{q\rightarrow g}^{\rm conv}(\ph\cdot \kh) \left(f_1^q(\k) - f_1^{\bar q}(\k) \right) \, ,
\stp
where $\C_{q\rightarrow g}^{\rm conv}(\ph\cdot\kh)$ (which is given explicitly below in \Eq{eq:convcqg}) is a squared matrix element which specifies the
angular structure of the conversion process. The angular average of
$\C_{q\rightarrow g}^{\rm conv} (\ph\cdot \kh)$ determines the
conversion coefficient $\Gamma_{q \rightarrow g}^{\rm conv}$,
\st
\label{eq:conv-sum}
p\, \Gamma_{q\rightarrow g}^{\rm conv} (p) = \int_\k
 \frac{f_0^q(k) (1 + f_0^g(k)) }{k} \mathcal
C_{q\rightarrow g}^{\rm conv}(\hat\p \cdot \hat \k) \,.
\stp
It is straightforward to  show that with the gain and loss terms the
total baryon number is conserved under the evolution specified by \Eq{eq:kinetic}.

To derive these results we return to the $\twotwo$ collision
integrals with soft fermion exchange. The phase space
integral and soft approximations are given  in the
previous section, \Eq{eq:softps}.
The relevant processes are Compton scattering and pair annihilation, $q g \lra g q$, $\bar q g\lra q g$, and $gg \lra q \bar{q}$.
The HTL-resummed matrix elements are
\cite{Arnold:2002zm,Arnold:2003zc}
\begin{eqnarray}
	\label{softann}
   \frac{\left|\M^{q\bar q}_{gg}\right|^2_{\mathrm{soft\, q}\,t}}{(2p)(2k)(2p')(2k') } &= &
    \frac{ 4d_F\cf^2 g^4}{16 pp'kk'} \mathrm{Tr}[i\slashed{P}S_R(Q)i\slashed{K}S_A(Q)]\,,\\
	\label{softcompton}
   \frac{\left|\M^{qg}_{qg}\right|^2_{\mathrm{soft\, q}}}{(2p)(2k)(2p')(2k')}&=&
    \frac{ 4 d_F\cf^2 g^4}{16 pp'kk'} \mathrm{Tr}[i\slashed{P}S_R(Q)i\slashed{K'}S_A(Q)]\,,
	\end{eqnarray}
where $S_R(Q)$ is the retarded HTL-resummed quark propagator (see App.~\ref{app_props}). 	
At leading order in the soft approximation the two become equal:	
\begin{align}
  \nn
  \frac{\left|\M^{q\bar q}_{gg}\right|^2_{\mathrm{soft\, q}\,t}}{(2p)^2 (2 k)^2 }=
  \frac{\left|\M^{qg}_{qg}\right|^2_{\mathrm{soft\, q}}}{(2p)^2(2k)^2 }=& \,\frac{d_F \cf^2 g^4}{4 p k}
  \mathrm{Tr}[i\slashed{v}_\p S_R(Q)i\slashed{v}_\k S_A(Q)] \\
  =&   \,    \frac{d_FC_F^2g^4}{2 p k}
  \left[\left(1-\frac{\omega}{q}\right)^2\left\vert S^+_R(Q)\right\vert^2+
    \left(1+\frac{\omega}{q}\right)^2\left\vert S^-_R(Q)\right\vert^2 \right.\nn\\
    &\hspace{2.0cm}\left.-\frac{\qp^2}{q^2}\cos\phi\left(
    S^+_R(Q)S^-_A(Q)+S^-_R(Q)S^+_A(Q)\right)\right].
  \label{softqmat}
\end{align}
Neglecting the small momentum exchange in evaluating the statistical functions,
the contributions to the quadratic functional from these two processes
are, for each light flavor\footnote{%
  Both processes occur four times for each light fermion flavor in the sum over species
  $\sum_{abcd}$. The pair annihilation process receives an extra
  factor of $2$ (which we place just in front of $|\M|^2$)  to account
for soft $u$-channel exchange~\cite{Arnold:2000dr}.}
\begin{multline}
\Big( f_1, \C^{\twotwo}_{\mathrm{Compton}} \, f_1\Big)_\mathrm{soft}
    \equiv
      \frac{\beta^3}{2} \int_{{\rm PS} }
      \frac{\left|\M^{q g}_{qg}\right|^2_{\mathrm{soft\, q}}}{(2p)^2(2k)^2} \>
   	f^q_0(p) \, f^g_0(k) \, [1 + f^g_0(p)] \, [1 - f^q_0(k)]\\
   \times  \Bigr[ (f_1^q(\p) + f_1^g(\k) - f_1^g(\p) - f_1^q(\k) )^2
    +  (q \rightarrow \bar{q} ) \Bigl]  \, ,
\label{eq:tchanncompton}
\end{multline}
\begin{multline}
\Big( f_1, \C^\twotwo_{\mathrm{annihilation}} \, f_1 \Big)_\mathrm{soft}
    \equiv
      \frac{\beta^3}{2} \int_{{\rm PS} }
      2 \,\frac{\left|\M^{q\bar q}_{gg}\right|^2_{\mathrm{soft\, q}\,t}}{(2p)^2 (2k)^2} \>
   	f^q_0(p) \, f^q_0(k) \, [1 + f^g_0(p)] \, [1 + f^g_0(k)]\\
   \times   (f_1^q(\p) + f_1^{\bar q}(\k) - f_1^g(\p) - f_1^g(\k) )^2   \, .
\label {eq:tchannannih}
\end{multline}
The quadratic functional for the conversion process is obtained by adding the
Compton and pair annihilation contributions, and sorting  the terms into direct (e.g.\ $(f_1^q(\p) - f_1^g(\p))^2$) and \gain
terms (e.g.\ $ (f_1^q(\p) - f_1^g(\p))(f_1^q(\k) - f_1^g(\k))$).
Minor manipulations lead
to the final form of the conversion functional%
\footnote{These manipulations include
 employing the identity $f_0^q(p)[1 + f_0^g(p)] = f_0^g(p) [1- f_0^q(p)]$,
symmetrizing the integrand over $\p,\k$, and using the definition $\nu_q \equiv 2 d_F$.
}
\begin{align}
  \label{22lossgain}
  (f_1, \C_{\rm conv}^\twotwo f_1) \equiv&
   \left. (f_1, \C_{\rm conv}^\twotwo f_1) \right|_{\rm loss} + \left. (f_1, \C_{\rm conv}^\twotwo f_1)  \right|_{\rm gain} \, .
\end{align}
Here the loss part stems from the direct terms
\begin{multline}
\left. (f_1, \C_{\rm conv}^\twotwo f_1) \right|_{\rm loss}
\equiv
   \beta^3 \sum_{q}^f \nu_q
	\int_\p \GammaC_{q\to g}(p)\,
	 f^q_0(p)
	(1{+} f_0^g(p))
   \left[(f_1^q(\p)-f_1^g(\p))^2 +(q\to\bar q)\right] ,
	\label{eq:conversion-loss}
\end{multline}
while the gain part stems from the cross terms
\begin{multline}
   \left. (f_1, \C_{\rm conv}^\twotwo f_1) \right|_{\rm gain}
\equiv
   -\, \frac{\beta^3}{2}  \sum_{q}^f \nu_q
   \int_{\p\k}
   \frac{ f^q_0(p)
   (1+ f_0^g(p)) }{p}
   \frac{ f^q_0(k)
   (1+ f_0^g(k)) }{k}
   \, {\mathcal C}_{q\rightarrow g}^{\rm conv}(\ph \cdot \kh)  \\
    \,\times \,  (f_1^q(\p)-f_1^{\bar q}(\p)) (f_1^q(\k) - f_1^{\bar q}(\k))  \, .
	\label{eq:conversion-cross}
\end{multline}
The conversion coefficient $\Gamma_{q\rightarrow g}^{\rm conv}$ is given by \Eq{eq:conv-sum}, and the conversion
kernel ${\mathcal C}_{q\rightarrow g}^{\rm conv}(\ph \cdot \kh)$ is given
by
 \begin{equation}
 \label{eq:convcqg}
    {\mathcal C}_{q\rightarrow g}^{\rm conv}(\ph\cdot \kh) = \frac{1}{4} g^4 C_F^2  \int\frac{d^4Q}{(2\pi)^4} 
	2\pi \delta(v_\k \cdot Q)\, 2\pi \delta(v_\p \cdot Q)\,
    {\rm Tr}\left[i\slashed{v}_\p S_R(Q) i\slashed{v}_\k S_A(Q) \right]   .
 \end{equation}
Varying the conversion functional according to \Eq{variation} yields the
kinetic equations given by \Eq{eq:kinetic}.

At a technical level, the loss terms arises when the
deviations from equilibrium are on  the same side of the
fermion exchange diagram, $(f_1^q(\p) -f_1^g(\p))^2$,
as illustrated by the black dots on \Fig{fig_softquark} (left).
The gain term, which records the correlations between the scattered
particles, arises through an exchange of quantum
numbers \emph{across} the fermion exchange diagram, \Fig{fig_softquark} (right).

Examining the expression for
$\left. (f_1,C_{\rm conv}^{2\lra2} f_1) \right|_{\rm loss}$, we see
it involves one light-like vector $v_\p^{\mu}$. Indeed,
the conversion coefficient, $\Gamma_{q\rightarrow g}^{\rm conv}$, can
be rewritten as a light-like Wightman
correlator of soft fermion fields~\cite{Ghiglieri:2015ala,Ghiglieri:2015zma}.
As shown in \cite{Ghiglieri:2015ala}, this correlator can also be
evaluated in closed form using light-cone techniques (see App.~D.2 in
\cite{Ghiglieri:2015ala}), yielding
\begin{equation}
	\GammaC_{q\to g}(p)
	=\frac{g^2C_Fm_\infty^2}{8\pi p}\ln\frac{\mu_\perp}{m_\infty}.
	\label{defconvrate}
\end{equation}
As in the previous section, the dependence on the cutoff $\mu_\perp$ cancels
against the hard region, $\omega, q \sim T$, where bare matrix elements may be
used.

In practice, for the shear viscosity ($\ell=2$) we solve for the
fermion sum $(f_1^{q} + f_1^{\bar q})/2$ and set the fermion
difference $(f_1^{q} - f_1^{\bar q})/2$ to zero, while
for the diffusion coefficient ($\ell=1$) we solve for the difference
and set the sum to zero. Thus, the fermion gain term only enters when calculating the diffusion coefficient.
For the numerical evaluation of the loss term, we substitute
the angular  form $f_1^a(\p) = \beta^2 \chi_\ij^a(\p) X_{\ij}$ into the
quadratic functional (\Eq{eq:conversion-loss}), use \Eq{rotinvar}, and sum over
flavors to find
   \begin{multline}
   \left. \Big( \chi_\ij, \C^\twotwo_{\mathrm{\lossfshort}} \, \chi_\ij \Big)\right|_{\rm loss} =
	\frac{2d_F \nf}{ T^3}
	\int_\p\GammaC_{q\to g}(p)
	 f^q_0(p)
	[1+ f_0^g(p)] \\
      \left[(\chi^q(p)-\chi^g(p))^2 +(\chi^{\bar q}(p)-\chi^g(p))^2\right].
	\label{eq:conversion-loss2}
   \end{multline}
For the gain terms (which are only relevant for $\ell=1$), we
substitute $f_1^a(\p) = \beta^2 \chi_{i}^a(\p) X_{i}$ into \Eq{eq:conversion-cross}
and find
   \begin{multline}
   \left. \Big( \chi_{i}, \C^\twotwo_{\mathrm{conv}} \, \chi_i\Big) \right|_{\rm gain}=
      -  \frac{d_F \nf}{4\pi^4T^3}
		 \int_0^\infty dp\,p\int_0^\infty dk\,k
  f^q_0(p) \,[1 + f^g_0(p)] \, f^g_0(k) \, [1 - f^q_0(k)]  \\
     \, \times c_1\,
   (\chi^q(p) -\chi^{\bar q}(p)) (\chi^{q}(k) - \chi^{\bar q}(k))
    	 \, ,
		 \label{eq:gainconv}
   \end{multline}
where
 \begin{align}
 \label{eq:c1fermion}
    c_{1} &\equiv \int \frac{d\Omega_{k}}{4\pi} \C_{q \rightarrow g}^{\rm conv}(\ph \cdot \kh) P_{1}(\ph \cdot \kh).
 \end{align}
Similarly to the momentum diffusion case, the gain coefficient
 must be evaluated numerically as worked out in \App{sub_gain}.

At NLO we expect the form of the quadratic functional
(\Eq{eq:conversion-loss2} and \Eq{eq:gainconv}) to remain valid,  but
we have been unable to evaluate the gain coefficient $c_1$ beyond
leading order. We will estimate the NLO modifications of this
coefficient in the next section, and evaluate its (numerically small)
contribution to the NLO diffusion coefficient.

\subsection{Diffusion and identity in collinear processes}
\label{diffcoll}

Consider the collinear process introduced in \Eq{eq:C12}.  Although it
is unnecessary to do so in a leading-order calculation, one can
interpret the $k\ll p$ and $(p-k)\ll p$ parts of the integration in
\Eq{eq:C12} as representing diffusion and identity-changing processes respectively
for the case $q \to qg$, as each representing identity changing
processes for the case $g\to q\bar{q}$, and as each representing diffusion
processes for the case $g\to gg$.  Specifically, for the case of
$q \to qg$, one can estimate \Eq{onetworate} with
\Eq{defimplfull}--\Eq{defdeltaE} for $k \ll p$ or $(p-k)\ll p$ as
\cite{Ghiglieri:2015ala} (see App.~\ref{sub_ir} for details)
\begin{equation}
  \label{gamma_approx}
  \gamma^q_{qg}(p;p-k,k) \sim \frac{g^4 p^2}{k} \,.
\end{equation}
Therefore the small $(p-k) \equiv p'$ region of the integration in \Eq{eq:C12}
is parametrically of form
\begin{equation}
  \label{C12id}
  \Big( f_1, \Cot f_1\Big) \sim \beta^3 g^4 \int d\Omega_{\hat\n}
  \int p\, dp \; f^q_0(p) (1{+}f^g_0(p)) \int_0 dp'
  \Big[ f_1^q(p\hat{\n}) - f_1^g(p \hat{\n}) - f^q_1(p' \hat{\n})
    \Big]^2 \,.
\end{equation}
The $(f_1^q - f_q^g)^2$ piece represents an identity-changing process.  There
is also a gain term due to $f^q_1(p')$ in the $\ell=1$ case (see
the considerations on the IR behavior of $f_1$ in App.~\ref{sub_ir}).
So the small $p'$ region of the
integral can be understood as identity-change.  However it is not
necessary to do so, since there is no $dp'/p'$ enhancement of this
region, so $p' \ll T$ gives rise to a suppressed contribution.

Similarly, for $k \ll p$, the integral is approximated by
\begin{equation}
  \label{12smallk}
  \Big( f_1, \Cot f_1\Big) \! \sim \beta^3 g^4 \!\!\! \int \!\!\! d\Omega_{\hat\n}
  \!\! \int \!\!p^2 dp \: f^q_0(p) (1{-}f^q_0(p))
  \!\int_0 \! dk \frac{f_0^g(k)}{k}
  \Big[ f_1^q(p\hat{\n}) - f_1^q([p{-}k] \hat{\n}) - f^g_1(k \hat{\n})
    \Big]^2 .
\end{equation}  
We can approximate $(f_1^q(p)-f_1^q(p-k))^2 \simeq k^2 (f_1^q(p)')^2$,
canceling the $f_0(k)/k \sim T/k^2$ behavior to give a nonzero
contribution to the longitudinal diffusion coefficient.  But again,
because the integration is then only $\int_0 dk$, the $k\ll T$ region
is suppressed.  In particular, if we take $k$ (or $p'$) to be
$\OO(gT)$, in each case we find a contribution which is $\OO(g)$
suppressed.  Therefore we do not technically need to consider these
regions as identity change or longitudinal momentum diffusion in a
leading-order treatment.  But it will be important in an NLO
investigation that these limiting regions can be described in this
way.

\section{NLO corrections}
\label{sec_nlo2}

Here we show how to incorporate next-to-leading order corrections into
the leading-order treatment discussed in the previous section.  We
begin by showing how to do so in a strict expansion in $g$.  Then we
show the problem with this method; the resulting collision integral is
not manifestly positive.  Arnold, Moore, and Yaffe already encountered
this problem in their leading-order treatment \cite{Arnold:2003zc},
which they then avoided by not using momentum cutoffs, instead
applying screening corrections at all  momentum transfer scales.  This
led to a manifestly positive collision operator which agreed to
$\OO(g)$ corrections with the strict leading-order form when $g$ is
held small. We show how to make a similar treatment of the $\OO(g)$
corrections, which leads to a stable numerical treatment.

\subsection{Strict NLO treatment}
\label{sec:strict}

In the last section we saw how to reorganize the leading-order
treatment of Arnold Moore and Yaffe \cite{Arnold:2003zc} into a
contribution from generic momenta without screening, cut off at a
transverse scale $\mu_\perp$, and effective diffusion and identity
changing processes.  The scale $\mu_\perp$ cancels when summing the
two contributions, providing that we choose this scale to be
sufficiently small.  This leads to a self-consistent definition of the
leading-order scattering operator.  Furthermore, under this definition
the linearized collision operator $\C$ is structured strictly as a
$g^4$ object times a log plus constant, and therefore contains no
formally subleading in $g$ content.  Our goal in this subsection is to
extend this treatment, capturing all $\OO(g)$ corrections.

The only way $\OO(g)$ corrections can arise is if the physics of
$q\sim gT$ degrees of freedom features in a calculation.  These are
highly occupied, and loop corrections are of order
$g^2 f_0(q) \sim g$ when bosons propagate at this energy scale.
Furthermore, the HTL effects which are essential at this momentum
receive the first non-HTL corrections at $\OO(g)$.

Among $\twotwo$ processes, the $gT$ scale appears only when the
exchange momentum becomes small -- in which case the process
degenerates into a diffusion or identity change process -- or when an
external particle becomes soft, $p \sim \OO(gT)$.  In the latter case
the other states are nearly collinear, and this possibility will be
part of what we call semi-collinear processes below.
Among $\onetwo$ processes, the $gT$ scale appears in the transverse
exchange momentum $\bqp$ and the screening mass $m_\infty$
appearing in \Eq{defimplfull} and \Eq{defdeltaE}.  Each will receive an
$\OO(g)$ correction \cite{Ghiglieri:2013gia}.
Furthermore, our treatment in \Eq{eq:C12} involved a collinear
approximation which breaks down when one of the splitting daughters
becomes soft, $k\sim gT$, $p{-}k \sim gT$, or when the transverse
momentum becomes larger, $h \sim \sqrt{g} T^2$.  We already showed
that the case of a soft splitting daughter can be treated as a
correction to the diffusion and identity change rates.  The large-$h$
region is what we call semi-collinear processes.

We showed in \cite{Ghiglieri:2015ala} how to handle each sort of
$\OO(g)$ correction, except for the gain terms which we discussed
above.  In summary, to perform an almost-NLO treatment (in the sense
of only missing these gain terms), we include the following:
\begin{itemize}
\item
  We shift the transverse momentum diffusion coefficient $\qhat^a$ by
  \cite{CaronHuot:2008ni}	
  \begin{equation}
    \delta\hat {q}^a=\frac{g^4C_{R_a} C_A \md T^2 }{32\pi^2}
    \left(3\pi^2+10-4\ln 2\right) \,.
    \label{qhatnlo}
  \end{equation}
\item
  We shift the longitudinal momentum diffusion coefficient
  $\ql^a$ by \cite{Ghiglieri:2015ala}
  \begin{equation}
    \delta\ql^a=-\frac{g^4C_{R_a}  C_A\md T^2 }{4\pi^2}
    \left[\ln\left(\frac{\mu_\perp^\NLO}{M_\infty}\right)-\frac12\right],
    \label{qlnlo}
  \end{equation}
  where $\mu_\perp^\NLO$ is a new separation scale between NLO soft
  and hard (semi-collinear) processes.
\item
  We correct the conversion process rate $\GammaC_{q\to g}$ to
  \cite{Ghiglieri:2015ala}
  \begin{equation}
    \delta	\GammaC_{q\to g}(p)=-\frac{g^4 C_F^2 \md T  }{16\pi^2 p}
    \left[\ln\left(\frac{\mu_\perp^\NLO}{m_\infty}\right)-\frac12\right].
    \label{nloconvrate}
  \end{equation}
\item
  We correct collinear $\onetwo$ processes via the incorporation of
  $\OO(g)$ corrections to $\bar C(q_\perp)$ and $m_\infty$, appearing
  in \Eq{Cbar} and \Eq{defdeltaE}.  The procedure is to modify the
  splitting rate $\gamma^a_{bc}(p;p-k,k)$ precisely as is described
  in Appendix E of Ref.~\cite{Ghiglieri:2015ala}:
  \begin{equation}
    \label{nlodeltaC}
    \gamma^a_{bc,\NLO}(p;p{-}k,k) \equiv \gamma^a_{bc}(p;p{-}k,k)
    + \delta \gamma^a_{bc}(p;p{-}k,k) \; \;
    \mbox{of Ref.~\cite{Ghiglieri:2015ala} Appendix E}.
  \end{equation}
  
\item
  We include corrections to the collinear approximation in
  $\Cot$ by incorporating the first non-collinear corrections.
  We postpone the details to subsection \ref{sec:reorg}.
  The short version is that,
  in \Eq{defimplfull}, we have made approximations which only hold
  when $\bh$ is sufficiently small, $\bh \sim gT^2$.  When it
  becomes larger, $\bh \sim T^2 \sqrt{g}$, the approximations break
  down and we must be more careful.  In treating this region we need
  an IR cutoff on $\bh$, which exactly compensates the (UV) cutoff
  $\mu_\perp^{\NLO}$ we need for the longitudinal momentum diffusion
  and identity change processes at NLO.
\end{itemize}
Adding these contributions to the strict leading-order contributions
of the previous section produces a collision operator which is fully
NLO except for NLO contributions to the gain terms.  Furthermore, it
again exists strictly as an $\OO(g^4)$ piece and an $\OO(g^5)$ piece,
each containing logs of the coupling, but with no formally higher
order content.

\subsection{Problem with strict order-by-order}
\label{notstrict}

Except for quite small coupling, the approach of the last subsection
fails in practice.  We already see why by considering its application
at leading order.  How small does the separation scale $\mu_\perp$
need to be to find a result which is $\mu_\perp$-independent?  The
answer is that we need $\mu_\perp \ll T$, since $T$ is the natural
scale for $f_0(p)$ and $\chi(p)$ to vary.
However, once $\mu_\perp < \md$, the momentum diffusion
and identity change contributions of \Eq{qhatql} and \Eq{defconvrate}
become negative.  But when $\qhat$ and $\ql$ are negative, the
collision operator is not strictly positive.  Within a finite basis of
relatively smooth functions, this nonpositivity may not manifest
itself, if the strictly positive contributions from
$\Ctt$ and $\Cot$ are large enough.  But as we consider functions with
large $p$-derivatives, the importance of $\ql$ grows relative to other
terms.  So too does $\qhat$ for functions which peak at very small
$p$.  So for a sufficiently large basis of test functions in
\Eq{Ansatz}, the collision operator will operate nonpositively within
our Ansatz subspace.

This problem was already recognized in Ref.~\cite{Arnold:2003zc}.  The
solution there was to abandon the strict leading-order methodology.
Rather than introducing a separation scale and replacing the screened
IR piece with a differential operator, they incorporated screening
corrections into the scattering matrix elements responsible for
diffusion and number change, at finite momentum exchange.  At
weak coupling this procedure is equivalent to the strict leading-order
treatment up to corrections which begin at $\OO(g)$, and which are
dependent on the exact methodology used for incorporating the
screening corrections (see \cite{Arnold:2003zc} section IIIB and
Figure 4).  Here we will adopt the precise prescription detailed in
Appendix A of the reference.

We can certainly use this methodology for the leading-order collision
term $\Ctt$.  However, we must check whether the resulting
leading-order collision operators then already incorporate formally
$\OO(g)$ subleading corrections, and if so, we must make a subtraction
to avoid a double counting.

To see how each approach works in practice, and to illustrate how
nonpositivity arises in the strict case and $\OO(g)$ corrections arise
in the AMY procedure, we will delve a little into the details of the
soft region at leading order.  The most convenient choice of phase
space integration variables for evaluating the leading-order $\twotwo$
process in the $t$ channel (suppressing particle-species labels and an
overall factor of $1/(2^8 \pi^5)$) is
\cite{Arnold:2003zc}
(see also App.~\ref{sub_phase} for details)
\begin{align}
  \Big( \chi_{\ij} , \Ctt \chi_{\ij} \Big) & =
  \int_0^\infty dq \int_{-q}^q d\omega
  \int_{\frac{q-\omega}{2}}^\infty dp
  \int_{\frac{q+\omega}{2}}^\infty dk
  \frac{|\M|^2}{16pkp'k'} f_0(p) f_0(k) [1{\pm}f_0(p')][1{\pm}f_0(k')]
  \nonumber \\ & \phantom{=} \times
  \Big( \chi_{\ij}(\p) + \chi_{\ij}(\k)
  - \chi_{\ij}(\p') - \chi_{\ij}(\k') \Big)^2 \,.
\label{22general}
\end{align}
Here $(p,k)$ are the incoming particle energies,
$(p'=p+\omega,k'=k-\omega)$ are the outgoing energies, and we
frequently reorganize the first two integrals,
$\int_0 dq \int_{-q}^{q} d\omega =
\int d\omega \int_0 dq_\perp (q_\perp/q)$, with
$q^2=q_\perp^2+\omega^2$.  Reducing the final line to a scalar
expression requires evaluating the angles between the momenta
$\p,\k,\p',\k'$; the relevant angles are listed in Eq.~(A21) of
Ref.~\cite{Arnold:2003zc}.

For small $q_\perp$ one must modify the
matrix element to reflect (HTL) screening effects.  The strict
leading-order procedure is to introduce an intermediate scale
$\mu_\perp$.  Above this scale we neglect changes to the matrix
element.  Below this scale, we systematically expand in
$(\omega,q_\perp) \ll (p,k)$ to obtain the
diffusion and conversion expressions of the previous section. 
 This affects the integration limits,
with the $(p,k)$ integrals extending to 0, and it affects the matrix
element and the departures from equilibrium, which can be expanded in
gradients for gluon exchange or replaced with their $p=p',k=k'$ limits
for quark exchange:
\begin{align}
  \Big( \chi_{\ij} , \Ctt_{\mathrm{strict}} \chi_{\ij} \Big) & =
  \int d\omega \int_{\mu_\perp} \frac{q_\perp\, dq_\perp}{q}
  \big( \mbox{\Eq{22general}, $\M=\M_{\mathrm{free}}$} \Big)
  \nonumber \\ & \phantom{=} +
  \int d\omega \int_0^{\mu_\perp} \frac{q_\perp\, dq_\perp}{q}
  \int_0 dp \int_0 dk \frac{|\M|^2_{\sss{\rm HTL}}}{16p^2k^2}
  \nonumber \\ & \phantom{=} \quad \times
  f_0(p)[1{\pm}f_0(p)] f_0(k)[1{\pm} f_0(k)]
  \Big[ \chi(p) - \chi(k) \;\mbox{or...} \Big]^2,
  \label{strict}
\end{align}
where the last square bracket means either the last line of
\Eq{diffterms} and \Eq{gainterms} or of \Eq{eq:conversion-loss2} and
\Eq{eq:gainconv}.  In the second and third lines the $q_\perp,\omega$
integrals and $p,k$ integrals factorize separately.  The
$\omega,q_\perp$ integrals are computed using sum rules,
giving rise to a logarithm
$\ln(\mu_\perp/\md)$.  In practice when $\md$ is not small this is
where lack of positivity enters.

On the other hand, the AMY procedure \cite{Arnold:2003zc} is to retain
the integration measure and distribution functions of
\Eq{22general}, and to replace the tree level $|\M|^2$ with
an HTL form, detailed at some length in the reference, at all
$q,\omega$ values.  The $p,k$ integration limits are not changed and
the statistical functions are not simplified.  That is,
\begin{align}
  \Big( \chi_{\ij} , \Ctt_{_\AMY} \chi_{\ij} \Big) & =
  \int_0^\infty \!\!dq \int_{-q}^q \! d\omega
  \int_{\frac{q-\omega}{2}}^\infty dp
  \int_{\frac{q+\omega}{2}}^\infty dk
  \frac{|\M_{_\AMY}|^2}{16pkp'k'} f_0(p)
  f_0(k) [1{\pm}f_0(p')][1{\pm}f_0(k')]
  \nonumber \\ & \phantom{=} \times
  \Big( \chi_{\ij}(\p) + \chi_{\ij}(\k)
  - \chi_{\ij}(\p') - \chi_{\ij}(\k') \Big)^2 \,.
\label{22AMY}
\end{align}

To identify $\OO(g)$ differences between these procedures, we must
understand where the differences occur when $g$ is genuinely small (so
$\md/T \ll 1$).  For generic $p,k \sim T$, the regime
$q_\perp \sim T$ differs by an $\OO(g^2)$ amount, because
$\M_{\AMY} = \M_{\mathrm{free}} + \OO(g^2)$ in this regime.
For $q_\perp \sim gT$, we have $\M_{\AMY} = \M_{\sss{\rm HTL}} + \OO(g^2)$
and the small-$\omega,q$ approximations to statistical functions and
angles are $\OO(g^2)$ after we symmetrize over positive and negative
$\omega$.

On the other hand, there is the region where one external
particle becomes soft and the exchanged four-momentum is
soft.  (The region where both external particles are soft is highly
suppressed.)  When $k \sim gT$, the two treatments differ in several
respects.  Clearly the phase space treatment is different; \Eq{strict}
integrates to $k=0$, while \Eq{22general} integrates to
$2k=q{+}\omega$, an important difference if $k \sim gT$.  Also the
approximations to the statistical functions and matrix element are no
longer reliable.  Therefore the two approaches have $\OO(1)$
differences in this region.  Provided that $k$ is a gluon, this region
is only suppressed by $\OO(g)$.  Therefore this region represents a
source of $\OO(g)$ differences between the strict and AMY
methodologies.

Fortunately, when $p$ is hard but $k$ is soft,
the process is well described by either momentum diffusion or identity
change from the point of view of the high-energy particle.  Therefore
the treatments differ at $\OO(g)$, but only because of the region
where one external particle is soft, and this region can be captured
in terms of diffusion and identity change processes.

In a leading-order calculation we are free to use either approach.
The AMY approach is preferred because it gives a positive collision
operator.  In an NLO treatment, we have calculated the NLO corrections
to diffusion and identity changing processes assuming that the strict
leading-order treatment is to be used.  If we use instead the AMY
approach, as we do, we need a calculation at $\OO(g)$ of the \textsl{difference}
between the two leading-order approaches, written in terms of
diffusion and identity-change rates.  These can then be included as
``counterterms'' in our NLO treatment.  We compute these counterterms
in detail in Appendix \ref{sub_double}.

\subsection{Semi-collinear contributions and reorganization}
\label{sec:reorg}

Returning to the NLO corrections we introduced in
\Eq{qhatnlo}, \Eq{qlnlo}, and \Eq{nloconvrate}, we see a similar
problem.  Two of these depend on an introduced intermediate scale
$\mu_\perp^{\NLO}$.  This scale must again be chosen in such a way
that its influence is small. Furthermore, even at small $g$, the 
$\delta\ql$ and $\delta\Gamma_\mathrm{conv}$ corrections and the
semi-collinear ones tend to represent a negative contribution to the 
NLO collision operator, as the proper $\OO(g)$ evaluation of these
regions is smaller than the naive one included at LO in e.g.\
\Eq{C12id} and \Eq{12smallk}. 
We will thus lose positivity of the
collision operator when we incorporate these corrections at not-so-small
values of $g$.  So again we need to find a
reorganization which reproduces these contributions in the sense of a
strict NLO expansion in $g$.  To do this we need to return to the
semi-collinear process in more detail.

\begin{figure}[ht]
	\begin{center}
		\includegraphics[width=7.2cm]{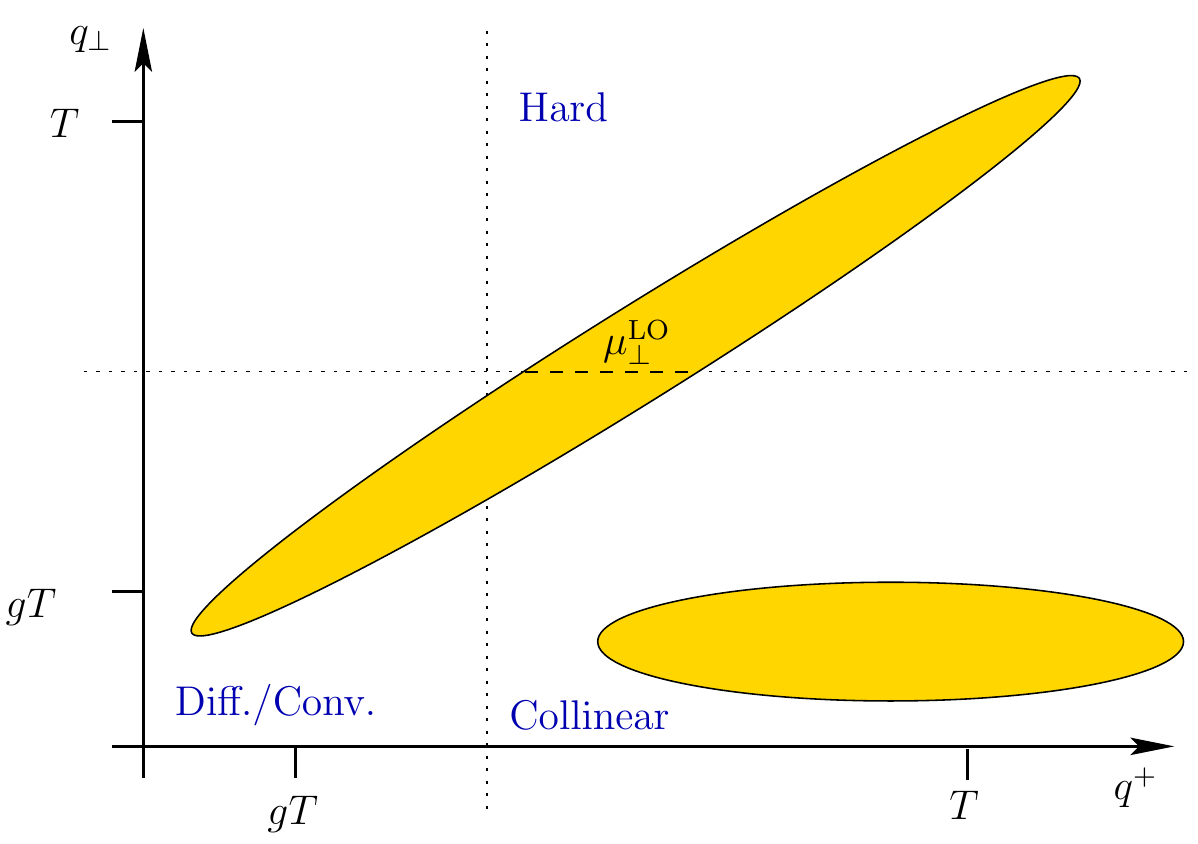}
                \hfill
		\includegraphics[width=7.2cm]{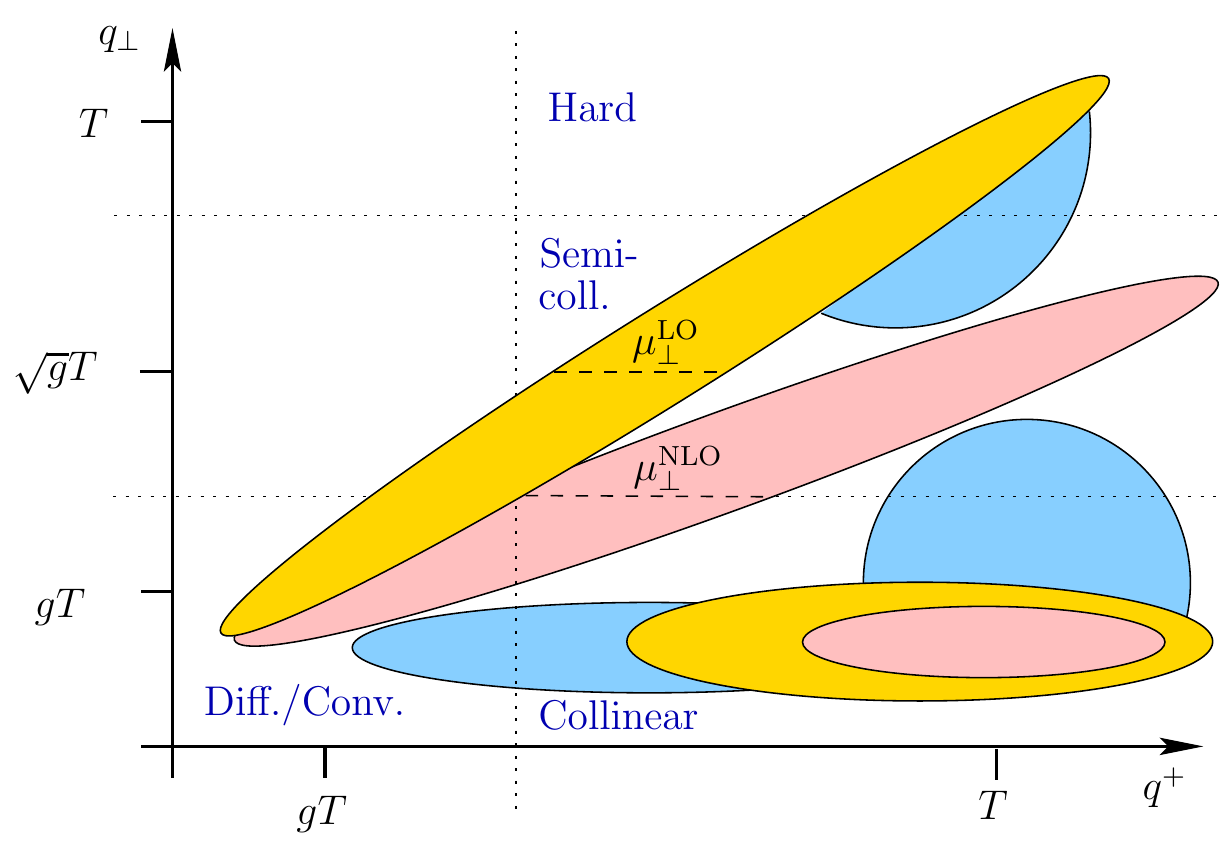}
	\end{center}
	\caption{Left: kinematic regions, in terms of exchanged
          transverse momentum and exchanged light-cone momentum $q^+$,
          which are relevant at
          leading order. Right: the same regions (yellow) plus regions
          which are relevant at next-to-leading order (pink) and
          where subtractions of leading-order effects are needed (blue).}
		  	\label{fig:kin}
\end{figure}

The relevant kinematics are summarized in Figure \ref{fig:kin}.
Collinear processes correspond to a particle making a large change in
energy but a small change in transverse momentum.  Elastic scattering
is a large change in both -- or, for soft processes, a small change in
both.  The semi-collinear region is where the exchanged transverse
momentum is intermediate between these two cases.  Therefore it
requires subtractions from each.  It also requires a subtraction of
its soft-exchange tail.  We implement this as a cutoff on $q_\perp$ at
the scale $\mu_\perp^{\NLO}$, but physically one could also see this
as a way to cut off small energy (really $q^+$) exchanges.

To understand this region better, consider \Eq{onetworate} and
\Eq{defimplfull}.  In deriving the equation we assumed that
$p,k\sim T$ and $\bh \sim gT^2$.  This allowed a collinear expansion;
in particular we could equate $q_\parallel = \omega$ in exchange
processes ($q^- = 0$).  But this breaks down as we consider larger
$\bh$ values.  Fortunately in this regime there is a new simplifying
approximation; the integral equation, \Eq{defimplfull}, can be solved
iteratively in large $\delta E$:
\begin{align}
  \label{Fseries}
  \bff_{b1}(\bh) & = 2\bh/i\delta E(\bh) \,, \\
  \nn
  \bff_{b2}(\bh) & = \frac{i}{\delta E(\bh)}
  \int \frac{d^2 q_\perp}{(2\pi)^2}
  \bar C(q_\perp) \left\{ \left( C_{R_b} - \frac{C_A}{2} \right)
       [\bff_{b1}(\bh) - \bff_{b1}(\bh-k\bqp)] \right.
       \\ & \quad
       \left. {}+ \frac{C_A}{2} [\bff_{b1}(\bh) - \bff_{b1}(\bh+p\bqp)]
       + \frac{C_A}{2} [ \bff_{b1}(\bh) - \bff_{b1}(\bh-(p{-}k)\bqp)]
       \right\} . \nn
  \end{align}
This is the same as treating the emission in the Bethe-Heitler
limit, ignoring LPM corrections.  We will make this approximation in
the following.  We can also assume that the $\bh^2$ term dominates in
the expression for $\delta E$, \Eq{defdeltaE}, so
$\delta E(\bh) = h^2/(2pk(p{-}k))$.

\begin{figure}[ht]
	\begin{center}
		\includegraphics[width=12cm]{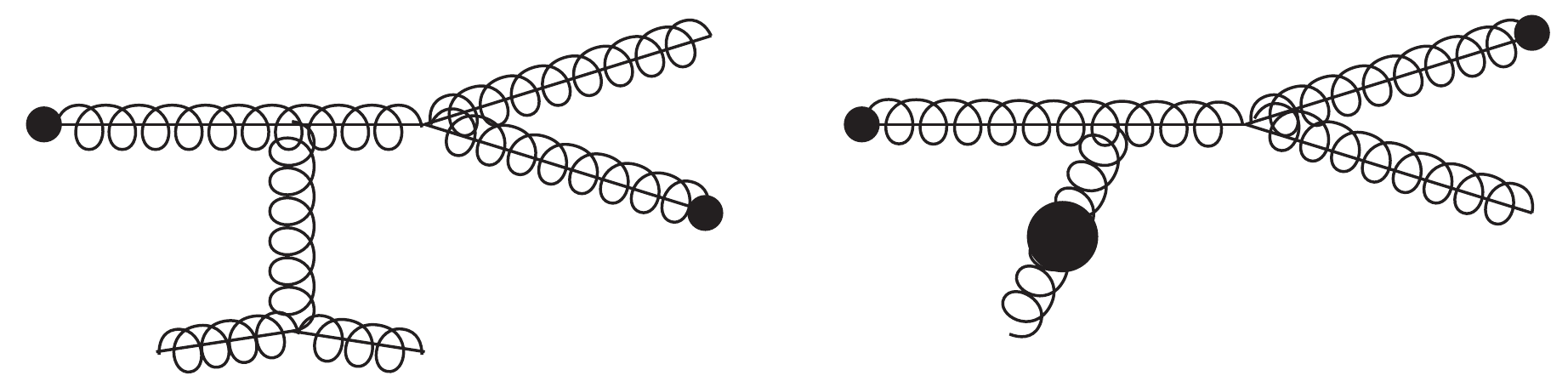}
	\end{center}
	\caption{Diagrams for typical semi-collinear processes.  The
          exchange of momentum $\bh/p$ and energy with the plasma can
          be due to elastic scattering or absorption of a soft
          quasiparticle, leading to the semi-collinear splitting
          process.}
		  \label{fig:semi}
\end{figure}

However, as noted above, it is no longer sufficient to neglect $q^-$
relative to $q_\perp$, because $q^- = \delta E \sim gT \sim q_\perp$.
Therefore the kinematics of scattering is changed and
$\bar C(q_\perp)$ must be recomputed.  A more accurate form for
$\bar C(q_\perp)$ in this regime, replacing \Eq{Cbar}, is
\cite{Ghiglieri:2013gia,Ghiglieri:2015ala}\footnote{%
\label{foot_qhatde}These references introduce $\qhat(\delta E)$. Here we
take $\bar \qhat(\delta E)=\int d^2\qp/(2\pi)^2\,\qp^2\, \bar C_{\NLO}(q_\perp,\delta E)$
and use this to infer $ \bar C_{\NLO}(q_\perp,\delta E)$.}
\begin{equation}
  \label{CLONLO}
  \bar C_{\NLO}(q_\perp,\delta E) =
  \frac{g^2 T \md^2}
       {(q_\perp^2 + \delta E^2)(q_\perp^2 + \delta E^2 + \md^2)}
       + \frac{2 g^2 T \delta E^2}{q_\perp^2(q_\perp^2 + \delta E^2)}
       \,.
\end{equation}
Physically this arises from two types of processes, in which the
splitting is either induced by an elastic scattering or by the
absorption of a soft on-shell particle, see Figure \ref{fig:semi}.
Therefore we need to make \textsl{two} subtractions, corresponding to
the already-computed LO $\onetwo$ contribution (the small $\delta E$
limit), \textsl{and} the already-included LO $\twotwo$ contribution
(the small $\md$ limit) \cite{Ghiglieri:2013gia}:
\begin{equation}
  \label{CNLO}
  \delta \bar C(q_\perp,\delta E) =
  \frac{g^2 T \md^2}
       {(q_\perp^2 + \delta E^2)(q_\perp^2 + \delta E^2 + \md^2)}
       - \frac{g^2 T \md^2}{q_\perp^2(q_\perp^2+\md^2)}
       \,.
\end{equation}
The second term is the LO collinear form for $\bar C$ (the small
$\delta E$ limit of \Eq{CLONLO}).  The other subtraction, of the
$\md\to 0$ limit, precisely removed the second term appearing in
\Eq{CLONLO}.

The semi-collinear contribution is found by substituting \Eq{CNLO}
into \Eq{Fseries} and using it to evaluate \Eq{onetworate} and hence
\Eq{eq:C12}.  But one further simplification can be made.  For generic
$p,k \sim T$, we have $\delta E \sim h^2/T^3$.  The two terms in
\Eq{CNLO} cancel
up to small corrections unless $\delta E \sim \md \sim gT$, which then
requires the semi-collinear value $h \sim T^2 \sqrt{g}$.  On the other
hand, $q_\perp \sim gT$; for larger values, $q_\perp^2 \gg \delta E^2$
and the two terms again cancel.  Therefore, we can make a systematic
expansion in $p\bqp \ll \bh$ in \Eq{Fseries}.  And to get a strict
NLO result, we also \textsl{need} to make such an expansion.
Averaging over directions for $\bqp$, we have
\begin{equation}
  \bff_{b1}(\bh) - \bff_{b1}(\bh - p\bqp) \simeq
  -\frac{p^2 q_\perp^2}{4} \nabla_{\bh}^2 \bff_{b1}(\bh)\,,
\label{F:angleavg}
\end{equation}
which we combine with the definition (see footnote~\ref{foot_qhatde})
\begin{equation}
  \delta  \bar{\qhat}(\delta E) \equiv
  \int \frac{d^2 q_\perp}{(2\pi)^2} \: q_\perp^2 \:
  \delta \bar{C}(q_\perp,\delta E)
  \label{qhatdE}
\end{equation}
to get an explicit expression for $\bff_{b2}(\bh)$, leading to \cite{Ghiglieri:2015ala}\footnote{There is
an unfortunate misprint in the $g\lra q\bar q$ rate in Eq.~(8.8) of \cite{Ghiglieri:2015ala}.
The term proportional to $C_A$ should be negative, as it is in \Eq{semi1} here.}
\begin{align}
  \label{semi1}
   \gamma^{a}_{bc}\bigg\vert_{\mathrm{semi}}^\mathrm{strict}
(p;p{-}k,k) & = \frac{g^2}{8\pi^4}\left\{
\begin{array}{cc}
	2d_AC_A^2\frac{(p^2+k^2-pk)^3}{pk(p-k)}& g\leftrightarrow gg\\
	d_F C_F \frac{p^2+(p-k)^2}{k} [C_Fk^2+C_Ap(p-k)]& q\leftrightarrow q g\\
	d_F C_F\frac{(p-k)^2+k^2}{p} [C_Fp^2-C_Ak(p-k)]& g\leftrightarrow q\bar q
\end{array} \right\}
\nonumber \\ & \quad {} \times
\int\frac{d^2h}{(2\pi)^2}\frac{\delta \bar\qhat(\delta E_s)}{h^4}.
\end{align}
When this result is inserted into
\Eq{eq:C12} it leads to logarithmic small $k$ and $(p{-}k)$
divergences unless we impose an IR cutoff on the
allowed value of $\bh$, namely $\bh \geq p\mu_\perp^{\NLO}$.  The
cutoff dependence cancels that in the NLO longitudinal momentum and
identity change rates \cite{Ghiglieri:2015ala}.

The problem with this procedure is the same as the problem with the
strict LO rate.  We need to insert a regulator scale which separates
regions with finite $k$-momentum exchange from regions which are
treated diffusively.  Neither side is necessarily positive and the
collision operator can have serious positivity problems when the
coupling is not small.  This necessitates a rewriting of the NLO
contributions along the lines of the AMY method at LO.
The problem arises because we made a strict $\bh \gg p \bqp$
expansion in \Eq{F:angleavg}.  Without this approximation, that is, by
using the full expression for $\delta \bar{C}$, \Eq{CNLO}, in
\Eq{Fseries}, we obtain instead the manifestly finite result
\begin{align}
  \nn	\gamma^{a}_{bc}\bigg\vert_\mathrm{semi}(p;p-k,k)
  & = \frac{g^2}{32\pi^4}\left\{
\begin{array}{cc}
	d_AC_A\frac{p^4+k^4+(p-k)^4}{p^3k^3(p-k)^3}& g\leftrightarrow gg\\
	d_F C_F \frac{p^2+(p-k)^2}{p^2(p-k)^2k^3} & q\leftrightarrow q g\\
	d_F C_F\frac{(p-k)^2+k^2}{(p-k)^2k^2p^3} & g\leftrightarrow q\bar q
\end{array}
\right.
\int\frac{d^2h}{(2\pi)^2}
\int\frac{d^2\qp}{(2\pi)^2} \delta\bar{C}(\qp,\delta E)\\
&\nn\hspace{-3cm}\times\left[\left(C_R-\frac{C_A}{2}\right)\left(\frac{\bh}{\delta E(\bh)}-
\frac{\bh-k\bqp}{\delta E(\bh-k\bqp)}\right)^2+\frac{C_A}{2}\left(\frac{\bh}{\delta E(\bh)}-
\frac{\bh+p\bqp}{\delta E(\bh+p\bqp)}\right)^2\right.\\
&\left.+\frac{C_A}{2}\left(\frac{\bh}{\delta E(\bh)}-
\frac{\bh-(p-k)\bqp}{\delta E(\bh-(p-k)\bqp)}\right)^2\right],
	\label{seminew}
\end{align}
with $(C_R-C_A/2)$ appearing on the $\bh + p\bqp$ term for
$g \leftrightarrow q\bar{q}$ processes.  This is then inserted into
\Eq{eq:C12}, resulting in
\begin{eqnarray}
	\nn
	\left(f_1,\mathcal{C}^\mathrm{semi}f_1\right)
	&\equiv&\frac{2\pi}{T^3}\sum_{abc}\int_0^\infty \!\! dp
        \int_0^p dk\,\gamma^a_{bc}\bigg\vert_\mathrm{semi} \!\! (p;p-k,k)
	f^a_0(p)[1\pm f^b_0(k)][1\pm f^c_0(p-k)]\\
	&&\times\left[f_1^a(\p)-f_1^b(k\hat\p)-
	f_1^c((p-k)\hat\p)\right]^2.
	\label{semicoll}
\end{eqnarray}
In the small $g$ limit, this
single expression reduces to the \textsl{sum} of the previous NLO
semi-collinear, $\delta \ql$, and $\delta \GammaC$ contributions, as we show
at some length in Appendix \ref{app_equiv}.  The appendix also
explains how this result is related
to the light-cone sum rules.

We conclude the illustration of this region with a remark.
Currently, we treat the collinear region with \Eq{eq:C12} and the
semi-collinear region, including a subtraction due to the collinear
one, using \Eq{semicoll}.  But we could combine them into a single
calculation by adding $\delta \bar C(\qp,\delta E)$ to
$\bar C(\qp)$ in \Eq{defimplfull}. In this way one would perform LPM
resummation with the $\delta E$-dependent kernel and thus would no
longer need to subtract the strictly collinear one.  The contribution
would also be manifestly positive.  However, this would be extremely
impractical.  Because $\bar C(\qp)$ is quite simple, we can Fourier
transform it analytically and solve \Eq{defimplfull} in
impact-parameter space as a differential equation.  But
$\delta \bar C$ does not have a simple Fourier expression, so
\Eq{defimplfull} would need to be solved as an integral equation in
$\bqp$ space, making a numerical solution very intricate.
Treating the parts separately as we do, with the expansion
\Eq{Fseries} used for the semi-collinear but not the collinear case,
does not lead to a manifestly positive result.  But the sum
nevertheless tends to remain positive, even for large $\md/T$, because
$\delta \bar C(\qp,\delta E)$ becomes smaller in that limit.

\subsection{Estimate of NLO gain terms}

We have now presented the NLO contributions \textsl{except} for
possible gain terms, as explained in Section \ref{sec_nlo}.  Although
we will not be able to compute these, we can at least estimate their
size, which allows us to assign a systematic error budget for their
exclusion.  NLO effects arise from soft momenta and from corrections
to collinear and semi-collinear physics.  For the latter, a small
momentum exchange induces a large change in the particle which splits,
and since we capture all aspects of this large change, only the soft
exchange partner is mistreated.  This is a subleading effect.
Therefore we only need concern ourselves about NLO gain terms due to
momentum diffusion and identity change.

To get an estimate for their magnitude, we compute the soft
contribution from gain terms in the LO calculation, where we know how
to compute and include them.  Then we estimate that the missing  NLO
gain terms are of order the same size, times a factor reflecting how much
smaller NLO effects are relative to the leading order.  In the
$\ell=2$ case, where, as we've seen in the previous section, \gain
terms are only mediated by soft gluon exchange, we shall use
\begin{equation}
	\Big( \chi_{ij}, \C^{\delta\mathrm{\gain}} \, \chi_{ij} \Big)
	= C_{\ell=2} \frac{\md}{T}\left.\Big(\chi_{ij}, 
	\C^\twotwo_{\mathrm{\lossshort}} \chi_{ij}\Big)\right|_{\rm gain},
	\label{nlogcross}
\end{equation}
where $C_{\ell=2}$ is a constant that we vary to incorporate our
ignorance of the actual size (and sign!) of the NLO gain terms.
We are thus making the Ansatz that the NLO corrections to the \gain
terms take the exact same form as at leading order, but rescaled by
$\md/T\sim g$ times an arbitrary constant. Similarly, for the $\ell=1$
case, where only fermionic \gain terms contribute, we assume
\begin{equation}
	\Big( \chi_i, \C^{\delta\mathrm{\gain}} \, \chi_i \Big)
	= C_{\ell=1} \frac{\md}{T}\left.\Big( \chi_{i},
        \C^\twotwo_{\mathrm{conv}} \, \chi_i\Big)\right|_{\rm gain}.
	\label{nlofcross}
\end{equation}
We evaluate each expression, as given in Eqs.~\eqref{gainterms}
and \eqref{eq:gainconv}, in Appendix \ref{sub_gain}.  Our results
will show that the impact of these gain terms is very modest.

\subsection{Summary}
We conclude this section by summarizing the form of the collisional part
of the quadratic functional. At LO it is given by \Eq{eq:Q2}.
At NLO we have
\begin{equation}
	\label{fullnlo}
    \Big( f_1, \C_\NLO \, f_1 \Big)
    {}=
    \Big( f_1, \C_\LO\, f_1 \Big) +
    \Big( f_1, \delta\C \, f_1 \Big) \,,
\end{equation}
where
\begin{align}
  \label{nlo1}
  \Big( f_1, \delta\C \, f_1 \Big)
  \equiv &
  \Big( f_1, \C^{\delta\qhat} \, f_1 \Big)-
  \Big( f_1, \C^\twotwo_{\OO(g)\,\mathrm{finite}} \, f_1 \Big)
  +\Big( f_1, \C^{\delta\mathrm{\gain}} \, f_1 \Big)\\
  \label{nlo2}
  &		  +
  \Big( f_1, \C^{\semi} \, f_1 \Big)  +
  \Big( f_1, \delta\C^{\onetwo} \, f_1 \Big).
\end{align}
Here the first contribution is found by inserting $\delta \qhat^a$ from
\Eq{qhatnlo} in place of $\qhat^a$ in \Eq{diffterms}. In the form 
most suited for numerical evaluation it reads 
 \begin{equation}
 	\Big( \chi_\ij, \C^{\delta\qhat} \, \chi_\ij \Big) =
 	 \frac{\beta^3}{2}\sum_a\nu_a\,\delta\hat {q}^a
 	 \int_\p   f^a_0(p)
 	[1\pm f_0^a(p)]
 	 \frac{\ell(\ell+1)\chi^a(p)^2}{2p^2} .
 	\label{cnloqhat}
 \end{equation}
The second term in \Eq{nlo1} is the
``counterterms'' discussed in Subsection \ref{notstrict}
and computed in Appendix \ref{sub_double},
the third is the estimate in \Eq{nlogcross} or \Eq{nlofcross},
the fourth is from \Eq{semicoll}, and the last is the result of the
modification of \Eq{nlodeltaC}, inserted in \Eq{eq:C12}.

\section{Results}
\label{sec_results}

As we have mentioned in Sec.~\ref{sec:ingredients}, we obtain the leading-order transport coefficients
by maximizing the functional $\mathcal{Q}[\chi]$, as given by \Eq{eq:Q1}, with the source
term given in \Eq{eq:Sij} and the LO collision operator given in
\Eq{eq:Q2}, with modifications described in Subsection
\ref{notstrict}, especially \Eq{22AMY}.
In the previous section we have derived the NLO collision operator,
as summarized in \Eq{fullnlo}. This operator is the sum of the LO one and of its
$\OO(g)$ corrections. In principle, one could treat the latter as a
perturbation.  One extremizes \Eq{eq:Q1} by series expanding
$1/(\C+\delta \C)$ as a geometric series,
\begin{equation}
 \beta^4 \mathcal{Q}_{\mathrm{max}}= \Big( \S ,
  \frac{1}{\C + \delta \C}\, \S \Big)
  \; \Longrightarrow \; \Big( \S, \frac{1}{\C}\, \S \Big)
  - \Big(\S , \frac{1}{\C}\, \delta \C\, \frac{1}{\C}\, \S \Big)
  +\ldots \,.
\label{C_expand}
\end{equation}
We have however chosen not to pursue this avenue because $\delta \C$
can be large and positive, and the above expression becomes negative
for intermediate $g$ values.  Hence, we instead define the NLO
transport coefficients as the expression before the $\Longrightarrow$
in \Eq{C_expand}, rather than the expression after the arrow.
One way to think of this is that we are
computing $1/\eta$ at NLO and then inverting it, similar to
resumming self-energy insertions into a Dyson sum so they appear
linearly in the denominator of the propagator.

We will plot the ratio $\eta/s$ using the leading-order,
Stephan-Boltzmann value for the entropy density with $\nf$
massless quarks, since the first
perturbative corrections to the partition function, from which all
thermodynamic observables are obtained, are of order $g^2$.  This is
self-consistent within our kinetic approach, and to do otherwise would
be treating particles' contribution to the entropy and to the stress
tensor differently and inconsistently.

We will start by presenting results in full QCD with fermions. Later
on, in Sec.~\ref{sub_gauge},
we will also present results for the pure Yang-Mills theory and in Sec.~\ref{sub_qed} for QED. 
Accurate fits for the
NLO results will be presented in App.~\ref{app_fits}.

\subsection{Results in full QCD}
\label{sub_qcd}

\begin{figure}
	\begin{center}
		\includegraphics[width=0.49\textwidth]{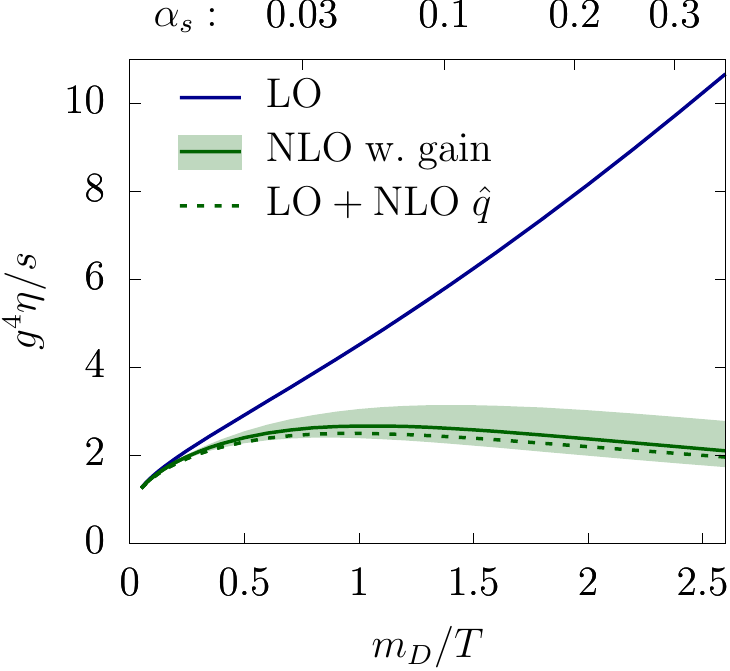}
      \hfill
		\includegraphics[width=0.49\textwidth]{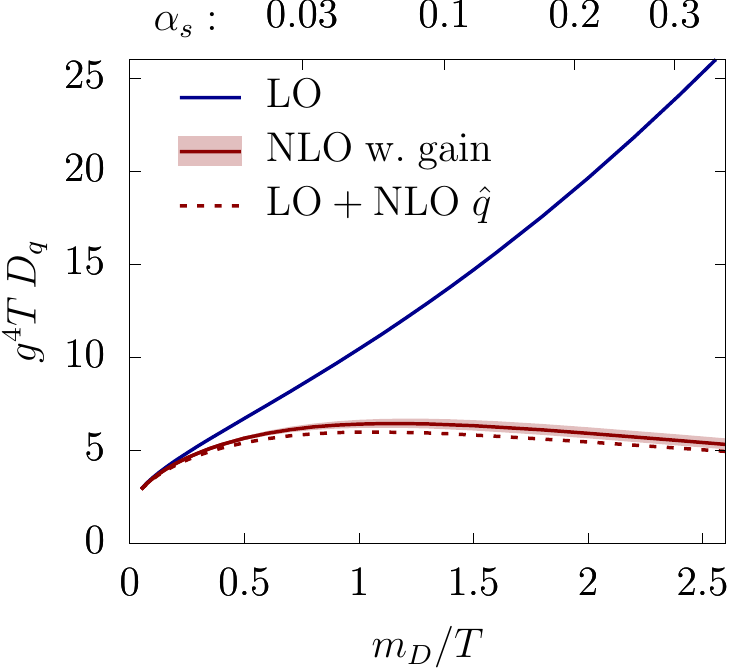}
	\end{center}
   \caption{
      (a) The shear viscosity to entropy ratio $\eta/s$, and (b) the baryon number diffusion coefficient $D_q$ (times temperature) normalized by the
	leading $1/g^4$  as a function
   of $\md/T$ for  QCD with $\nf=3$ light flavors. (The corresponding value
   of $\als$ is shown on the upper horizontal axis.) The LO result is
   from \cite{Arnold:2003zc}. The uncertainty from the unknown gain
   terms is shown by the bands; it is estimated from the leading-order gain terms (which 
   have the same structure as their NLO counterparts) by changing an
   unknown coefficient $C_\ell$ through the range $[-2,2]$ as specified by
   \Eq{nlogcross}  and \Eq{nlofcross}.
   The dashed lines represent an estimate in which we include only the
   NLO $\qhat$ to the LO collision operator  -- see \Eq{nloqhatonly} and surrounding text.
      \label{fig_visc_md} 
   }
\end{figure}
In Fig.~\ref{fig_visc_md} we show our results for the shear viscosity over entropy $\eta/s$  and
quark number diffusion $D_q$ as a function of $\md/T$ for QCD with $\nf=3$ flavors. 
We plot the LO results \cite{Arnold:2003zc} in solid blue, and for
NLO we plot our result in solid green (for $\eta/s$) and red (for $D_q$). 
To estimate the uncertainty from the undetermined NLO gain terms we provide
bands in the same color around these central $C_\ell=0$ NLO values. The bands
are obtained from taking $C_{\ell}$ in the range $[-2,2]$. This apparently arbitrary
choice  is motivated by comparing LO and NLO
momentum diffusion rates; the NLO $\qhat$ to LO $\qhat$ ratio is
$\delta \qhat^a/(\qhat^a\md/T/\ln(\mu_\perp/\md))$, which we can read off
from Eqs.~\eqref{qhatql} and \eqref{qhatnlo}; it ranges from $\sim 1$
for $\nf=6$ to $\sim 2.2$ in the pure Yang-Mills case.  Therefore
$|C_{\ell}|= 2$ appears to be a conservative value in estimating
resulting errors. As we point out in App.~\ref{sub_gain}, we have
made another conservative choice there in the evaluation of the
gain terms. The uncertainty arising from the \gainf terms is smaller 
for $D_q$ than for the
shear viscosity; in the former case we are dealing with the soft-fermion, $\ell=1$ term given
by \Eq{nlofcross}, whose LO value (Eqs.~\eqref{eq:gainconv} and \eqref{c1fermionfinal}) 
is numerically smaller than its $\ell=2$ counterpart (Eqs.~\eqref{gainterms} 
and \eqref{eq:cigluongain}).

In both cases the main difference between LO and NLO results arises from
$\delta \qhat$. 
This is reinforced by the dashed lines in Fig.~\ref{fig_visc_md},
which shows  results obtained from a collision operator containing,
beyond leading order, only the NLO corrections to $\qhat$:
\begin{equation}
    \Big( f_1, \C^{\delta\qhat}_\mathrm{only} \, f_1 \Big)
    \equiv
    \Big( f_1, \C_\LO\, f_1 \Big) +
	\label{nloqhatonly}
	    \Big( f_1, \C^{\delta\qhat} \, f_1 \Big)-
		\Big( f_1, \C^\twotwo_{\OO(g)\,\mathrm{finite}\,\qhat} \, f_1 \Big)\,,
\end{equation}
with the pertinent  ``counterterm'' $( f_1,
\C^\twotwo_{\OO(g)\,\mathrm{finite}\,\qhat} \, f_1 )$ given by
\Eq{finalogqhat}.  We see that this curve lies quite close to the
($C_{\ell}=0$) full NLO result, indicating that other corrections
are small or largely cancel each other.  But the $\delta \qhat$
contribution is so large that it starts to overtake the leading-order
collision operator before $\md = 1 T$ and represents a factor-5
modification for $\als=0.3$.
We present an accurate fit for the NLO curves (at $C_{\ell}=0$) in
App.~\ref{app_fits}, namely in Eqs.~\eqref{etafit}
and \eqref{viscfit3} for $\eta/s$ and in Eqs.~\eqref{dfit}
and \eqref{difffit3} for $D_q$.

\begin{figure}
	\begin{center}
		\includegraphics[width=0.65\textwidth]{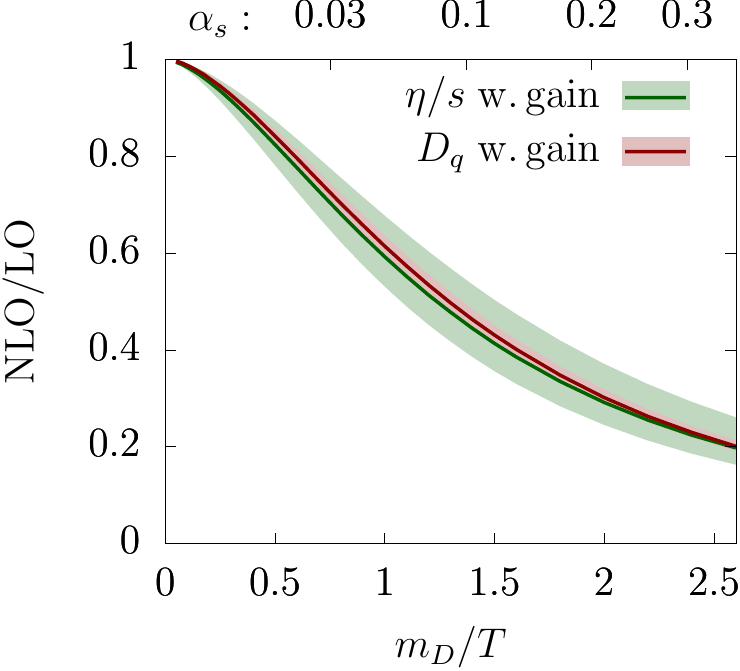}
	\end{center}
	\caption{The NLO/LO ratio for the shear viscosity and diffusion constant
	for QCD with $\nf=3$. The uncertainty bands from the unknown  gain terms are described in \Fig{fig_visc_md}.
	}
	\label{fig_ratio}
\end{figure}
In order to study more quantitatively the observed similar trend
between the NLO $\eta$ and $D_q$, compared to their respective leading
orders, we plot the NLO/LO ratios in Fig.~\ref{fig_ratio}, complete
with \gain uncertainty bands, as a function of $\md/T$ for QCD with
$\nf=3$. As the plot shows, the two central values fall within the
uncertainty bands.  Each transport coefficient is dominated by elastic
scattering and in each case the ratio of $\delta \qhat$ to the
leading-order elastic effect is about the same; therefore the trend
with coupling is very similar.

True vacuum renormalization effects will first arrive at NNLO (at
$\OO(g^2)$), so we do not yet see effects of coupling
renormalization.  This makes it difficult to use any internal
consistency to set the scale in our calculations.
Nevertheless, we are clearly very interested in plotting the
temperature dependence of the LO and NLO transport coefficients,
which requires picking a prescription for $g(T)$ and for the quark mass thresholds, with
the understanding that the different choices might differ starting parametrically
from NNLO. Various choices
are commonly employed in the literature. One widely used
prescription is to simply take the $\overline{\mathrm{MS}}$
coupling at $n$ loops, with threshold matching at $n-1$ loops, and
choose the renormalization scale $\mu$ to be a multiple of the
Matsubara frequency $2\pi T$ (usually a set of values such as
$\mu=\{0.5,1,2,4\}\pi T$ is employed to estimate the scale
setting uncertainty).
Another choice is to use the ``effective QCD coupling'', introduced
in \cite{Laine:2005ai} as the matching coefficient appearing in the dimensionally reduced
effective theory EQCD
(Electrostatic QCD, \cite{Braaten:1994na,Braaten:1995cm,Braaten:1995jr,Kajantie:1995dw,Kajantie:1997tt}).
The two-loop expression for this matching coefficient, as computed in \cite{Laine:2005ai}, is better
suited to describe the coupling in settings where contributions from the soft scales play a major role,
as the computation of the spatial string tension and comparison with lattice data in \cite{Laine:2005ai}
display. Since the LO results are dominated by the logarithmically enhanced diffusion and conversion
processes \cite{Arnold:2003zc},
which are very sensitive to the soft scale, and the NLO results are dominated by the large corrections
to $\qhat$, which are in turn determined from EQCD, we argue that the
EQCD coupling is the best choice for these transport
coefficients. Hence we will mostly use the EQCD coupling from
\cite{Laine:2005ai} in our plots.

\begin{figure}[ht]
	\begin{center}
		\includegraphics[width=0.65\textwidth]{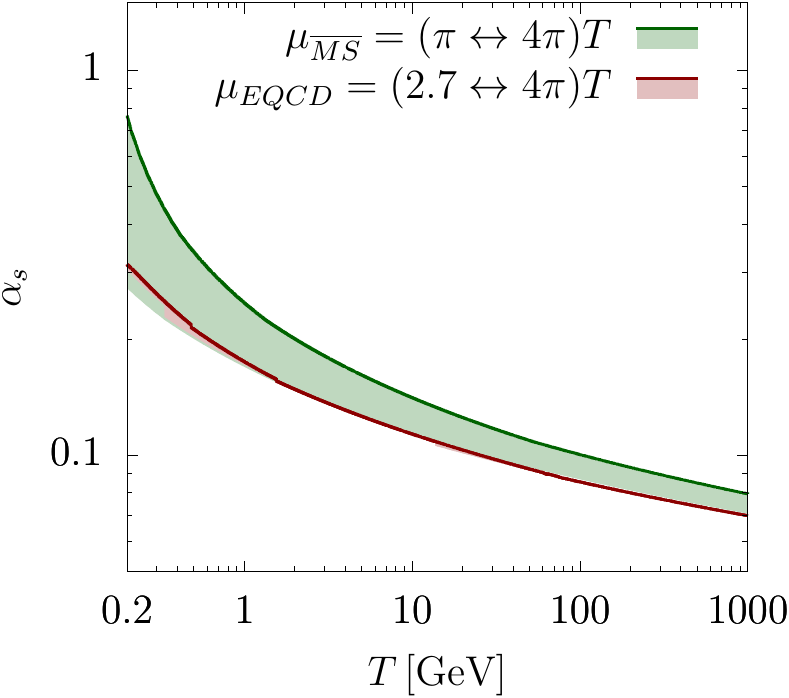}
	\end{center}
	\caption{The different choices for the coupling as a function
	of the temperature adopted in this work. 
   The green band (and bounding line) is obtained
   from a two-loop QCD $\MS$ running with one-loop threshold matching (and  hence is continuous at the thresholds) with $\mu_{\overline{MS}}=(\pi \leftrightarrow 4\pi)T$. 
   The green bounding line is for the smallest scale $\mu_{\overline{MS}}=\pi T$.  
   The corresponding  red band and bounding line are obtained from the EQCD
   effective coupling  with $\mu_{EQCD}=(2.7\leftrightarrow
   4\pi)\, T$~\cite{Laine:2005ai}, which is discontinuous at the quark
   thresholds.  }
	\label{fig_grun}
\end{figure}
We start by plotting the coupling itself, as shown in Fig.~\ref{fig_grun}. The detailed definitions
for the two choices of the coupling are given in App.~\ref{app_coupling}. The green line and band
 represent
the QCD $\MS$ coupling as given by \Eq{qcdevol},
 obtained from a numerical two-loop evolution from $\als(M_z)=0.1185$, with  the
$\overline{\mathrm{MS}}$ renormalization scale $\mu$ in the range
$[\pi T,4\pi T]$ and with one-loop
 quark threshold matching at $\mu=m_q$, hence the continuity.
The red line and band are instead obtained from the effective
EQCD coupling, as given by \Eq{eqcdevol}, with
threshold matching at $\mu=m_q$ and with renormalization scale $\mu$ in the range
$[2.7 T,4\pi T]$.
The lower bound ($\mu=2.7 T$) is at the
quark mass value ($m_q=2.7 T$) where a quark contributes half as much
(Stephan-Boltzmann) entropy as a massless quark, which we therefore
pick as our criterion for the quark's decoupling temperature (for
instance, the $b$ quark decouples at $T_b\approx 1.55$ GeV under this
choice). As we remark in
App.~\ref{app_coupling}, the matching to the EQCD coupling cancels
the leading renormalization point dependence, which is why the EQCD
curves are nearly identical. 

\begin{figure}[ht]
	\begin{center}
		\includegraphics[width=0.49\textwidth]{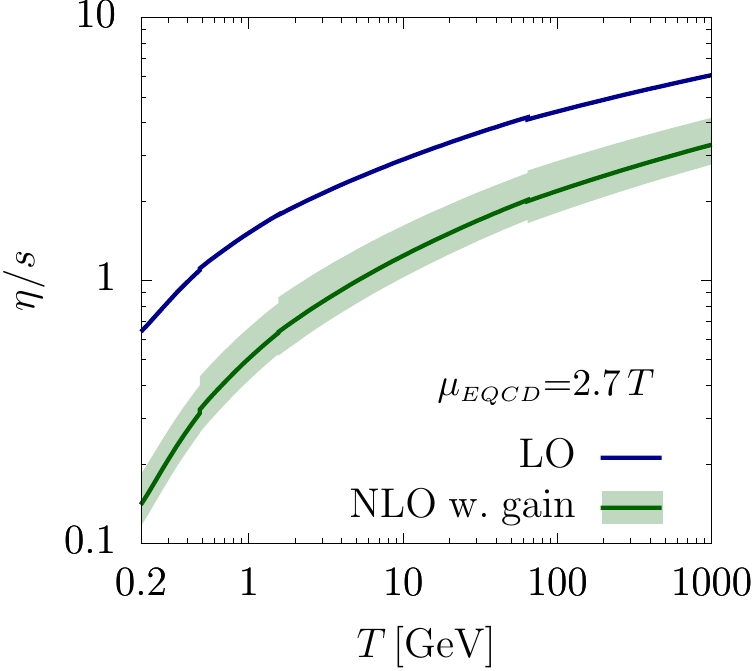}
		\includegraphics[width=0.49\textwidth]{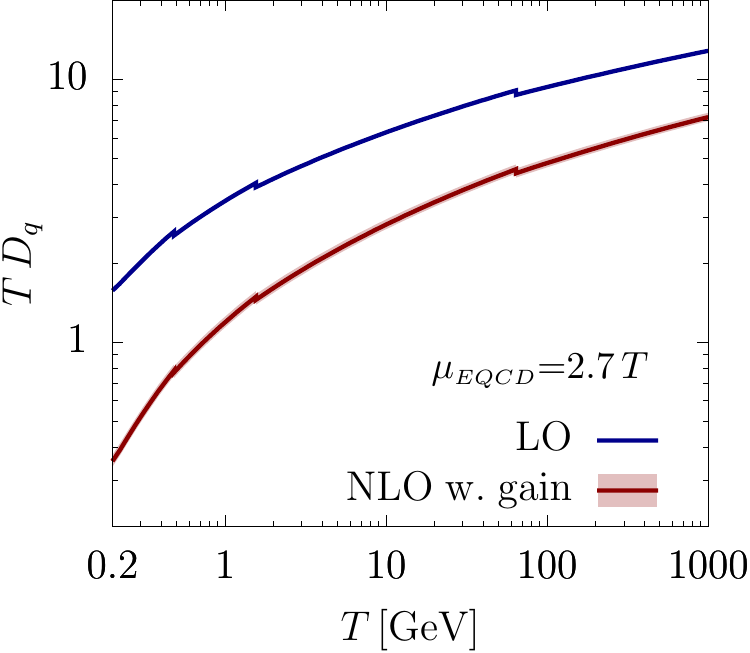}
	\end{center}
   \caption{(a) The QCD  shear viscosity to entropy density ratio
     $\eta/s$, and (b) diffusion coefficient $D_q$ as a function of
     $T$. All curves in this plot are obtained using the effective
     EQCD coupling with $\mu_{EQCD}=2.7\,T$. The uncertainty due to
     the gain terms which are estimated (and not computed) is shown by
     the shaded green and red bands respectively.   This uncertainty
     estimate is described in \Fig{fig_visc_md}.
   \label{fig_run} }
\end{figure}
Fig.~\ref{fig_run} shows the LO (blue) and NLO  results for $\eta/s$ (green) and $D_q$ (red)  with the effective EQCD coupling,
set at the entropy-motivated prescription $\mu=2.7 T$. At the quark mass
thresholds we switch from describing a system with $\nf+1$ massless
quark flavors to describing a system with $\nf$ massless flavors,
leading to a discontinuity in the coupling, the entropy density
 and the transport coefficients, and therefore in each curve.  Our treatment is
insufficient near each threshold because we have not developed an
$\eta$ (or $D_q$) calculation which correctly treats massive quarks.
We show the uncertainty bands corresponding to the previous values for the arbitrary
constant in \Eq{nlogcross}: $C_{\ell}=\pm 2$. As the plot shows,
in the $\eta/s$ case
the uncertainty band due to missing NLO (gain) contributions grows
larger as $\nf$ increases with increasing temperature. This is
because the LO \gain term, which multiplies $C_{\ell=2}$ in \Eq{nlogcross},
has terms proportional to $\nf$ and to $\nf^2$, as can be inferred from \Eq{gainterms}.
We remark that, as expected from Fig.~\ref{fig_visc_md}, the NLO results are much smaller than
the leading order: at temperatures of the order of the QCD
transition the NLO $\eta/s$ is smaller by a factor of 5, which becomes a factor
of two for $T\sim1$ TeV.%
\footnote{%
  We present these high-temperature
  results only to analyze the convergence of the perturbative
  series. They do not apply to the early universe at these
  temperatures, where electroweak and leptonic degrees of freedom,
  absent from this calculation, would play a major role.  In fact, for
  early universe applications, electroweak degrees of freedom will
  \textsl{always} play a dominant role. }
The \gain uncertainty band, on the other
hand, represents a $+30\%$, $-20\%$ correction to the NLO result.
In terms of the strong-coupling results \cite{Policastro:2001yc,Policastro:2002se,CaronHuot:2006te}, 
 the NLO results for $\eta/s$ ($D_q$) can get smaller than $2/(4\pi)$  ($2/(2\pi T)$)
 at the lowest temperatures,
corresponding to couplings of the order of $\als\sim0.35$.

In Fig.~\ref{fig:intro} we analyzed another source of theoretical uncertainty, arising from
a different scheme for the running coupling. Besides the LO and NLO results with the EQCD
effective coupling, already presented in Fig.~\ref{fig_run},
we also show results obtained from the two-loop QCD $\MS$ coupling
discussed above. As the plot shows, the LO and NLO uncertainty bands
introduced by the  different choices adopted for
the renormalization scale are well separated (except
at the lowest temperatures, where, as Fig.~\ref{fig_grun} shows, the $\MS$ coupling
for $\mu=\pi T$ is $\OO(1)$). This is consistent
with the expectation that the running coupling is an NNLO effect and should thus be smaller than NLO
corrections.

\subsection{Results in pure Yang-Mills}
\label{sub_gauge}

Pure Yang-Mills theory is only of interest for academic reasons.
Nevertheless, since it is straightforward, and since most lattice
results for the viscosity \cite{Nakamura:2004sy,Meyer:2007ic,Astrakhantsev:2015jta,%
  Astrakhantsev:2017nrs,Mages:2015rea,Pasztor:2016wxq,Pasztor:2018yae},
as well as analytical studies \cite{Keegan:2015avk},
are actually for pure Yang-Mills theory and not full QCD, we will
present results for this case.
\begin{figure}
	\begin{center}
		\includegraphics[width=0.65\textwidth]{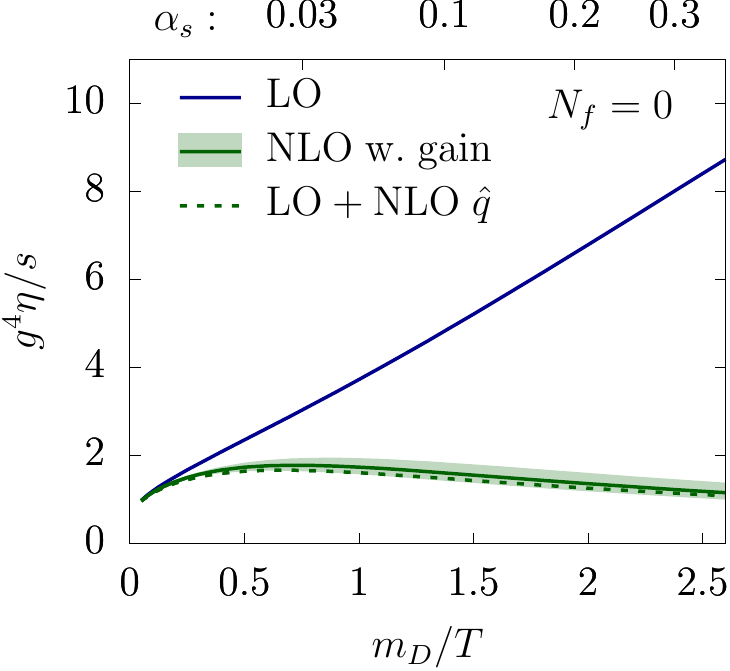}
	\end{center}
	\caption{The $\eta/s$ ratio in units of $1/g^4$ as a function of $\md/T$
	for $\nc=3$ Yang-Mills theory.  The uncertainty band is due 
   to the unknown gain terms described in \Fig{fig_visc_md}.  
	A fit for the solid green curve is available in Eqs.~\eqref{etafit} and
	\eqref{viscfitquench}.
   The dashed line shows a partial NLO result with only the NLO modifications
   to $\hat q$.
   }
	\label{fig_visc_quench_md}
\end{figure}
In Fig.~\ref{fig_visc_quench_md} we show the $\eta/s$ ratio in pure Yang Mills for $\nc=3$. The
general trends are the same as in full QCD but, interestingly, the NLO/LO ratio is smaller as a function
of $\md/T$ than it is for full QCD. When examined in terms of $g$ (see the upper scale in $\als$) they are however
similar.  It is also worth noting that the absolute values for
$\eta/s$ are larger than for $\nf=3$.

\begin{figure}
  \begin{center}
     \includegraphics[width=0.65\textwidth]{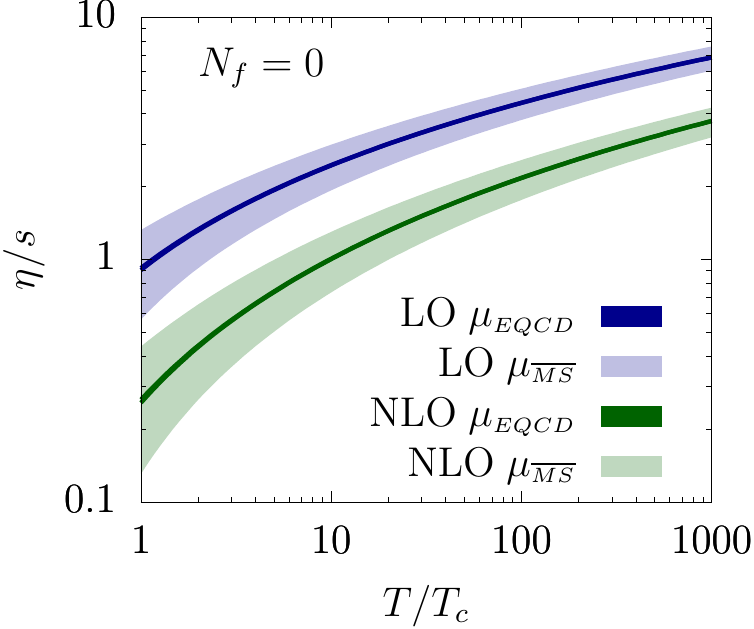}
  \end{center}
  \caption{The $\eta/s$ ratio as a function of the temperature
    for $\nc=3$, $\nf=0$ pure Yang-Mills theory.
   The thin darker bands and the thicker lighter bands respectively show 
   the EQCD and $\overline{MS}$ 
   renormalization schemes  with $\mu=(\pi \leftrightarrow 4\pi) T$. 
   \label{fig_visc_quench} }
\end{figure}
In Fig.~\ref{fig_visc_quench} we plot the $\eta/s$ ratio in $\nc=3$
Yang-Mills theory as a function
of the temperature. The coupling is fixed as follows:
\begin{itemize}
\item
  At a sufficiently high scale we impose the two-loop asymptotics
  $\als(\mu)/\pi=-8/(\beta_0 t)-16\beta_1\ln(t)/(\beta_0^3t^2)$,
  with $t=\ln(\mu^2/\Lambda^2_{\overline{\mathrm{MS}}})$, $\beta_i$ as
  given by \Eq{betafunction} and
  $\Lambda_{\overline{\mathrm{MS}}}=1.24 T_c$ \cite{Francis:2015lha},
  with $T_c$ the critical temperature.
\item
  For the two-loop QCD $\MS$ coupling, this asymptotic value is then evolved
  down to lower scales using the two-loop $\beta$-function in
  \Eq{qcdevol}. We present LO and NLO results as wide blue and green
   bands respectively, reflecting the renormalization
  scale uncertainty.
  Uncertainties arising from the
  $\Lambda_{\overline{\mathrm{MS}}}/T_c$ ratio or from the two-loop
  truncation of the $\beta$-function should be smaller than the
  large bands arising from the variation of the renormalization
  scale.
\item
  For the effective EQCD coupling we use \Eq{eqcdevol} as before. The displayed
  darker blue (LO) and green (NLO)
  bands, for the the same $\mu$ interval as in the $\MS$ case, 
  are much narrower, given that the dependence on
  $\mu$ is very small in the absence of the discontinuities at the
  quark mass thresholds. 
\end{itemize}
In this case one observes again two non-overlapping bands
for the LO and NLO shear viscosity. Due  to the smaller values of the
couplings%
\footnote{%
  For comparison, in pure glue and for $\mu=\pi T$, the effective EQCD
  coupling is $\als{}_\mathrm{EQCD}(T=T_c)=0.1945$, corresponding to
  $\md/T=1.563$, while in QCD for
  $\mu=2.7T$ $\als{}_\mathrm{EQCD}(T=177\,\mathrm{MeV})=0.3244$,
  corresponding to $\md/T=2.473$. (The EQCD coupling in QCD with fermions
   for $\mu=2.7T$ breaks down shortly below $T=177\,\mathrm{MeV}$.)}
the shear viscosity over entropy density
is larger than in full QCD in the transition region.

\subsection{Results in QED}
\label{sub_qed}
We have also obtained the shear viscosity for QED. In this 
theory the large NLO $\qhat$ contribution is absent, due to its
non-abelian nature, and the coupling is small. We have taken
 $\alpha=0.0072973525664$ and one massless Dirac fermion,
 so as to describe an electron-positron-photon plasma
 at $m_\mu\gg T\gg m_e$. In QED  $m_D^2=e^2T^2/3$ and 
hence $m_D/T\approx 0.17$.
At leading order we obtain
\begin{equation}
	\label{loqed}
	\frac{\eta}{s}\bigg\vert_\mathrm{QED}^\mathrm{LO}=2779.2,
\end{equation} 
whereas our next-to-leading order results, for three values of the gain constant
$C_{\ell=2}$, are
\begin{equation}
	\label{nloqed}
	\frac{\eta}{s}\bigg\vert_\mathrm{QED}^\mathrm{NLO,\,C_{\ell=2}=0}=2818.5,\qquad
	\frac{\eta}{s}\bigg\vert_\mathrm{QED}^\mathrm{NLO,\,C_{\ell=2}=+2}=2836.1,\qquad
	\frac{\eta}{s}\bigg\vert_\mathrm{QED}^\mathrm{NLO,\,C_{\ell=2}=-2}=2801.2.
\end{equation} 
Hence, the NLO central value ($C_{\ell=2}=0$) corresponds
to a 1.4\% increase over the LO shear viscosity, while the upper
and lower values correspond to a 2\% and a 0.8\% increase respectively.
We can thus conclude that the abelian NLO corrections tend to decrease
the collision operator and that in the case of QED perturbation theory works
very well.

\section{Conclusions}
\label{sec:conclusion}

The main aim of this paper has been to compute the shear viscosity and
quark diffusion coefficient of QCD at ``almost'' NLO in $g$.  This
involved partially resumming some $\OO(g)$ effects in the
leading-order treatment, and some $\OO(g^2)$ effects in the NLO
treatment, in
order to maintain positivity of the collision operator.  Also we
invert the full $\C+\delta \C$ (leading plus next-to-leading order)
collision operator, rather than expanding in $\delta \C$ as suggested
in \Eq{C_expand}.
It also
involved neglecting \gain terms%
\footnote{%
  We emphasize again that despite the name, these terms are not
  manifestly positive and it is unclear whether their correct
  inclusion would increase or decrease $\eta$.}
which we could not compute at NLO,
but which proved to be small at leading order.  We have estimated the
possible effects of these missing contributions and found they are
likely quite small. From a technical standpoint, the most important result
of this paper is the methodology introduced in Sec.~\ref{sec:reorg},
where we introduce a $\onetwo$ rate, \Eq{seminew}, which smoothly extends into
regions of soft or semi-collinear (less collinear) radiation, without
the need for intermediate regulators. In this paper we have only needed
to treat this new equation in the single-scattering (Bethe-Heitler) regime, but
it would be interesting to try to solve it as an integral equation, thereby
incorporating LPM interference when needed. We leave this, together
with applications of this
approach to thermalization or jet quenching, to future studies.

The qualitative trend observed for the shear viscosity and the light quark diffusion coefficients
as a function of the coupling $\md/T$,
both in QCD with three light fermions and in the pure gauge theory, is as follows
(see Figs.~\ref{fig_visc_md}, \ref{fig_ratio},
\ref{fig_visc_quench_md}): the
NLO curves in green ($\eta/s$) or red ($D_q$) start to diverge significantly from the LO
ones in  blue for $\md/T\sim 0.5$, with the NLO transport coefficients
becoming as small as one fifth of the LO for values of $\md/T$ corresponding to $\als\sim0.3$.
Furthermore, the uncertainty band introduced by considering a rather large
value for the \gain terms at NLO only modify the NLO transport
coefficients by 30\% at most.
In Figs.~\ref{fig:intro}, \ref{fig_run} and \ref{fig_visc_quench}, we plot
instead the transport coefficients as functions of the temperature, which requires picking a prescription
for the coupling as a function of the temperature and for the decoupling of heavy quarks. The LO and NLO
curves do not overlap, even accounting for the uncertainties arising
from the choice of the running prescription, renormalization scale and
decoupling point.  Therefore the limitations of perturbation theory
are much more severe than simply the question of what to choose for
the renormalization point.  Indeed, even at temperatures of order one
TeV, where perturbation theory would be expected to work well, the NLO
transport coefficients are smaller than the LO value by about a factor
of 2.

The dashed curves in Figs.~\ref{fig_visc_md}
and \ref{fig_visc_quench_md} show that by far the dominant NLO effect
is the large NLO correction to $\qhat$, first derived in
Ref.~\cite{CaronHuot:2008ni}.  This should perhaps not be too
surprising.  The corrections to splitting rates are not small, but they
tend to be compensated by the semi-collinear ones \cite{Ghiglieri:2013gia}.
And as emphasized in \cite{Arnold:2003zc}, elastic scattering
is the principal contributor to shear viscosity and number diffusion,
with splitting processes amounting to 10--20\% effects.  Further, the
NLO contributions to $\qhat$ represent new physical processes not
included at leading order; the inclusion of additional soft emissions
in the course of scattering and interference between different
scattering processes.  Unfortunately the Euclidean methods used to
compute $\delta \qhat$ do not allow us to evaluate these contributions
separately. In order to test a theory that is not sensitive to this
large $\delta\qhat$ contribution, we examined QED in Sec.~\ref{sub_qed},
finding that the remaining abelian NLO contributions are a percent-level correction.

One important question to be addressed is what should we make of a perturbative expansion
that does not converge above $\md/T\sim 0.5$, or equivalently below temperatures well above the
TeV scale. Taken at face value, the results plotted in Sec.~\ref{sec_results} would perhaps
suggest a grim answer to this question. However, one could optimistically think that,
if we were to correctly identify the physics responsible for these
large corrections, and rearrange the perturbative expansion
by resumming it, possibly in the form of an Effective Field Theory, then
the outlook on convergence would be quite different; a redefined LO somehow incorporating most
of the NLO corrections to $\qhat$ would not look so different from the dashed lines
in Fig.~\ref{fig_visc_md}, so that the deviation from the NLO in solid green/red would be much
less pronounced. Of course, much action is needed to move this scenario
from the realm of wishful thinking into physically motivated perturbative
schemes.  One possible direction
would be to treat the problematic soft sector non-perturbatively. The mapping to the Euclidean 3D
theory makes a lattice determination of the soft contribution to $\mathcal{C}(\qp)$ and
$\qhat$ possible, with first results reported in \cite{Panero:2013pla}. Refinements of this
measurement, together with calculations of the shift in the dispersion relation $\delta\mmf$
and of $\qhat(\delta E)$, seem within reach, due to their Euclidean
nature.  We also need a better understanding of how such Euclidean
measurements can be systematically included into transport
calculations within a rigorous Effective Field Theory framework.
Other needed ingredients, such as the longitudinal momentum
broadening, conversion rates and \gain terms, on the other hand,
cannot be mapped to the 3D Euclidean theory and cannot thus be
currently determined on the lattice. Therefore we should view it as
good news that these effects appear to be much smaller than $\delta
\qhat$.  One might hope that, with enough nonperturbative Euclidean
contributions, the perturbative approach might work down closer to
experimentally realizable temperatures.

\section*{Acknowledgments}
JG would like to thank Aleksi Kurkela, Marco Panero and P\'eter Petreczky
for useful conversations.
GM would like to acknowledge support by the Deutsche
Forschungsgemeinschaft (DFG) through the grant CRC-TR 211
``Strong-interaction matter under extreme conditions.''
DT would like to acknowledge support by the U.S. Department of Energy
through the grant DE-FG02-88ER40388.
%
%

\appendix

\section{Hard Thermal Loop propagators}
\label{app_props}

In the next appendices we will look at matrix elements with soft
exchange momenta in more detail.  Therefore we need to specify the
hard thermal loops, which appear in the expressions for these soft,
screened matrix elements.  We start with the fermionic HTLs,
which are most easily written in terms of components with positive and
negative chirality-to-helicity ratio. The retarded fermion propagator
reads
\begin{equation}
\label{htlfermiondef}
S_{R}(P)=h^+_\bp S^+_{R}(P)+h^-_\bp S^-_{R}(P)\,,
\end{equation}
where
\begin{equation}
\label{htlfermion}
S^{\pm}_R(P)=\frac{i}{p^0\mp (p+\Sigma^\pm(p^0/p))}
= \left.\frac{i}{\displaystyle p^0\mp\left[p+\frac{\mmf}{2p}
\left(1-\frac{p^0\mp p}{2p}\ln\left(\frac{p^0+p}{p^0-p}
\right)\right)\right]}\right\vert_{p^0=p^0+i\epsilon},
\end{equation}
where the upper (lower) sign refers to the positive (negative)
chirality-to-helicity component. The  projectors are
$h^\pm_\bp\equiv(\gamma^0\mp\vec\gamma\cdot\hat p)/2$. Here $\mmf$ is
the fermionic asymptotic mass squared, defined such that the
large-momentum dispersion relation for helicity=chirality fermions is
$p_0^2 = p^2 + \mmf$.  We similarly define the asymptotic gluonic
mass $\mmg$.  At leading order, their values are
\begin{equation}
\label{loasym}
\mmg=\frac{\md^2}{2}=\frac{g^2T^2}{6}\left(\nc+\frac{\nf}{2}\right),
\qquad
\mmf=2 m_q^2 = \cf\frac{g^2T^2}{4},
\end{equation}
where we have also shown the relations to the more commonly used Debye
mass $\md$ and quark ``mass'' $m_q$.

Gluons are described in the strict Coulomb gauge by
\begin{eqnarray}
\label{htllong}
G^{00}_R(Q)&=&\frac{i}{\displaystyle q^2+\md^2\left(1-\frac{\omega}{2q}\ln\frac{\omega+q+i\epsilon}{\omega-q+i\epsilon}\right)},\\
\nn G^{ij}_R(Q)&=&(\delta^{ij}-\hat q^i\hat q^j)G^T_R(Q)=
\left.\frac{i(\delta^{ij}-\hat q^i\hat q^j)}
     {\displaystyle \omega^2-q^2-\mmg \left(\frac{\omega^2}{q^2}
       -\left(\frac{\omega^2}{q^2}-1\right)\frac{\omega}{2q}
       \ln\frac{\omega{+}q}{\omega{-}q}\right)}\right\vert_{\omega=\omega+i\epsilon}.\\
&& 	\label{htltrans}
\end{eqnarray}

\section{Gain terms and finite order-g subtractions in 2 to 2 processes}
\label{sec_double}
In this section we first provide some details on the phase
space integration coordinates in Sec.~\ref{sub_phase}. 
We then evaluate numerically the \gain
terms at leading order in Sec.~\ref{sub_gain}.
In Sec.~\ref{sub_double} we will instead address the $\OO(g)$ contributions
in the $2\leftrightarrow 2$ collision operator that need to be subtracted, i.e.\
$( \chi_\ij, \C^\twotwo_{\OO(g)\,\mathrm{finite}} \, \chi_\ij )$ in \Eq{nlo1}.

\subsection{Phase space}
\label{sub_phase}
In Sec.~\ref{sec:softglue} we provided the phase
space integration in the soft approximation in \Eq{eq:softps}.
We now set out to briefly justify that equation and provide more 
elements for the evaluations that will be performed in Sec.~\ref{sub_gain}
and \ref{sub_double}. One starts by eliminating a variable through the three-momentum
$\delta$-function,
 \begin{align}
    \int_{PS} \equiv \int_{\p\k\p'\k'} (2\pi)^4 \delta(P + K - P' - K')
    =  \int_{\p\k\p'}  2\pi \delta(p + k - p' - k') \, .
    \label{softpsstart}
 \end{align}
In the $t$ channel,  $\p'$ can then be shifted to $\q=\p'-\p$ and an extra $\omega$ integral
is introduced, i.e.
 \begin{align}
    \int_{PS}     =  \int_{\p\k\q}\int d\omega \, 2\pi \delta(p  - p' +\omega)\,\delta(  k  - k'-\omega) 
	    \simeq  \int_{\p\k} \int \frac{d^4Q}{(2\pi)^4} \, 2\pi \delta(v_\p \cdot Q) \, 2\pi  \delta(v_\k \cdot Q)\, ,
    \label{softpinter}
 \end{align}
where we have introduced the four-vector $Q=(\omega,\q)$ and expanded the arguments of
the $\delta$-functions for $\omega,q\sim g T\ll p,k$, recovering \Eq{eq:softps}.
Using the coordinate parameterization of \cite{Arnold:2003zc}, \Eq{softpinter}
becomes
 \begin{align}
   \nn \int_{PS} \bigg\vert_\mathrm{soft}    &=  \frac{1}{32\pi^6} \int_0^\infty dp\,p^2  \int_0^\infty dk\, k^2 \int_0 dq 
	\int_{-q}^q d\omega \int_0^{2\pi}d\phi\\
	&= \frac{1}{32\pi^6} \int_0^\infty dp\,p^2  \int_0^\infty dk\, k^2 \int_0 d\qp\,\qp 
	\int_{-\infty}^\infty \frac{d\omega}{q} \int_0^{2\pi}d\phi\, ,
    \label{softpamy}
 \end{align}
 where $\phi$ is the angle between the $\p,\q$ and $\k,\q$ planes and in going from
 the first to the second line we have used the change of variables discussed in
 footnote~\ref{foot_coords}, with $q=\sqrt{\omega^2+\qp^2}$. 
  For future convenience we recall that in these
 coordinates and in the soft approximation
 \begin{equation}
	 \label{softangle}
 	\hat\p\cdot\hat\k=\frac{\omega^2}{q^2}+\left(1-\frac{\omega^2}{q^2}\right)\cos\phi\,.
 \end{equation} 
\subsection{LO \gain terms}
\label{sub_gain}
Let us begin with the gluon exchange contribution at leading order.
We recall that it only contributes for $\ell=2$.  Starting
from \Eq{gainterms} and \Eq{eq:cigluongain}, using the results in Sec.~\ref{sub_phase}
for the phase space and the $\hat\p\cdot\kh$ angle, as well as the explicit
form of the propagators in App.~\ref{app_props}, we have for $c_1$
\begin{align}
    \nn c_1 &=  \int \frac{d\Omega_{\k}}{4\pi}\int \frac{d^4Q}{(2\pi)^4}  \, |G_{\mu\nu}^R(Q) v_\p^{\mu} v_{\k}^{\nu} |^2  \,
    2\pi \delta(v_\p \cdot Q) \, 2\pi  \delta(v_{\k} \cdot Q)\, \omega^2 P_2(\hat\p\cdot\hat\k)\\
	&= \frac{1}{8\pi^2} \int_0^\infty d\qp\,\qp 
	\int_{-\infty}^\infty \frac{d\omega}{q} \int_0^{2\pi}d\phi \left\vert G_L^R(Q)+\frac{\qp^2\cos\phi}{q^2}G_T^R(Q)\right\vert^2
	\omega^2 P_2(\hat\p\cdot\hat\k)=\frac{0.3066}{4\pi},
	\label{c1final}
\end{align}
where the $\qp$ integration has been extended to infinity,\footnote{%
\label{foot_gaincoords}
The integral converges without $q_\perp$ cutoff because the result of
the $\omega$ integration vanishes faster than $1/q_\perp^2$.  However,
this happens due to cancellations; the \textsl{absolute} convergence
of the $\omega$ integral is slower.  Therefore the result is actually
dependent on our integration choice (like many convergent but not
absolutely convergent integrals).  If we integrate
$\int_0^{\mu} dq \int_{-q}^q d\omega$ then we also get a valid
$\mu\to \infty$ limit but with a different answer.  The difference
between integration choices is the integral over a region lying
between a sphere and the superscribed cylinder; for instance, for
$c_1$ and $\ell=2$, and performing the integrals from innermost to
outermost, we find
\begin{equation}
	\int_0^{\infty} q_\perp dq_\perp \int \frac{d\omega}{q}
        \int_0^{2\pi}d\phi \,\Theta(q_\perp^2 + \omega^2 - \mu_\perp^2)
	\frac{(1-\cos\phi)^2}{q^4}
	\omega^2 P_2(\hat\p\cdot\hat\k) = \frac{27\pi}{70},
	\label{coordterm}
\end{equation}
where we have used  bare matrix elements for this UV integral.
This is an ambiguity in the soft part of the leading-order gain term,
which is canceled by a matching ambiguity in the hard part.  The gain
term found with the $q_\perp$ coordinates is larger, leading to a more
conservative estimate for the gain-term uncertainty.
}
and has been performed numerically, together with the 
$\omega$ integration. The other coefficients in \Eq{eq:cigluongain} can be computed analogously,
yielding
\begin{equation}
	\label{c2c3final}
	4\pi\, c_2=0.1360\,,\qquad 4\pi\, c_3=0.1833\,.
\end{equation}

The fermion exchange contribution only arises at $\ell=1$.  Starting
with \Eq{eq:c1fermion} and using Eqs.~\eqref{eq:convcqg} and \eqref{softqmat},
together with the same techniques as
the gluonic case, we have
 \begin{align}
\nn c_{1} &= \int \frac{d\Omega_{k}}{4\pi} \C_{q \rightarrow g}^{\rm conv}(\ph \cdot \kh) P_{1}(\ph \cdot \kh)\nn\\
	&=\frac{g^4 \cf^2}{16\pi^2} \int_0^\infty d\qp\,\qp 
	\int_{-\infty}^\infty \frac{d\omega}{q} \int_0^{2\pi}d\phi\left[\left(1-\frac{\omega}{q}\right)^2\left\vert S^+_R(Q)\right\vert^2
	+\left(1+\frac{\omega}{q}\right)^2\left\vert S^-_R(Q)\right\vert^2\right.\nn\\
    &\hspace{5.1cm}\left.
     -\frac{\qp^2}{q^2}\cos\phi \left(
    S^+_R(Q)S^-_A(Q)+S^-_R(Q)S^+_A(Q)\right)\right]P_{1}(\ph \cdot \kh)\nn\\
	&=0.9283\,\frac{g^4\cf^2}{8\pi},
	\label{c1fermionfinal}
 \end{align}
 where again the $\omega$ and $\qp$ integrations have been carried out numerically.

\subsection{Order-g terms}
\label{sub_double}
In \Eq{nlo1}
we have introduced
$( f_1, \C^\twotwo_{\OO(g)\,\mathrm{finite}} \, f_1)$
as the $\OO(g)$ region of the $\twotwo$ processes that needs to be subtracted.
As we have argued, both gluon and quark exchange processes contribute to it. Let us
then write it in terms of $\chi_\ij$ as
\begin{equation}
	\Big( \chi_\ij, \C^\twotwo_{\OO(g)\,\mathrm{finite}} \, \chi_\ij \Big)=
	\Big( \chi_\ij, \C^\twotwo_{\OO(g)\,\mathrm{finite}\,g} \, \chi_\ij \Big)+
	\Big( \chi_\ij, \C^\twotwo_{\OO(g)\,\mathrm{finite}\,q} \, \chi_\ij \Big),
\end{equation}
where the $g$ and $q$ labels stand for gluon and quark (and antiquark) exchange
contributions.

Let us then begin by evaluating the gluon exchange
contribution. As we have stated in
Sec.~\ref{notstrict}, we need to consider the region where
 $\omega$ and $q$ and an external gluon line ($p$ or $k$) are soft.
Let us then take these assumptions in \Eq{22general} and simplify
accordingly:
\begin{align}
\Big( \chi_\ij, \C^\twotwo \, \chi_\ij \Big)_{\mathrm{soft\, g}\,k}
    &=
       \frac{\beta}{(4\pi)^6}
       \sum_{a}
	   \int_{-\infty}^{+\infty} d\omega
       \int_0^{\mu_\perp} d\qp\frac{\qp}{q}
       \int_{0}^\infty dp 
       \int_{\frac{q+\omega}{2}}^{\mu_k} \frac{dk}{k(k-\omega)}
       \int_0^{2\pi} d\phi
   \nonumber \\ & \hspace {1cm} {}\times
   	2(2-\delta^{ag})\left|{\cal M}^{ag}_{ag}\right|^2_{\mathrm{soft\, g}\,t\,k}\>
   	f^a_0(p) \,  [1 \pm f^a_0(p)] 
   \nonumber \\ & \hspace {1cm} {}\times
	\Bigl[\chi^a_\ij(\p)+\chi^g_\ij(\k)-\chi^a_\ij(\p')
		-\chi^g_\ij(\k')\Bigr]_{\strut}^2.
\label {eq:tchannogstart}
\end{align}
We have already switched to $\omega,\qp$ coordinates;
 $\mu_\perp$ and $\mu_k$ are cutoffs separating the soft and hard scales.
The three soft integrations in $\qp$, $\omega$ and $k$ contribute to
a factor of $g^3$, the soft expansion of the Bose-Einstein distributions
contributes a factor of $1/g^{2}$, which is compensated by the $g^2$
behavior of the matrix element squared (see \Eq{softmatgluegeneric})
and finally the departure from equilibrium contributes another $g^2$,
bringing the total to $g^5$. The last line becomes
 \begin{align}
    	\left[
    	    \chi^a_\ij(\p) + \chi^g_\ij(\k) - \chi^a_\ij(\p') - \chi^g_\ij(\k')
    	\right]^2_{\mathrm{soft}\,k}
	 &=\omega^2\big[ (\chi^{a}(p)')^2+(\chi^{g}(0)')^2\big]\nn\\
	&\hspace{-8cm}+\frac{\ell(\ell+1)}{2}\qp^2\left(\frac{\chi^{a}(p)}{p}\right)^2
	-2k(k-\omega)\left[P_\ell\left(1-\frac{\qp^2}{2k(k-\omega)}\right)-1\right](\chi^{g}(0)')^2+\OO(g^3),
	 \label{offeqg}
 \end{align}
where the $(\chi^{g}(0)')^2$ arises due to the infrared
nature of the $\ell=2$ departure from equilibrium, as illustrated in
App.~\ref{sub_ir}. We have not included \gain terms, where 
contributions proportional to $\chi(p)\chi'(0)$ or $\chi'(p)\chi'(0)$ would arise.
Since we do not know the NLO corrections to the \gain terms, it makes little sense to subtract
this contribution: in our current Ansatz, \Eq{nlogcross}, it just amounts
to picking a different arbitrary constant $C_{\ell=2}$.
Finally, as we argue in Sec.~\ref{notstrict}, the matrix element in this scaling can be obtained from 
App.~A of \cite{Arnold:2003zc}. It reads
\begin{equation}
	 \label{softmatog}
	\left|{\cal M}^{ag}_{ag}\right|^2_{\mathrm{soft\, g}\,t\,k}=16 d_A\ca T_{R_a}g^4 p^2\left\vert
	(2k-\omega)G^L_R(Q)+\frac{\qp^2}{q^2} \cos(\phi)
	\sqrt{4k(k-\omega)-\qp^2}G_R^T(Q)\right\vert^2,
  \end{equation}
 which correctly reduces to \Eq{softmatglue} for $k\gg \omega,q$. 
For what concerns the symmetry factors, the $gg\lra gg$ process receives a factor of 2
from the identical $u$-channel contribution and a factor of 2 from the $p\sim gT$,
$k\sim T$ region. The $qg\lra qg$ process receives a factor of 4 from the initial
and final state symmetries, so that the full contribution is
\begin{align}
\Big( \chi_\ij, \C^\twotwo_{\OO(g)\,g} \, \chi_\ij \Big)
    &=
       \frac{	d_A \ca g^4}{32 \pi^5 T}
       \int_{-\infty}^{+\infty} d\omega
	   \int_0^{\infty} d\qp \frac{\qp}{q}
       \int_{0}^\infty dp \,p^2
       \int_{\frac{q+\omega}{2}}^{\mu_k}
      \frac{dk}{k(k-\omega)}
   \nonumber \\ & \hspace {0.5cm} {}\times
	\bigg[(2k-\omega)^2\left\vert G^L_R(Q)\right\vert^2+\frac{\qp^4}{2q^4}
	(4k(k-\omega)-\qp^2)\left\vert G_R^T(Q)\right\vert^2\bigg]
   \nonumber \\ & \hspace {0.5cm} {}\times\sum_{a}T_{R_a}
  f^a_0(p) \, [1 \pm f^a_0(p)] \bigg[ \omega^2\big[ (\chi^{a}(p)')^2+(\chi^{g}(0)')^2\big]\nn\\
	&\hspace{-1cm}+\frac{\ell(\ell+1)}{2}\frac{\qp^2}{p^2}[\chi^{a}(p)]^2
	-2k(k-\omega)\left[P_\ell\left(1-\frac{\qp^2}{2k(k-\omega)}\right)-1\right](\chi^{g}(0)')^2\bigg]
   	 \, ,
\label {eq:tchannog}
\end{align}
where we have not used the ``finite'' label, as this equation
contains also power-law UV divergences. Indeed,
 performing the $k$ integral with $\mu_k\gg gT$
yields a linear-in-$\mu_k$ divergent term plus a finite
part, the latter responsible for the genuine, double-counted $\OO(g)$ contribution.
Keeping only the aforementioned finite contribution
 and dropping the
odd-in-$\omega$ terms we have
\begin{align}
\Big( \chi_\ij, \C^\twotwo_{\OO(g)\,g} \, \chi_\ij \Big)
    &=
       \frac{	d_A \ca g^4}{32 \pi^5 T}
       \int_0^{\infty} d\qp\, \qp
       \int_{-\infty}^\infty d\omega
       \int_{0}^\infty dp \,p^2  \sum_{a}T_{R_a}f^a_0(p) \, [1 \pm f^a_0(p)]
   \nonumber \\ & \hspace {-3cm} \times\bigg\{\bigg[
	   -2 \left\vert G^L_R(Q)\right\vert^2-\frac{\qp^4}{q^4}
	   	\left\vert G^T_R(Q)\right\vert^2 {}\left.+\frac{1}{2\omega q }\left(2\omega^2 \left\vert G^L_R(Q)\right\vert^2-\frac{\qp^6}{q^4}
	\left\vert G^T_R(Q)\right\vert^2\right)\ln\frac{q+\omega}{q-\omega}
\right]
   \nonumber \\ & \hspace {-3cm} {}\times
 \bigg[ \omega^2\big[ (\chi^{a}(p)')^2+(\chi^{g}(0)')^2\big]
  +\frac{\ell(\ell+1)}{2}\frac{\qp^2}{p^2}\big[(\chi^{a}(p))^2+(p\chi^{g}(0)')^2\big]\bigg]\nn\\
  &\hspace{-3cm}-\delta_{\ell 2}\frac{3\qp^2}{2\omega^2}\left[\frac{\qp^2}{2\omega q}
  \left(2\omega^2 \left\vert G^L_R(Q)\right\vert^2
  +\frac{\qp^4(\qp^2+2\omega^2)}{q^4}
  \left\vert G^T_R(Q)\right\vert^2\right)\ln\frac{q+\omega}{q-\omega}\right.\nn\\
  &\hspace{-1cm}\left.+2\omega^2 \left\vert G^L_R(Q)\right\vert^2-\frac{\qp^6}{q^4}
  \left\vert G^T_R(Q)\right\vert^2 \right](\chi^{g}(0)')^2\bigg\}
   	 \, .
\label {eq:tchannog2}
\end{align}
The terms on the second and third line contribute to both $\ell=1$ and $\ell=2$ (and to any $\ell$ in general),
whereas those on the final two lines, are specific for the $\ell=2$ case. We recall that in the diffusion case
gluons are in equilibrium, so that $\chi^g(p)=0$.
This expression is moreover still not UV finite. Indeed,
by using the bare propagators $G^{L(0)}_R(Q)=i/q^2$,
$G^{T(0)}_R(Q)=-i/\qp^2$
and performing the $\omega$ integrations we obtain
\begin{align}
	\Big( \chi_\ij, \C^\twotwo_{\OO(g)\,\mathrm{UV}\,g} \, \chi_\ij \Big) =&- \frac{d_AC_Ag^4\mu_\perp}{64\pi^3 T}
	\sum_a T_{R_a}
	 \int_{0}^\infty dp\,  f^a_0(p)[1\pm f_0^a(p)]\left\{
	\frac{\ell(\ell+1)}{2}[\chi^{a}(p)]^2\nn\right.\\
	&\hspace{4cm}\left.+\left(\frac{\ell(\ell+1)}{2}+\frac{9}{4}\delta_{\ell2}\right)[p\chi^{g}(0)']^2\right\}.
	\label{subtr}
\end{align}
As a consistency check, let us remark that the form of $\delta \qhat$
we have written in \Eq{qhatnlo} includes the \emph{finite part only}.
In its evaluation \cite{CaronHuot:2008ni}, Caron-Huot found also a linearly-divergent
part in $\mu_\perp$, which cancels against a corresponding term in the IR expansion of the hard
gluon exchange at NLO. Including such a term, \Eq{qhatnlo} turns into
\begin{equation}
	\delta \hat {q}=\frac{g^4C_R C_A T^2 }{32\pi}\left[-\mu_\perp+\md
	\frac{3\pi^2+10-4\ln 2}{\pi}\right].
	\label{nloqhat}
\end{equation}
When plugging its UV-divergent part in \Eq{cnloqhat} this agrees
with the transverse diffusion part of \Eq{subtr}. The calculation of $\delta\ql$ in \cite{Ghiglieri:2015ala},
on the other hand, does not contain
linear divergences in $\mu_\perp$, which also agrees with \Eq{subtr}.\footnote{
\label{foot_uv_ql}
That calculation contains a linear divergence in an analogue of $\mu_k$. These are related
to the discussion of App.~\ref{app_equiv}.}

We can then subtract the bare, UV-divergent contribution \Eq{subtr} from
\eqref{eq:tchannog2}.
The resulting $d\omega$ integrations do not seem doable by means of analyticity
techniques.\footnote{
It is possible to do some manipulations so that some terms become amenable to
analytical methods,
 but others remain non-analytical
due to branch cuts on the imaginary axis starting at $\omega=\pm i\qp$.}
Upon numerical integration\footnote{%
In this case there is no coordinate ambiguity, contrary to what was encountered
in footnote~\ref{foot_gaincoords}.} we obtain
\begin{eqnarray}
	\Big( \chi_\ij, \C^\twotwo_{\OO(g)\,\mathrm{finite}\,g} \, \chi_\ij \Big) &=&
	\frac{d_AC_Ag^4 \md}{32\pi^5 T}\sum_a T_{R_a}
	  \int_{0}^\infty dp\, p^2\,f^a_0(p)[1\pm f_0^a(p)]\nn\\
	  &&\hspace{-2cm}
	  \times\bigg\{4.2695\big[(\chi^{a}(p)')^2+(\chi^{g}(0)')^2\big]
	  +7.1769\frac{\ell(\ell+1)}{2p^2}\big[(\chi^{a}(p))^2+(p\chi^{g}(0)')^2\big]\nn\\
	  &&\hspace{5cm}+
	  18.0669\,\delta_{\ell2}[\chi^{g}(0)']^2
	  \bigg\}.
	\label{finalog}
\end{eqnarray}
In Sec.~\ref{sec_results} we needed the $(\chi(p))^2$ part of \Eq{finalog}, i.e.
\begin{align}
  	\Big( \chi_\ij, \C^\twotwo_{\OO(g)\,\mathrm{finite}\,\qhat} \, \chi_\ij \Big) =&
  	\frac{d_AC_Ag^4 \md}{32\pi^5 T}7.1769\sum_a T_{R_a}
	\int_{0}^\infty dp\,f^a_0(p)[1\pm f_0^a(p)]\nn\\
		\label{finalogqhat}
	&\hspace{3.1cm}\times\frac{\ell(\ell+1)\,\big[(\chi^{a}(p))^2\big]}{2},
\end{align}
Upon comparing with \Eq{cnloqhat} we see that
$( \chi_\ij, \C^\twotwo_{\OO(g)\,\mathrm{finite}\,\qhat} \, \chi_\ij )$
is approximately 1/8 of $( \chi_\ij, \C^{\delta\qhat} \, \chi_\ij )$.

Let us now look at the fermion exchange processes, i.e.\ Compton scattering
and $q\bar q$ annihilation.
We start again from \Eq{22general} and
we need to expand for $\omega,q,p\sim gT$, $k\sim T$. In both cases
there will also be an equivalent contribution for $p\sim T$, $k\sim gT$.
The deviation from equilibrium for Compton processes becomes
\begin{equation}
    	\left[
    	    \chi^q_\ij(\p) + \chi^g_\ij(\k) - \chi^g_\ij(\p') - \chi^q_\ij(\k')
    	\right]^2_{\mathrm{soft}\,p}
	 =(\chi^g(k)-\chi^q(k))^2+(\chi^q(0))^2+\OO(g).
	 \label{offeqgq}
 \end{equation}
 The annihilation case is equivalent.
As we shall see in App.~\ref{sub_ir}, 
in the $\ell=1$ case the quark departure from
equilibrium approaches a constant at LO in the IR, due  to the action of the 
$\onetwo$ processes, while it vanishes for $\ell=2$, 
so that the $(\chi^q(0))^2$ term needs to be considered only when computing
quark number diffusion. We have also neglected gain terms of the form
$\chi^q(k)\chi^q(0)$.
Given the $\hat\p\cdot\hat\k$-  (and hence $\phi$-)
independence of that expression, we can directly compute 
the $\phi$-averaged expansion of Eqs.~\eqref{softcompton} and \eqref{softann}, which is
\begin{eqnarray}
	&&\hspace{-5mm}\nn\int_0^{2\pi}\frac{d\phi}{2\pi}	\left\vert\mathcal{M}_{qg}^{qg}\right\vert^2_{
	\mathrm{soft\,q}\,t}=
	\int_0^{2\pi}\frac{d\phi}{2\pi}	\left\vert\mathcal{M}_{qg}^{qg}\right\vert^2_
	\mathrm{soft\,q}=-
	\frac{8 d_F C_F^2g^4 k}{q^2}\left\{(p+\omega)\left[(\omega-q)^2S^+_R(Q)S^+_A(Q)
	\right.\right.\\
&&\left.\left.+(\omega+q)^2S^-_R(Q)S^-_A(Q) \right]-\frac12\left[(\omega-q)^3S^+_R(Q)S^+_A(Q)
+(\omega+q)^3S^-_R(Q)S^-_A(Q)\right]
\right\},
	\label{softmatqog}
\end{eqnarray}
so that the $\mathcal{O}(g)$ contribution from soft $p$ becomes, summing the Compton
and annihilation contributions
\begin{eqnarray}
	\nn \Big( \chi_\ij, \C^\twotwo_{\OO(g)\,\mathrm{soft}\,p} \, \chi_\ij \Big) &=& -\frac{ d_F C_F^2\nf g^4  }{16\pi^5T^2}
	\int_0^\infty \frac{d\qp\,\qp}{q^3}\int_{-\infty}^\infty d\omega \int_{0}^\infty dk\,k
	\int_{\frac{q-\omega}{2}}^{\mu_p} \frac{dp }{2(p+\omega)}
\\
	\nn &&\times  \left\{(p+\omega)\left[(\omega-q)^2S^+_R(Q)S^+_A(Q)
	+(\omega+q)^2S^-_R(Q)S^-_A(Q) \right]\right.\\
\nn&&\left.-\frac12\left[(\omega-q)^3S^+_R(Q)S^+_A(Q)
+(\omega+q)^3S^-_R(Q)S^-_A(Q)\right]
\right\}\\
&&\times f^q_0(k)	[1+ f_0^g(k)]\big[(\chi^q(k)-\chi^g(k))^2+(\chi^q(0))^2\big],
	\label{softkfermionstart}
\end{eqnarray}
where we have included the factor of $8 \nf$ to account for the initial and final state
 symmetries,  the antiquark contribution in the Compton
 case and the $u$-channel contribution in the annihilation case. In the $\ell=2$ case
we have used the fact that $\chi^q=\chi^{\bar q}$, whereas in the $\ell=1$ case
we have used the fact that $\chi^g=0$ to sum the quark and antiquark contributions.
We can perform the $dp$ integration with a UV cutoff and discard
linearly divergent terms as in the gluon exchange case. Keeping only the terms
that are even in $\omega$ we get to
\begin{align}
	\nn  \hspace{-1mm} \Big( \chi_\ij, \C^\twotwo_{\OO(g)\,\mathrm{soft}\,p} \, \chi_\ij \Big) &=
	 \frac{ d_F C_F^2\nf g^4  }{64\pi^5T^2}
\int_{0}^\infty dk\,k\,
 f^q_0(k)	[1+ f_0^g(k)]\big[(\chi^q(k)-\chi^g(k))^2{+}(\chi^q(0))^2\big]\\
\nn &\hspace{-2cm}\times \int_{-\infty}^\infty d\omega	\int_0^\infty \frac{d\qp\,\qp}{q^3}  \left\{q\left[(\omega-q)^2S^+_R(Q)S^+_A(Q)
	+(\omega+q)^2S^-_R(Q)S^-_A(Q) \right]\right.\\
	&\hspace{-1.5cm}\left.-\tanh^{-1}\left(\frac{\omega}{q}\right)
	\left[(\omega-q)^3S^+_R(Q)S^+_A(Q)
+(\omega+q)^3S^-_R(Q)S^-_A(Q)\right]
\right\}.	\label{softkfermion2}
\end{align}
The two-dimensional $\omega,\qp$ integration is finite, as expected, since there would be
nothing to absorb UV divergences otherwise, given that the $\OO(g)$
correction to the conversion rates is free of linear UV divergences in
the transverse integrals.\footnote{It has UV divergences similar to those
discussed in footnote~\ref{foot_uv_ql}.} The numerical integration yields
\begin{align}
\Big( \chi_\ij, \C^\twotwo_{\OO(g)\,\mathrm{soft}\,p} \, \chi_\ij \Big) =& \,\frac{ d_F C_F^2\nf g^4
m_\infty  }{64\pi^5T^2}\,9.95268\,
\int_{0}^\infty dk\,k\,
 f^q_0(k)	[1+ f_0^g(k)]\nn\\
 &\hspace{3.5cm}\times\big[(\chi^q(k)-\chi^g(k))^2+(\chi^q(0))^2\big]\,.
	\label{softkfinal}
\end{align}
This was just the contribution from having $p$ soft and $k$ hard.
The opposite case gives the same
contribution, as can be inferred from the symmetries of the integrand, so that
the total double-counted contribution amounts to 2 times \Eq{softkfinal}, i.e.\
\begin{align}
\Big( \chi_\ij, \C^\twotwo_{\OO(g)\,\mathrm{finite}\,q} \, \chi_\ij \Big) =& \,\frac{ d_F C_F^2\nf g^4
m_\infty  }{32\pi^5T^2}\,9.95268\,
\int_{0}^\infty dk\,k\,
 f^q_0(k)	[1+ f_0^g(k)]\nn\\
 &\hspace{3.5cm}\times\big[(\chi^q(k)-\chi^g(k))^2+(\chi^q(0))^2\big]\,.
	\label{softktot}
\end{align}

\section{Equivalence of semi-collinear implementations}
\label{app_equiv}

In subsection \ref{sec:reorg} we argued that the semi-collinear
regions and NLO contributions to longitudinal diffusion and identity
change could all be treated simultaneously by evaluating the
semi-collinear corrections without approximating $p\q \ll \bh$ and
without IR regulation.  Here we verify this claim.
We also analyze in greater detail the IR form
of the $\onetwo$ processes and its consequences on the
departures from equilibrium in Sec.~\ref{sub_ir}.

\subsection{IR limits}
\label{sub_ir}
Let us start from examining the IR behavior of the
the $\onetwo$ rate given by \Eq{onetworate},
which determines the IR tail of the  departures
from equilibrium. To do so, let us start from
 the single soft scattering (Bethe-Heitler)
limit of the $\onetwo$ rate.
It can be easily obtained by solving \Eq{defimplfull}
by substitution, as shown in \Eq{Fseries}, 
under the assumption that $\delta E$ is much
larger than the effect of collisions.
We then have
\begin{eqnarray}
\nn	\gamma^{a}_{bc}\bigg\vert_\mathrm{BH}(p;p-k,k)&=&\frac{g^2}{32\pi^4}\left\{
\begin{array}{cc}
	d_AC_A\frac{p^4+k^4+(p-k)^4}{p^3k^3(p-k)^3}& g\leftrightarrow gg\\
	d_F C_F \frac{p^2+(p-k)^2}{p^2(p-k)^2k^3} & q\leftrightarrow q g\\
	d_F C_F\frac{(p-k)^2+k^2}{(p-k)^2k^2p^3} & g\leftrightarrow q\bar q
\end{array}
\right.
\int\frac{d^2h}{(2\pi)^2}
\int\frac{d^2\qp}{(2\pi)^2}\bar C(\qp)\\
&&\nn\hspace{-3cm}\times\left[\left(C_{R_b}-\frac{C_A}{2}\right)\left(\frac{\bh}{\delta E(\bh)}-
\frac{\bh-k\bqp}{\delta E(\bh-k\bqp)}\right)^2+\frac{C_A}{2}\left(\frac{\bh}{\delta E(\bh)}-
\frac{\bh+p\bqp}{\delta E(\bh+p\bqp)}\right)^2\right.\\
&&\left.+\frac{C_A}{2}\left(\frac{\bh}{\delta E(\bh)}-
\frac{\bh-(p-k)\bqp}{\delta E(\bh-(p-k)\bqp)}\right)^2\right],
	\label{betheheitler}
\end{eqnarray}
where we remind that the $g\lra q\bar q$ process has $\cf-\ca/2$
multiplying the second, rather than the first, term in square brackets.
Let us first remark that for generic $p,k,p-k\sim T$ \Eq{betheheitler}
is not, parametrically, a good approximation to \Eq{onetworate}, since
it is missing the relative $\OO(1)$ effect of LPM suppression. On the other hand,
in the region of interest, i.e.\
when the final states $k$ or $p-k$ become soft, it becomes accurate,
 as LPM suppression becomes
negligible. (It is easy to see that, in that limit, the effect of $\delta E$
in \Eq{defimplfull} does become much larger.)
Hence, in the soft gluon radiation limit for $k\to 0$ we can
reduce \Eq{betheheitler} to leading order in $k$ as
\begin{align}
	\nn\gamma^{a}_{ag}\bigg\vert_\mathrm{BH}(p;p,k)=&\frac{g^2d_{A}C_AT_{R_a}p^2}{64\pi^8k}
\int d^2h'd^2\qp \bar C(\qp)\left(\frac{\bh'}{h'^2+\mmg}-
\frac{\bh'+\bqp}{(\bh'+\bqp)^2+\mmg}\right)^2\\
=&\frac{g^4d_{A}C_AT_{R_a}p^2T}{32\pi^6k}\ln\left(\frac{e^2}{2}\right),
	\label{betheheitlersoftg}
\end{align}
where $h'\equiv h/p$ and the two transverse integrations are finite, as shown.
 When plugged in the relevant corners of \Eq{eq:C12}, it
turns it into
\begin{eqnarray}
	\nn\left(\chi_\ij,\mathcal{C}^{1\leftrightarrow 2}
	_{\mathrm{soft}\,g}\chi_\ij\right)
	&=&\frac{g^4d_AC_A}{8\pi^5T}\ln\left(\frac{e^2}{2}\right)
	\sum_{a}T_{R_a}\int_0^\infty dp\int_0^{\mu_k} dk\,p^2\,
	f^a_0(p)[1\pm f^a_0(p)]\\
	&&\hspace{6.3cm}\times\left[\chi^a(p)'-\chi^g(0)'\right]^2,
	\label{bhcollsoftg}
\end{eqnarray}
where we have introduced a factor of 2 to accounts for the $k\sim gT$ and
$p-k\sim gT$ corners in the $g\lra gg$ process
and for the final state symmetry in the $q\lra qg$ process. We have furthermore
assumed $\chi^g(k\to 0)=k\chi^g(0)'$. That is because, even though the $k$ integration (with
$gT\ll\mu_k\ll T$) might seem finite, since the soft $k$ expansion
of the departures from equilibrium in \Eq{eq:C12} yields a factor of $k^2$
which compensates the $1/k$ in the rate and the $1/k$ from the Bose distribution,
 one should however recall that in writing the quadratic functional in the form
of \Eq{eq:C12} we have performed a symmetrization by shifting
some integrations, which is allowed only as long as these
integrations are finite. If one were to work with
 $(\mathcal{C}^\onetwo\chi_{ij})^a(p)$,
entering in \Eq{eq:Boltz1}, in the soft gluon limit, i.e.\
 $(\mathcal{C}^\onetwo\chi_{ij})^g(p\to 0)$, one would see
a fixed point arising, enforcing
\begin{equation}
	\label{irfixedpoint}
	\sum_{a}T_{R_a}\int_0^\infty dk\,k^2\,
		f^a_0(k)[1\pm f^a_0(k)]\left[\chi^a(k)'-\chi^g(0)'\right]=0\,,
\end{equation}
i.e.\ giving rise to a boundary term which indeed forces a linear behavior
for the gluonic departure from equilibrium in the IR (see also \cite{Hong:2010at}).
The variational solution of the  LO quadratic functional (Eqs.~\eqref{eq:Q1} and \eqref{eq:Q2})
 is sensitive to this effect: with an
Ansatz that allows a single test function with a linear IR behavior, one sees that its coefficient
approximately satisfies \Eq{irfixedpoint}. 

For what concerns the quark departure from equilibrium, one has for
a soft quark
\begin{equation}
	\gamma^{a}_{qc}\bigg\vert_\mathrm{BH}(p;p-k,p)=\frac{g^2d_F
 C_F^2p}{64\pi^8}
\int d^2h'd^2\qp \bar C(\qp)\left(\frac{\bh'}{h'^2+\mmf}-
\frac{\bh'-\bqp}{(\bh'-\bqp)^2+\mmf}\right)^2.
	\label{betheheitlersoftq}
\end{equation}
The analysis of $(\mathcal{C}^\onetwo\chi_\ij)^q(p\to 0)$
then shows that in the $\ell=2$ case
\begin{equation}
	\label{irfixedpointql2}
\int_0^\infty dk\,k\,
		f^q_0(k)[1+ f^g_0(k)]\left[\chi^g(k)'+\chi^q(k)'-2\chi^q(0)'\right]=0\,,
\end{equation}
so that a linear behavior is enforced for $\chi^q(p\to0)$. In the $\ell=1$ case one
has instead
\begin{equation}
	\label{irfixedpointql1}
\int_0^\infty dk\,k\,
		f^q_0(k)[1+ f^g_0(k)]\left[\chi^q(k)-\chi^q(0)\right]=0\,,
\end{equation}
which enforces a constant behavior. Again, these constraints are approximately
satisfied by the LO variational solution.

Let us now look at the semi-collinear implementation
in \Eq{seminew}. In the soft gluon radiation limit one has to replace
$\bar C(\qp)$ with $\delta \bar C(\qp,\delta E)$ in \Eq{betheheitlersoftg},
yielding\footnote{%
The leading region for $k\to 0$ for the $\bar C_\NLO$
part in $\delta\bar C(\qp,\delta E)$ in the two transverse integrations is $\delta E\sim \qp$.}
\begin{equation}
	\gamma^{a}_{ag}\bigg\vert_\mathrm{semi}(p;p,k)=-\gamma^{a}_{ag}\bigg\vert_\mathrm{BH}(p;p,k)+
	\frac{g^4d_AC_AT_{R_a}  k p^2T}{24\pi^6\md^2}+\OO(k^3),
	\label{seminewsoftg}
\end{equation}
where the negative $1/k$ contribution arises from the subtracted collinear limit
in $\delta \bar C(\qp,\delta E)$. It would seem that, once \Eq{seminewsoftg}
 is plugged into the quadratic functional, it would generate a contribution opposite
 to \Eq{bhcollsoftg}, canceling it and removing
the linear behavior for the ($\ell=2$) infrared gluonic departure from equilibrium.
However, \Eq{seminewsoftg} is valid when $k\sim g^2T$. It is easy to see
from Eqs.~\eqref{tj}  that the
LO contribution comes from $k\sim T$, $\chi^a(p\sim T)\sim T^{\ell-1}/g^4$.
The soft region ($k\siml gT$) has a large phase-space suppression, so that,
even accounting for the Bose enhancement of soft gluons and the
linear or constant LO form of $\chi^a(p\to0)$, it always contributes
beyond NLO for all transport coefficients under consideration. Hence,
we only need to know the functional form of
the deviations from equilibrium no further down than $T\gg k\gg gT$, and one
can show that the semi-collinear implementation
in \Eq{seminew} does not alter the linear behavior found at NLO 
in that region, so that we may keep the functional form
given in \Eq{Ansatz} for the test functions.

For quarks one has instead
\begin{equation}
	\gamma^{a}_{qc}\bigg\vert_\mathrm{semi}(p;p-k,p)=
	-\gamma^{a}_{qc}\bigg\vert_\mathrm{BH}(p;p-k,p)+\OO((p-k)^2),
	\label{seminewsoftq}
\end{equation}
which is equally valid only for $k\sim g^2T$. The behavior at the 
interface $T\gg k\gg gT$ remains unaltered in this case too.

\subsection{Equivalence}
\label{sub_equiv}
Let's look at \Eq{semicoll}.
The leading-order contribution to it would naively come from the strictly
collinear scaling, i.e.\ $\qp\sim gT$, $h\sim gT^2$, $p,k,(p-k)\sim T$. This however
implies that $\delta E(\bh)\sim g^2T$ and that it can thus
safely be dropped from the denominators
in the collision kernel in \Eq{CNLO}, as we have argued in Sec.~\ref{sec:reorg}, i.e.\
\begin{equation}
\bar \delta C(\qp,\delta E)=
g^2 T\bigg[\frac{\md^2}{(\qp^2+\delta E^2)
(\qp^2+\delta E^2+\md^2)}
-\frac{\md^2}{\qp^2
(\qp^2+\md^2)}\bigg]\stackrel{\delta E\sim g^2T}{=}\OO\left(\frac{g^2}{ T}\right),
\label{transsemivanish}
\end{equation}
which, when plugged into \Eq{seminew}, makes it of order $g^6$ and hence
beyond NLO.

At relative $\OO(g)$, three regions contribute.
These are
\begin{enumerate}
	\item The \emph{diffusion} region, where a final-state gluon becomes soft. There,
	assuming $k$ is the gluon's momentum, $p\sim T$, $k\sim gT$  and $h/T\sim\qp\sim gT$.
	\item The analogous \emph{conversion} region,  where a final-state quark
	(or antiquark) becomes soft. Assuming now $p-k$ is the quark's momentum,
	the scaling is the same: $p,k\sim T$, $p-k\sim gT$  and $h/T\sim\qp\sim gT$.
	\item The \emph{semi-collinear} region, where $p,k,(p-k)\sim T$, $h\sim\sqrt{g}T^2$
	and $\qp\sim gT$.
\end{enumerate}
$g\leftrightarrow gg$ and $q\leftrightarrow qg$ processes contribute to the diffusion region.
Upon accounting for the $k$ and $p-k$ soft regions in the all-glue case
and for the final state symmetry in the $q\leftrightarrow qg$ case we have
\begin{eqnarray}
	\nn\left(\chi_\ij,\mathcal{C}^{\mathrm{semi}}_\mathrm{diff}\chi_\ij\right)
	&=&\frac{d_AC_Ag^2}{\pi^3T^2}\sum_{a}T_{R_a}\int_0^\infty dp\int_0^{\mu_k} dk\,
p^2\,f^a_0(p)[1\pm f^a_0(p)]\left[\chi^{a}(p)'-\chi^g(0)'\right]^2
\\
&&\hspace{-2.5cm}\times\int\frac{d^2h'}{(2\pi)^2}\int\frac{d^2\qp}{(2\pi)^2}\delta \bar C(\qp,\delta E_d(\bh'))
\left(\frac{\bh'}{h'^2+\mmg}-
\frac{\bh'-\bqp}{(\bh'-\bqp)^2+\mmg}\right)^2,
	\label{semicollsoft2}
\end{eqnarray}
where we have again rescaled $h=h'p$ and $\delta E_d(\bh')\equiv(h'^2+\mmg)/(2k)$
is the diffusion (soft gluon) limit of \Eq{defdeltaE}.
Following the arguments of the previous section, we have kept a linear
 $\chi^g(k<\mu_k)\approx k\chi^g(0)'$
term in the square brackets on the first line.
The $k$ integration to the cutoff $gT\ll \mu_k\ll T$ yields  cutoff-linear
and  cutoff-independent terms, i.e.
\begin{align}
\int_0^{\mu_k}dk\,\delta \bar C(\qp,\delta E_d(\bh'))=
g^2 T\bigg[&\frac{\mu_k \md^2}{\qp^2
(\qp^2+\md^2)}+\pi\frac{h'^2+\mmg}{4}\left(\frac{1}{(\qp^2+\md^2)^{3/2}}+\frac{1}{\qp^3}
\right)\nn\\
&-\frac{\mu_k \md^2}{\qp^2
(\qp^2+\md^2)}-\pi\frac{h'^2+\mmg}{2\qp^3}
\label{kintcutoff}
\bigg],
\end{align}
where the terms on the first line arise from the $\bar C_\NLO$ terms in \Eq{CNLO},
whereas those on the second line result from the subtracted collinear and hard pieces
of \Eq{CNLO} respectively.
Hence, the linearly divergent piece cancels out and
\begin{align}
	\nn&\left(\chi_\ij,\mathcal{C}^{\mathrm{semi}}_\mathrm{diff}\chi_\ij\right)
	=\frac{d_AC_Ag^4}{4\pi^2T}\sum_{a}T_{R_a}\int_0^\infty dp\,
p^2\,f^a_0(p)[1\pm f^a_0(p)]\left[\chi^{a}(p)'-\chi^g(0)'\right]^2
\\
&\times\int\frac{d^2h'}{(2\pi)^2}
\int\frac{d^2\qp}{(2\pi)^2}\left(h'^2{+}\mmg\right)
\left(\frac{1}{(\qp^2{+}\md^2)^{3/2}}-\frac{1}{\qp^3}
\right)
\left(\frac{\bh'}{h'^2{+}\mmg}-
\frac{\bh'-\bqp}{(\bh'-\bqp)^2{+}\mmg}\right)^2.
	\label{semicollsoft2b}
\end{align}
This can be further simplified with a few manipulations
in the $h'$ integral. By shifting it to $\bh'\to\bh'+\bqp$, many terms in the
final round brackets either cancel one another or vanish in the azimuthal integration, leaving us with
\begin{align}
	\nn\left(\chi_\ij,\mathcal{C}^{\mathrm{semi}}_\mathrm{diff}\chi_\ij\right)
	=&\frac{d_AC_Ag^4}{4\pi^2T}\sum_aT_{R_a}\int_0^\infty dp\,
p^2\,f^a_0(p)[1\pm f^a_0(p)]\left[\chi^{a}(p)'-\chi^g(0)'\right]^2
\\
&\times\int\frac{d^2h'}{(2\pi)^2}\int\frac{d^2\qp}{(2\pi)^2}\frac{h'^2\qp^2}{(h'^2+\mmg)^2}
\left(\frac{1}{(\qp^2+\md^2)^{3/2}}-\frac{1}{\qp^3}\right)\nn\\
=&\frac{d_AC_Ag^4\md}{4\pi^2T}
\ln\left[\frac{\sqrt{e}M_{\infty}}{\mu_\perp^\NLO}\right]
\sum_aT_{R_a}\int_\p
f^a_0(p)[1\pm f^a_0(p)]\left[\chi^{a}(p)'-\chi^g(0)'\right]^2 \! ,
\label{semicollsoft3}
\end{align}
where the $d^2h'$ integration has been regulated with a $gT\ll\mu_\perp^\NLO
\ll\sqrt{g}T$
UV cutoff.\footnote{This might seem to be conflicting with our previous shift.
However, let us point out that the transverse integrations in \Eq{seminew}
are finite and that, if the shift were performed there, no effect would be observed.}
Upon accounting for the factor of $2\nf$ from
the sum over $a$ in the $q\leftrightarrow q g$, the $[\chi^a(p)']^2$-proportional
part of \Eq{semicollsoft3} agrees with the contribution
one would obtain by plugging $\delta \ql$, as given
by \Eq{qlnlo}, into \Eq{diffterms}. For what concerns the terms
proportional  to $\chi^g(0)'$ in the $\ell=2$ case, it can be shown
that the coefficient $c_1$ in \Eq{gainterms}
contains a term proportional to $\ql$ (just rewrite
$P_\ell(\ph\cdot\kh)$ as $1+(P_\ell(\ph\cdot\kh)-1)$
and compare with \Eq{qhatintegral}). It reads
 \begin{align}
\left. (\chi_\ij, \C_{\mathrm{\lossshort}} \chi_\ij)\right|_{{\rm gain},\ql}
    &=  - \beta^4  \, \sum_{ab} \ql^a \frac{g^2 \nu_a  \nu_b C_{R_b}}{2 \md^2  d_A}
	  \int_{\p}
	 \int_0^\infty dk\,k^2\,  
	 f_0^a(p) (1 \pm f_{0}^a(p))\nn \\
	 &\hspace{4cm} f_0^b(k) (1 \pm f_{0}^b(k))   
    \chi^{a}(p)'\chi^{b}(k)' \,.
  \label{gainterms2}
 \end{align}
When substituting $\ql$ with $\delta \ql$, as given
by \Eq{qlnlo}, and using \Eq{irfixedpoint},
this can be brought into agreement with the terms
proportional to to $\chi^g(0)'$ in \Eq{semicollsoft3}.

In the conversion region the relevant processes are the
$q\leftrightarrow qg$  and $g\leftrightarrow q\bar q$ ones. In the latter one
there is an identical contribution from $k\sim gT$. It is easy to see that the resulting
contribution is altogether similar to what we just found for the diffusion limit,
yielding
\begin{eqnarray}
	\nn\left(\chi_\ij,\mathcal{C}^{\mathrm{semi}}_\mathrm{conv}\chi_\ij\right)
	&=&\frac{\nf d_FC_F^2g^2}{8\pi^3T^3}\int_0^\infty dp\int_0^\mu dk\,
\frac{p}{k^2}f^q_0(p)[1+ f^g_0(p)]\left[\chi^{q}(p)-\chi^{g}(p)-\chi^q(0)\right]^2
\\
&&\hspace{-1.5cm}\times\int\frac{d^2h'}{(2\pi)^2}\int\frac{d^2\qp}{(2\pi)^2}
\delta \bar C(\qp,\delta E_c(\bh'))\left(\frac{\bh'}{\delta E_c(\bh')}-
\frac{\bh'-\bqp}{\delta E_c(\bh'-\bqp)}\right)^2,
	\label{semicollconv}
\end{eqnarray}
where we have relabeled $k$ to be the soft quark's momentum and
 $\delta E_c(\bh')\equiv(h'^2+\mmf)/(2k)$ is the conversion
 limit of \Eq{defdeltaE}. This then results in
\begin{align}
	\left(\chi_\ij,\mathcal{C}^{\mathrm{semi}}_\mathrm{conv}\chi_\ij\right)
&=\frac{\nf d_FC_F^2g^4\md}{2T^2}\ln\left(\frac{\sqrt{e}m_{\infty}}{\mu_\perp^\NLO}\right)
\int_\p
\frac{f^q_0(p)
[1+ f^g_0(p)]}{p}\left[\chi^{q}(p)-\chi^{g}(p)\right.\nn\\
&\hspace{9cm}\left.-\chi^q(0)\right]^2,
\label{semicollconv3}
\end{align}
which agrees with the contribution that would arise form inserting
$\delta\GammaC$, as given by \Eq{nloconvrate},
in \Eq{eq:conversion-loss2}. For what concerns the terms proportional
to $\chi^q(0)$ in the $\ell=1$ case, they match those obtained
from \Eq{eq:gainconv} through \Eq{irfixedpointql1}:
as in the case of \Eq{gainterms2}, we can rewrite $P_1(\ph\cdot\kh)$
as $1+(P_1(\ph\cdot\kh)-1)$ in \Eq{eq:c1fermion}. 

Finally, as we mentioned in Sec.~\ref{sec:reorg} in the main text,
in the semi-collinear region, where
$p,k,(p-k)\sim T$, $\qp\sim gT$ and $h\sim \sqrt{g}T^2$,
\Eq{seminew} can be expanded back yielding, at $\OO(g^5)$,
\Eq{semi1}. Subleading terms in the expansion contribute at higher orders.
This completes the proof of equivalence of the two approaches.

We conclude by commenting on the relation of this new approach
with the contour sum rules used to obtain $\delta\ql$ and
$\delta\Gamma_\mathrm{conv}$ in \cite{Ghiglieri:2013gia,Ghiglieri:2015ala}.
Take for instance the computation of $\delta\ql$ described 
in App.~F of \cite{Ghiglieri:2015ala}. There we used the analytical
properties of the light-cone amplitudes to deform the contour of the $k$
integration (it is called $q^+$ there). In doing so, we encounter poles in 
the $q^-$ variable (called $k^-$ there) that can be pinched or not, of the form
	$1/(k^--\delta E)$ (see for instance (F.6) in \cite{Ghiglieri:2015ala}). 
	Upon deforming the integration contour and expanding for large, complex $k$ ($q+$),
	the non-pinched poles contribute (see (F.7)) to a $\delta(k^-)+\delta E/(k^-)^2$ structure
	(in the variables of \cite{Ghiglieri:2015ala}).
	The first term is responsible for the linear-in-$\mu_k$ term on the first line of
	\Eq{kintcutoff}, whereas the second gives rise to the $\mu_k$ independent term 
	(recall that $\delta E\propto 1/q^+$ in the variables of \cite{Ghiglieri:2015ala}).
	The same reasoning applies to the subtracted terms on the second line of \Eq{kintcutoff}.
	In obtaining \Eq{kintcutoff} we have essentially inverted the order of the $q^-$ ($k^-$ in
	\cite{Ghiglieri:2015ala})
	and $k$ ($q^+$) integrations. As a consequence, it is 
 important to note that, between the new and old approaches, the rates themselves are different as functions
	of $p$ and $k$ (in the coordinates of this paper), it is only their integral,
the collision operator, which agrees (at LO and NLO).

\section{Fits of the NLO results}
\label{app_fits}
In this section we will present fits that reproduce the NLO results
 by smoothly interpolating between the NLL behavior at small
values of $\md/T$ and the $\delta\qhat$-dominated one at the opposite end.
The former is given by \cite{Arnold:2003zc}
\begin{equation}
	\label{nll}
	\frac{g^4}{T^3}\eta_\mathrm{NLL}=\frac{\eta_1}{\ln(\mu_*/\md)},\qquad
	g^4 T D_{q\,\mathrm{NLL}}=\frac{D_1}{\ln(\mu_*/\md)},
\end{equation}
where $\eta_1$ and $D_1$ are the leading-log coefficients \cite{Arnold:2000dr}
 and $\mu_*$ is the next-to-leading-log one \cite{Arnold:2003zc}.

To obtain the $\delta\qhat$-dominated behavior at large $\md/T$
 we first briefly show that the collision operator composed uniquely by
\Eq{cnloqhat} can be inverted analytically. Let us define
\begin{equation}
    \mathcal{Q}^{\delta\qhat}[\chi]
    \equiv
    \Big( \chi_\ij, \S_\ij\Big)
    - \half \, \Big( \chi_\ij, \C^{\delta\qhat} \, \chi_\ij \Big),
	\label{deltaqmax}
\end{equation}
which is simply the limiting form of \Eq{eq:Q1} under the assumption that at large enough
values of $\md/T$ it becomes dominated by \Eq{cnloqhat}. Figures~\ref{fig_visc_md}
and \ref{fig_visc_quench_md} already provide a motivation for this assumption, which will be reinforced later on.

The maximization of $\mathcal{Q}^{\delta\qhat}[\chi]$ can be done by functional differentiation
with respect to the $\chi^g(p)$ and $\chi^q(p)$ (see \Eq{variation}), 
which, since the source term is linear in these (see \Eq{eq:Sij})
and the collision operator is quadratic, leads to a simple solution:
\begin{equation}
	 \chi^a(p)=\frac{4 p^{\ell+1}T}{\ell(\ell+1)\delta\qhat^a},
	 \label{qhatsol}
\end{equation}
where $\delta\qhat^a$, the $\OO(g)$ correction to $\qhat$, is given by \Eq{qhatnlo}.
Upon plugging this in \Eq{deltaqmax} and recalling that $\eta=2\mathcal{Q}_\mathrm{max}/15$,
$D_q=2 \mathcal{Q}_\mathrm{max}/(\nc T^2)$, we obtain
\begin{equation}
	\eta^{\delta\qhat}=\frac{16\pi^4T^6}{945}\left(\frac{2d_A}{\delta\qhat^g}
	+\frac{31}{32}\frac{4\nf d_F}{\delta\qhat^q}\right),\qquad
	D_q^{\delta\qhat}=\frac{14\pi^2 T^2}{15\,\delta\qhat^q},
	\label{transportasym}
\end{equation}
where the factor of $31/32$ arises from the fermionic, rather than bosonic, integrations, similarly to the
factor of $7/8$ in the Stephan-Boltzmann contribution of fermions to the pressure.

With these ingredients we can obtain simple fits for the NLO curves of Figures~\ref{fig_visc_md}
and \ref{fig_visc_quench_md} at $C_\ell=0$.
We fit the shear viscosity as
\begin{equation}
	\frac{g^4}{T^3}\eta^\mathrm{fit}_\mathrm{NLO}=\frac{\eta_1}{b^{-1}\ln\left(a+\left(\frac{\mu_*}{\md}\right)^b\right)+
	\frac{T^3 \eta_1}{g^4\eta^{\delta\qhat}}\frac{\md/T}{c+\md/T}+\frac{d}{(1+\md/T)^3}},
	\label{etafit}
\end{equation}
where $a$, $b$, $c$ and $d$ are fit parameters. As one can see, for small $\md/T$ the curve approaches the NLL
approximation~\eqref{nll},
while at large $\md/T$ they approach \Eq{transportasym}. Using the latter, the numerical values for
$\eta_1$ and $\mu_*$ from \cite{Arnold:2003zc} and fitting the parameters we obtain, for $\nc=3$
\begin{align}
	\nn \nf=3\;(\mathrm{Fig.}~\ref{fig_visc_md}):\quad&\eta_1=106.66,\,\mu_*/T=2.957,\,a=4.45096,\,
	b= 1.2732,\,c=1.91568,\\
	\label{viscfit3}
	&d= -0.0777985,\\
	\nf=0\;(\mathrm{Fig.}~\ref{fig_visc_quench_md}):\quad&\eta_1=27.126,\,\mu_*/T=2.765,\,a=8.5176,\,
	b= 1.38936,\,c=1.66144,\nn\\
	&d= -0.100421.
	\label{viscfitquench}
\end{align}
The fits are accurate to below 0.5\% for $\md/T<5$. We have tested that they remain below $2\%$ up to
$\md/T<10$.

With the same philosophy we can fit the NLO curve for $D_q$ (Fig.~\ref{fig_visc_md}) as
\begin{equation}
	g^4T\,D^\mathrm{fit}_{q\,\mathrm{NLO}}=\frac{D_1}{b^{-1}\ln\left(a+\left(\frac{\mu_*}{\md}\right)^b\right)+
	\frac{D_1}{g^4TD_q^{\delta\qhat}}\frac{\md/T}{c+\md/T}+\frac{d}{1+\md/T}},
	\label{dfit}
\end{equation}
with
\begin{align}
	\nn \nf=3\;(\mathrm{Fig.}~\ref{fig_visc_md}):\quad&D_1=11.869,\,\mu_*/T=2.949,\,a=1.33534,\,
	b= 1.28963,\,c=0.0378486,\\
	\label{difffit3}
	&d= -0.0769937.
\end{align}
The fit is accurate to 0.5\% for $\md/T<4$. We have tested that it remains below $4\%$ up to
$\md/T<10$.

\section{Running coupling prescriptions}
\label{app_coupling}
The two-loop QCD $\MS$ coupling is defined as the solution of
\begin{align}
	\label{qcdevol}
	\mu \frac{d}{d\mu}g^2_{{\sss\mathrm{QCD}}\,(\nf)}(\mu)=&\,\frac{\beta_0^{(\nf)}}{(4\pi)^2}g^4_{{\sss\mathrm{QCD}}\,(\nf)}(\mu)+
	\frac{\beta_1^{(\nf)}}{(4\pi)^4}g^6_{{\sss\mathrm{QCD}}\,(\nf)}(\mu)\,,\qquad m_q^{(\nf)}\le\mu\le m_q^{(\nf+1)}\,,\\
	g^2_{{\sss\mathrm{QCD}}\,(\nf)}\left(m_q^{(\nf+1)}\right)=&\,g^2_{{\sss\mathrm{QCD}}\,(\nf+1)}\left(m_q^{(\nf+1)}\right)\,,
	\label{threshold}
\end{align}
where $\beta_i^{(\nf)}$ are the coefficients of the QCD $\beta$-function with $\nf$ massless flavors. Thus,
\Eq{qcdevol} evolves the coupling with $\nf$ massless flavors from the mass scale of the heaviest of these,
$m_q^{(\nf)}$, up to the mass scale of the first heavier quark, $m_q^{(\nf+1)}$, where the one-loop threshold matching
\Eq{threshold} imposes continuity. For reference and to fix conventions
\begin{equation}
	\beta_0=\frac{-22\ca+4\nf}{3},\qquad\beta_1=\frac{-68\ca^2+20\ca\nf+12\cf\nf}{3}.
	\label{betafunction}
\end{equation}

The effective EQCD coupling reads instead \cite{Laine:2005ai}
\begin{equation}
	g^2_{{\sss\mathrm{EQCD}}\,(\nf)}(\mu)=g^2_{{\sss\mathrm{QCD}}\,(\nf)}(\mu)+\alpha_{\mathrm{E}7}^{(\nf)}
	\frac{g^4_{{\sss\mathrm{QCD}}\,(\nf)}(\mu)}{(4\pi)^2}+\gamma_{\mathrm{E1}}^{(\nf)}\frac{g^6_{{\sss\mathrm{QCD}}\,(\nf)}(\mu)}{(4\pi)^4}
	\,,
	\label{eqcdevol}
\end{equation}
where \cite{Huang:1994cu,Kajantie:1995dw,Laine:2005ai}
\begin{align}
	\nn\alpha_{\mathrm{E}7}=&-\beta_0\ln\left(\frac{\mu e^{\gamma_E}}{4\pi T}\right)+\frac{\ca}{3}-\frac{8}{3}\nf\ln\,2\,,\\
	\gamma_{\mathrm{E}1}=&-\beta_1\ln\left(\frac{\mu e^{\gamma_E}}{4\pi T}\right)+\alpha_{\mathrm{E}7}^2
	-\frac{1}{18}\bigg\{\ca^2\big[-341+20\zeta(3)\big]\nn\\
	&+2\ca\nf\big[43+24\ln\,2+5\zeta(3)\big]
	+3\cf\nf\big[23+80\ln\,2-14\zeta(3)\big]\bigg\}.
	\label{eqcdcoeffs}
\end{align}
\Eq{eqcdevol} holds for $m_q^{(\nf)}\le\mu\le m_q^{(\nf+1)}$. At the fermion thresholds we switch to the values
of the coefficients in \Eq{eqcdcoeffs} with $\nf\pm 1$. Hence, the EQCD coupling is not continuous at the thresholds.
Corrections to Eqs.~\eqref{qcdevol} and \eqref{eqcdevol} are of order $g^8$.

We conclude by noting that it is easy to see how, in the one-loop approximation, i.e.\ neglecting
$\beta_1$ and $\gamma_{\mathrm{E}1}$ in Eqs.~\eqref{qcdevol} and \eqref{eqcdevol}, the $\mu$ dependence
drops out of the EQCD coupling at order $g^4$.

\bibliographystyle{JHEP}
\bibliography{eloss.bib}

\end{document}